\documentclass[prd,amsmath,amssymb,aps,twocolumn,nofootinbib,floatfix]{revtex4-2}

\usepackage[dvipsnames]{xcolor}
\usepackage{graphicx}
\usepackage[colorlinks=true,citecolor=blue]{hyperref}
\usepackage{epsfig}
\usepackage{rotating}
\usepackage{slashed}
\usepackage{float}
\usepackage[normalem]{ulem} 
\usepackage[caption=false]{subfig}
\newcommand{\nn}{\nonumber}
\makeatletter
\renewcommand{\p@subsection}{}
\renewcommand{\p@subsubsection}{}
\def\l@subsubsection#1#2{}
\makeatother
\DeclareMathOperator\artanh{artanh}
\begin{document}
	
	
	\title{Response of an Unruh-DeWitt detector near an extremal black hole}
	
	\author{Aindri\'u Conroy}
	\email{aindriu.conroy@dcu.ie}
	\author{Peter Taylor}
	\email{peter.taylor@dcu.ie}
	\affiliation{Centre for Astrophysics and Relativity,
		School of Mathematical Sciences,
		Dublin City University,
		Glasnevin,
		Dublin 9, Rep. of Ireland.
	}
	\date{\today}
	
	\begin{abstract}
		We consider the response of an Unruh-DeWitt detector near an extremal charged black hole, modeling the near-horizon region of this extremal spacetime by the Bertotti-Robinson spacetime. The advantage of employing the Bertotti-Robinson limit is that the two-point functions for a massless scalar field are obtainable in closed form for the field in a number of quantum states of interest. We consider the detector coupled to a massless field in both the Boulware vacuum state and arbitrary thermal states, including the Hartle-Hawking state, and analyse the detector's response for a broad range of trajectories. Particular attention is paid to the thermalization of the detector, the anti-Unruh and anti-Hawking effect.
	\end{abstract}
	
	\pacs{}
	\maketitle
	\section{Introduction}
	Quantum field theory in curved spacetimes (QFTCS) remains one of the most robust approximations to a quantum theory of gravity which respects the standard paradigms of quantum mechanics and Einstein's GR. Quantum Field Theory on black hole spacetimes has played a particularly important role in this approximation to Quantum Gravity since both gravity and quantum effects turn out to be essential for an accurate description of the black hole's evolution. In particular, since Hawking's discovery \cite{Hawking:1975} that black holes evaporate by emitting a very low-energy quantum-stimulated radiation, the study of quantum effects near black holes has been a fruitful theoretical arena in which to explore signatures of quantum gravity. 
	
In QFTCS, there is no well-defined notion of a particle since, in general, global symmetries are absent. In particular one cannot, in general, identify a global time function to distinguish between positive frequency and negative frequency modes leading to an ambiguity in the particle concept. The standard procedure of QFTCS is to circumvent this particle ambiguity altogether by treating fields as the fundamental object of interest, not particles. However, in a seminal paper, Unruh \cite{Unruh:1976} offered a well-defined operational meaning to the concept of a particle by coupling a quantum field to a two-level idealized atom and considering the absorption and emission of field quanta by the atom. This is the so-called Unruh-DeWitt detector. In an operational sense then, we define a particle as what a particle detector detects. 

Application of particle detectors has been a very useful tool in QFTCS yielding two of the most well-known effects, that the spectrum of transitions is thermal for an accelerating detector in Minkowski spacetime with the temperature proportional to the acceleration \cite{Unruh:1976}, and also thermal for a static detector in Schwarzschild spacetime with the temperature proportional to the surface gravity of the black hole \cite{HartleHawking1976}. These of course are the Unruh effect and the Hawking effect, respectively. The Unruh-DeWitt particle detector model continues to provide new perspectives on these well-studied phenomena \cite{Fewster:2016ewy, SmerlakSingh, HodgkinsonLoukoOttewill, LoukoToussaint, JuarezAubry2018} as well as finding utility in the firewall proposal \cite{Louko2014, MartinezLouko} and quantum information \cite{Hu2012}. Moreover, the particle detector framework has recently facilitated the exploration of novel phenomena known as the anti-Unruh \cite{Brenna2016, Garay2016} and anti-Hawking effect \cite{Mann2020, Campos2021, CamposDappiaggi2021, robbins2021antihawking}. 
	
	In general, computing the response of an Unruh-DeWitt detector is technically challenging. The detailed dependence of the response on the switching function, which governs how the interaction between the field and the detector is switched on, presents some subtleties \cite{Schlicht2004, LoukoSatz2006, Satz2007, LoukoSatz2008}. Moreover, in the sharp-switching limit, the probability of registering a transition from one level to another diverges. For this reason it is easier to consider the rate of transition rather than the transition probability itself since the rate is finite in this sharp-switching limit. Even still, the transition rate is computationally challenging to compute, especially for non-stationary detectors. Mathematically, it involves integrating the Wightman Green function for the quantum scalar field over the history of the detector. When the two arguments of the Wightman Green function are evaluated at the same spacetime point, the distribution is singular and requires a regularization prescription to be rendered meaningful. While the Hadamard regularization scheme provides a general framework that solves this problem conceptually (see, for example, Refs.~\cite{WaldBook, FullingWald:1981}), there are still technical subtleties associated with its implementation. The problem is that while the short-distance singularity structure is easy to identify, the globally valid Wightman distribution is typically only known as a mode-sum representation with the modes often only solvable numerically. The Hadamard divergence in this mode-sum representation is not exhibited as a simple closed-form geometrical singularity, but is instead manifest as a mode-sum that no longer converges when the two spacetime points of the Wightman Green function coincide. Usually one proceeds by trying to express the Hadamard singularities as an appropriate mode-sum and subtracting from the Wightman distribution mode-by-mode. This procedure is often referred to as a mode-sum regularization prescription (MSRP); its implementation on black hole spacetimes originated with an influential paper by Candelas and Howard \cite{CandelasHoward:84}, though not in the context of Unruh-DeWitt detectors but of renormalized expectation values of field operators. More recent developments of mode-sum regularization prescriptions have proven to be very efficient numerically \cite{taylorbreen:2016, taylorbreen:2017, LeviOri:2015, LeviOri:2016, amoslevi:2016}. Nevertheless, using these methods to compute transition rates is computationally non-trivial which has limited the scope of this study in the literature. Certainly it is challenging to compare and contrast a large gamut of trajectories, quantum states and detector parameters in black hole spacetimes.
	
	From previous work on the response of an Unruh-DeWitt detector in black hole spacetimes in four dimensions, we know only of the cases of a detector on a static and circular-geodesic trajectories in Schwarzschild \cite{HodgkinsonLoukoOttewill} and Schwarzschild anti-de Sitter spacetimes \cite{Ng2014}, and the case of a static detector on massless topological black hole spacetimes \cite{CamposDappiaggi2021}, though there is considerable relevant work on the response of a particle detector in the three-dimensional BTZ black hole spacetime \cite{HodgkinsonLouko2012, Mann2020,  Campos2021, CamposDappiaggi2021, robbins2021antihawking}.
	
	In this work, in an attempt to explore a broader range of the parameter space, we consider an Unruh-DeWitt detector in the near-horizon region of an extremal charged black hole. In a certain approximation limit, there is an enhanced symmetry which allows one to map the problem to that of a scalar field on a Bertotti-Robinson spacetime. The advantage is that the Wightman distribution is known in closed-form which allows for an immediate exploration of the transition rate on arbitrary trajectories, different quantum states and arbitrary energy gap for our detector, all without recourse to significant numerical work. The disadvantage of course is that it is not clear how representative our results are of those of a particle detector in a generic Reissner-Nordstrom black hole spacetime. On the other hand, it is rare in black hole spacetimes to be able to so easily and fully explore the phenomenology associated with a particular effect so these results are novel and useful. In particular, we study the detector on a range of geodesic and accelerated trajectories for the field in both a vacuum state and thermal states. Moreover, by zooming in on the near-horizon throat region, our results may be indicative of what a detector registers in the late-stage plunge into the black hole in a more generic scenario, which is certainly one of the more interesting cases to consider. We further explore the parameter space to see in what regions, if any, the detector thermalizes, as well as whether or not the anti-Unruh or anti-Hawking effects are present.
	
	The layout of our paper is as follows. In Section \ref{sec:theory} we review the basic theoretical framework for the Unruh-DeWitt detector. In Section \ref{sec:BertottiRobinson} we examine the Bertotti-Robinson limit of the Reissner-Nordstrom black hole, as well as reviewing the two-point functions in this spacetime for a quantum field in several quantum states. In Section \ref{sec:rates}, we present and analyse the results of the transition rates for a broad range of trajectories and parameters. We finish with some discussion and conclusions in Section \ref{sec:conclusions}.
	
	\section{Particle Detector Model}
	\label{sec:theory}
	As a mathematical model for our particle detector, we consider a two-level idealized atom interacting with a massless quantum scalar field. Absorption of field quanta by the atom can promote the atom from ground state to excited state and we interpret this atomic excitation as a detector registering a particle. Conversely, the detector can de-excite by emitting quanta.
	
	To be more specific, let the interaction Hamiltonian between the detector (i.e., the two-level atom) and the quantum scalar field $\hat{\varphi}(x)$ be
	\begin{equation}
		H_{\textrm{int}}=\alpha\,\chi(\tau)\hat\mu(\tau)\hat\varphi(x(\tau)),
	\end{equation}
	where $x(\tau)$ are the coordinates of the detector's worldline with $\tau$ its proper time, $\hat{\mu}(\tau)$ is the detectors monopole moment operator, $\chi(\tau)$ is the switching function which governs how the interaction between the field and the detector is switched on and off, and $\alpha$ is the coupling strength of the interaction.
	
	Before the detector and the quantum field interact, we suppose that the field $\hat\varphi(x)$ is in some initial Hadamard state $|\Phi_{\textrm{i}}\rangle$ on a given background, while the detector is in its ground state $|E_{\textrm{i}}\rangle$. When interaction takes place, the field $\hat\varphi(x)$ transitions from its initial state $|\Phi_{\textrm{i}}\rangle$ to a final state $|\Phi_{\textrm{f}} \rangle$. Concurrently, the detector will undergo a transition from ground state $|E_{\textrm{i}}\rangle$ to excited state $|E_{\textrm{f}}\rangle$. The probability of this transition occurring will not depend on the individual eigenvalues of these energy states but only on the difference $\omega=E_{\textrm{f}}-E_{\textrm{i}}$. When $\omega>0$, the detector has absorbed a field quanta and is in an excited state, while for $\omega<0$ the detector has de-excited by emitting a field quanta. If we further assume that the coupling strength $\alpha$ is small, then we can treat the interaction as a small perturbation around the free Hamiltonian. To first order in this perturbative expansion, and tracing over the field degrees of freedom since we are only interested in measuring the detector's state, the probability of measuring the detector in an excited state $|E_{\textrm{f}}\rangle$ is
	\begin{equation}
		\label{transitionprob}
		P(\omega)= \alpha^2\rvert\langle E_{\textrm{f}}|\hat{\mu}(0)|E_{\textrm{i}}\rangle\rvert^2 {\cal F}(\omega),
	\end{equation}
where $\mathcal{F}(\omega)$ is known as the response function and is given by
	\begin{align}
		\label{eq:response}
		\mathcal{F}(\omega)=2&\,\lim_{\epsilon\to 0^{+}} \Re \int_{-\infty}^{\infty}du\,\chi(u)\nonumber\\
		& \times\int_{0}^{\infty}ds\,\chi(u-s)e^{-i \,\omega\,s}W_{\epsilon}(u, u-s).
	\end{align}
The bi-distribution $W_{\epsilon}(u,u-s)\equiv W_{\epsilon}(x(u),x(u-s))$ appearing in this integral is a one-parameter family of Wightman Green functions for the quantum scalar field evaluated at the spacetime points $x=x(u)$ and $x'=x(u-s)$. There is an implicit `$i\,\epsilon$' prescription in this expression which is required to render the Wightman Green function a well-defined distribution on the lightcone (see, for example, Ref.~\cite{BirrellDavies}). It is also assumed that the switching function $\chi$ is smooth and of compact support so that the integral above is well-defined. All of the complicated dependence of the transition probability on the trajectory of the detector and on the quantum state of the field is contained in the response function, so it is typical in the literature to compute only this quantity and refer to it as the transition probability, albeit a slight abuse of nomenclature. 

For practical computations, it is better to have an explicit regularization for the Wightman Green function. This is possible provided the field is in a quantum state that satisfies the Hadamard condition \cite{WaldBook}, where the response function is given by \cite{LoukoSatz2008}
\begin{align}
		\label{responsereg}
		\nn\mathcal{F}(\omega) & =-\frac{\omega}{4\pi}\int_{-\infty}^{\infty}\chi^{2}(u)du
		\\& +2\int_{-\infty}^{\infty}du\,\chi(u)\int_{0}^{\infty}ds\,\chi(u-s)
		\nn\\& \times\left(\cos\omega s\;W(u,u-s)+\frac{1}{4\pi^{2}s^{2}}\right)
		\nn\\& +\frac{1}{2\pi^{2}}\int_{0}^{\infty}ds\frac{1}{s^{2}}\int_{-\infty}^{\infty}du\,\chi(u)\left[\chi(u)-\chi(u-s)\right],
	\end{align}
and $W(u,u-s)=\lim_{\epsilon\to 0}W_{\epsilon}(u,u-s)$. Now, $W(u,u-s)$ is a well-defined distribution everywhere except at the vertex of the lightcone (when $x(u)=x(u-s)$ or equivalently when $s=0$) but this pathology is now explicitly regularized by the counterterm $1/(4 \pi^{2}s^{2})$. 

We wish to consider the case of sharp-switching when the detector is switched on and off instantaneously. While this violates the assumption that the switching function is smooth, it is possible to consider sharp-switching as a limit of increasingly steep smooth switching functions. The result is regular except at the limit of infinite detection time. Notwithstanding, the \emph{rate} of detection for sharp-switching is regular even at this limit and is given by \cite{LoukoSatz2008}
\begin{align}
		\label{eq:transitionratesharp}
		\dot{ \mathcal{F}}_{\tau}(\omega)=&
		2\int_{0}^{\Delta\tau}ds\,\left(\cos\omega s\;W(\tau,\tau-s)+\frac{1}{4\pi^{2}s^{2}}\right)\nonumber\\
		&-\frac{\omega}{4\pi}+\frac{1}{2\pi^{2}\Delta\tau}.
	\end{align}
This is the quantity of primary interest for the remainder of this article. This quantity is proportional to the transition rate (though we will refer to it as the transition rate) of particles registered by our detector while the interaction is still turned on. The detection time is given by $\Delta \tau=\tau-\tau_{0}$, where the interaction is turned on at time $\tau_{0}$ and $\tau$ is the detector's proper time. The integral above is then tantamount to integrating the (regularized) Wightman Green function for the field over the worldline of the detector with a weighting that depends on the energy gap of the detector's states.  

While Eq.~(\ref{eq:transitionratesharp}) is elegant and succinct, it hides some problematic technical subtleties with its practical implementation in black hole spacetimes. The main issue is that the Wightman Green function is not known in closed-form for any black hole spacetimes in dimensions greater than three; instead one represents the Wightman Green function by a Fourier and multipole decomposition where even the individual modes are typically not known functions but must be obtained numerically. In this mode-sum representation, the Hadamard singularity structure near $s=0$ is not exhibited by the simple geometrical form of the counterterm $1/(4\pi^{2}s^{2})$ but rather as the non-convergence of the mode-sum in this coincidence limit. To circumvent this problem, one must try to express the counterterm as an appropriate mode-sum and subtract from the Wightman function mode-by-mode. All of this must be done with sufficient accuracy to numerically integrate the result. This is a challenge numerically since the mode-sum typically converges slowly and many modes are needed to obtain a result with sufficient accuracy. Moreover, the parameter space that one wants to explore is rather large since the transition rate depends sensitively on the detector's trajectory and energy gap, the detection time and the quantum state of the scalar field. All of this amounts to quite a large numerical undertaking.

To avoid these numerical challenges, we exploit an enhanced symmetry in the near-horizon region of an extremal black hole which enables us to express the Wightman Green function in closed form. This makes the evaluation of Eq.~(\ref{eq:transitionratesharp}) quite straightforward so that the parameter space can be explored in full. While it is difficult to ascertain to what extent our results may be extrapolated to more generic scenarios, e.g. to non-extremal black holes or beyond the near-horizon throat, this calculation may provide a means to simplify the numerical endeavour involved in computing the particle detector response in these more generic cases. This is made possible by the fact that knowing the transition rate for a given reference scenario\footnote{i.e. a given reference trajectory for the same spacetime and quantum state, or a given reference state for the same trajectory and spacetime, or indeed a given reference spacetime for the same trajectory and quantum state.} allows us to practically compute the transition rate in another scenario of interest through the identity \cite{Campos2021}
	\begin{align}
		\dot{ \mathcal{F}}_{\tau}(\omega)=  \dot{ \mathcal{F}}_{\tau}^{\textrm{ref}}(\omega)+ 2\Re\int_{0}^{\Delta\tau}ds \,e^{-i\, \omega s}\Big[W(\tau,\tau-s)\nonumber\\
		-W_{\textrm{ref}}(\tau,\tau-s)\Big].
	\end{align}
The integral contained above is ostensibly amenable to an efficient mode-by-mode numerical implementation.

One set of quantum states for the scalar field we will consider are those whose Wightman Green function are periodic in imaginary time. These are known as KMS \cite{Kubo, MartinSchwinger} or thermal states and the periodicity is identified with the inverse of the temperature of the field. In the limit of infinite detection time, we can define the temperature of the detector itself by considering the excitation to de-excitation ratio. If we let
    \begin{equation}
    \label{eq:ratioprob}
    {\cal R}=\frac{{\cal F}(\omega)}{{\cal F}(-\omega)}    
    \end{equation}
and if there exists a $T$ that satisfies the detailed-balance form of the KMS condition
    \begin{equation}
\label{KMS}
{\cal R}=e^{-\omega /T},    
    \end{equation}
then we identify $T=T_{\textrm{EDR}}$ with the temperature of the detector given by
\begin{align}
\label{eq:TEDR}
    T_{\textrm{EDR}}=-\frac{\omega}{\ln\mathcal{R}}.
\end{align}
For a static detector coupled to a scalar field in the Hartle-Hawking state, this temperature is independent of the energy gap and equals the locally-measured Hawking temperature in the limit of infinite detection time. For finite detection times, $T_{\textrm{EDR}}$ becomes dependent on the energy gap but this dependence can be sufficiently weak \cite{Garay2016} so that Eq.~(\ref{eq:TEDR}) remains a suitable temperature estimator for the detector.

Normally, in a black hole spacetime, one expects a positive correlation between the temperature of the quantum field and the temperature of the detector, in the sense that we expect hotter fields to correspond to hotter detectors. However, recently a number of authors \cite{Henderson2020, Campos2021, robbins2021antihawking} have found that for small field temperatures, it is possible to have the detector's temperature decrease as the local field temperature increases. This phenomenon has been dubbed the anti-Hawking effect, or more specifically the strong anti-Hawking effect. Correspondingly, one expects the transition probability and transition rate to increase as the field temperature increases. When the opposite occurs, it has been labelled the weak anti-Hawking effect. In an analogous way, we can define the strong anti-Unruh effect as the anti-correlation of the detector's temperature with the detector's acceleration and the weak anti-Unruh effect as the anti-correlation of the transition probability (or rate) with the detector's acceleration. 

The anti-Hawking effect has been reported for small temperatures in the BTZ and rotating BTZ black holes \cite{Henderson2020, Campos2021, robbins2021antihawking}, but as far as we know has not yet been reported for any four-dimensional black holes. Ref.~\cite{CamposDappiaggi2021} did not find evidence of the effect for massless topological black holes in four dimensions. The anti-Unruh effect has also been reported in lower-dimensional spacetimes \cite{Brenna2016, Garay2016, CamposDappiaggi2021}, but not yet in dimensions greater than three. In addition, we note that the distinction between the anti-Unruh and anti-Hawking effect for accelerating detectors in black hole spacetimes can be subtle.

A further complication comes from the fact that  the response function diverges in the sharp-switching limit. In this case, instead of the ratio of excitation to de-excitation probabilities as in Eq.~(\ref{eq:ratioprob}), we take the ratio of the rates 
\begin{align}
    T_{\textrm{EDR}}=-\frac{\omega}{\ln\tilde{\mathcal{R}}},\qquad \tilde{\mathcal{R}}=\frac{\dot{\mathcal{F}}_{\tau}(\omega)}{\dot{\mathcal{F}}_{\tau}(-\omega)}.
    \label{TEDR}
\end{align}
For a static detector coupled to a field in the Hartle-Hawking state in the limit of infinite detection, this definition  gives the expected $T_{\textrm{EDR}}=T_{\textrm{loc}}$ where $T_{\textrm{loc}}$ is the red-shifted Hawking temperature of the black hole. Indeed for static detectors, Eq.~(\ref{TEDR}) remains a suitable temperature estimator for finite but sufficiently long times in the sense that the dependence on $\omega$ is weak and $T_{\textrm{EDR}}$ asymptotes to the locally measured field temperature as the detection time is increased. Moreover, Eq.~(\ref{TEDR}) was employed in Ref.~\cite{HodgkinsonLoukoOttewill} for a detector on a circular geodesic in Schwarzschild in the limit of infinite detection time. When Eq.~(\ref{TEDR}) is a constant or approximately constant function of energy gap $\omega$, we will say the detector has thermalized at a temperature $T_{\textrm{EDR}}$. 
	\section{The Near-Horizon Extremal Approximation}
	\label{sec:BertottiRobinson}
We are interested in the behaviour of a particle detector near an extremal black hole and wish to exploit the enhanced symmetry which emerges in the near-horizon regime. To this end, we begin by considering the Reissner-N{\"o}rdstrom spacetime with
	line element
	\begin{equation}
		ds^{2}=-\frac{f(r)}{r^{2}}dt^{2}+\frac{r^{2}}{f(r)}dr^{2}+r^{2}d\Omega^{2},
	\end{equation}
	where $f(r)\equiv(r-r_{+})(r-r_{-})$. The metric describes a static,
	spherically symmetric charged black hole solution to the Einstein-Maxwell equations with event horizon $r_{+}$ and Cauchy horizon $r_{-}$ given by
	\begin{equation}
		r_{\pm}=M\pm\sqrt{M^{2}-Q^{2}}.
	\end{equation}
	Here, $Q$ signifies charge (assumed to be positive), $M$ is the
	mass of the black hole and we have adopted units whereby $c=\hbar=G=1$
	and hence Planck length is also unity. In the extremal limit $M=Q$, the horizons coincide and the spacetime is given by
	\begin{equation}
		\label{RNex}
		ds^{2}=-\frac{(r-r_{\textrm{H}})^{2}}{r^{2}}dt^{2}+\frac{r^{2}}{(r-r_{\textrm{H}})^{2}}dr^{2}+r^{2}d\Omega^{2},
	\end{equation}
	where $r_{\textrm{H}}=M=Q$. We can associate a temperature to each of the horizons in the Reissner-Nordstrom black hole by examining the periodicity in Euclidean time required to avoid a conical singularity. The corresponding temperatures are
	\begin{equation}
		T_{-}=\frac{r_{-}-r_{+}}{4\pi r_{-}^{2}},\quad T_{+}=\frac{r_{+}-r_{-}}{4\pi r_{+}^{2}}.
	\end{equation}
$T_{+}$ is the Hawking temperature which we associate with the black hole itself which vanishes in the extremal limit. 
	
Turning now to the near-horizon limit of the extremal black hole \eqref{RNex}. We
employ the coordinates
	\begin{equation}
		y=\frac{r-r_{\textrm{H}}}{\varepsilon},
	\end{equation}
	for some small parameter $\varepsilon$ to obtain
	\begin{equation}
		ds^{2}=-\frac{y^{2}}{(y+Q/\varepsilon)^{2}}dt^{2}+\frac{(\varepsilon y+Q)^{2}}{y^{2}}dy^{2}+(\varepsilon y+Q)^{2}d\Omega^{2}.
	\end{equation}
	The radial coordinate $r$ approaches the extremal horizon $r_{\textrm{H}}$
	as $\varepsilon$ tends to zero so that in the near-horizon regime
	$Q/\varepsilon\gg y$ and $Q\gg\varepsilon y$, yielding
	\begin{equation}
		ds^{2}=-\frac{y^{2}}{Q^{2}}dt^{2}+\frac{Q^{2}}{y^{2}}dy^{2}+Q^{2}d\Omega^{2},
	\end{equation}
	where, having served its usefulness, we have set $\varepsilon=1$.
	
	The resulting geometry is described by the direct product spacetime $\textrm{CAdS}\times \mathbb{S}^{2}$ where CAdS is the covering space of Anti-de Sitter space, i.e. an AdS spacetime where the periodic time coordinate has been `unwrapped'. This direct product spacetime is known as the Bertotti-Robinson solution \cite{Bertotti1959, Robinson1959}. By choosing coordinates $(\mathsf{t},\rho)$ related to $(t,y)$ by
	\begin{align}
		t\pm\frac{Q^{2}}{y}=\tanh\left[\tfrac{1}{2}\left(\mathsf{t}\pm\tfrac{1}{2}\ln\left(\frac{\rho-1}{\rho+1}\right)\right)\right],
	\end{align}
	we can more readily see the black hole interpretation of the spacetime from the resulting line element
	\begin{align}
		\label{BRmetric}
		\frac{ds^{2}}{Q^{2}}=-(\rho^{2}-1)d\mathsf{t}^{2}+(\rho^{2}-1)^{-1}d\rho^{2}+d\Omega^{2}.
	\end{align}
	These new coordinates do not cover the entire manifold since they are clearly singular at $\rho=\pm 1$. In this near-horizon throat, the asymptotically flat exterior has decoupled revealing a solution where `spatial' infinity is in fact timelike. The coordinates used in Eq.~(\ref{BRmetric}) cover only a patch of this timelike boundary and it is this fact that admits the black hole interpretation of this spacetime. This is related to the fact that we have inherited the time-coordinate from the Reisnner-N{\"o}rdstrom solution which in the limit considered results in the covering space of AdS, not AdS itself. This can be seen more clearly from the Penrose diagram in Ref.~\cite{TaylorOttewill2012}. That we no longer have an asymptotically flat spacetime is important for another reason, the near-horizon throat is no longer globally hyperbolic and hence solving the scalar field equation requires boundary conditions to be imposed at spatial infinity. We impose Dirichlet boundary conditions throughout but we note that other boundary conditions are possible, indeed generic Robin boundary conditions appear to be very interesting in the context of the anti-Unruh and anti-Hawking effect \cite{Henderson2020, Campos2021, CamposDappiaggi2021, robbins2021antihawking}.
	
Despite arriving at this spacetime from an extremal spacetime with vanishing Hawking temperature, there is now a natural, non-vanishing Hawking temperature associated with the near-horizon regime
	\begin{equation}
	\label{eq:HHTemp}
		T_{\textrm{H}}	=\frac{1}{2\pi}.
	\end{equation}
	
	One of the central ideas of this article is to push the black hole interpretation of the metric \eqref{BRmetric} to see if we can learn anything new about how an Unruh-DeWitt detector responds in the near-horizon regime of a black hole.  As mentioned above, the rationale behind such an approach stems from the enhanced symmetry of the metric \eqref{BRmetric}, which yields closed-form representations of the Wightman Green function in various quantum states. This offers the opportunity to probe deeper into the parameter space without recourse to a significant numerical endeavour. We spend the rest of this section looking at the Wightman function for a massless quantum scalar field which satisfies the Dirichlet boundary conditions in the Bertotti-Robinson spacetime for the Boulware, Hartle-Hawking and arbitrary thermal KMS states.

	We consider first the Boulware state. The Boulware vacuum is defined by requiring normal modes to have positive frequency with respect to $\partial/\partial \mathsf{t}$ -- the Killing vector for which the exterior region $\rho>1$ is static. This state corresponds to the familiar notion of an empty state at large radii. However, the state becomes singular on the Killing horizon $\rho=1$, implying that the Wightman Green function diverges even when $x\ne x'$. In other words, the state is not Hadamard on the horizon and our expression for the rate (\ref{eq:transitionratesharp}) ought to diverge there.
	
	A closed form representation for the Wightman Green function in the Bertotti-Robinson spacetime for a massless scalar field satisfying the Dirichlet boundary conditions in the Boulware vacuum state was found in Ref.~\cite{TaylorOttewill2012},
	\begin{align}
		\label{WightB}
		W_{\textrm{B}}(x,x')=\frac{\eta}{4\pi^{2}\sqrt{R}}\frac{1}{(-\Delta\mathsf{t}^{2}+\eta^{2})},
	\end{align}
	where $\eta$ and $R$ are given by
	\begin{eqnarray}
		\label{eq:defetaR}
		\cosh\eta&=&\frac{\rho\,\rho'-\cos\gamma}{\sqrt{\rho^{2}-1}\sqrt{\rho'^{2}-1}},\nonumber\\
		R&=&\rho^{2}+\rho'^{2}-2\rho\,\rho'\cos\gamma-\sin^{2}\gamma\nonumber\\
		&=&(\rho^{2}-1)(\rho'^{2}-1)\sinh^{2}\eta,
	\end{eqnarray}
	and $\gamma$ is the geodesic distance on the $2$-sphere given by
		\begin{eqnarray}
		\label{eq:defgamma}
		\cos\gamma=\cos\theta\cos\theta^{\prime}+\sin\theta\sin\theta^{\prime}\cos(\phi-\phi^{\prime}).
	\end{eqnarray}
If we take $x=x(\tau)$ and $x'=x(\tau-s)$ to be two points on a given worldline, then it is straightforward to show that the leading order term near coincidence (i.e. near $x=x'$, or equivalently, near $s=0$) is $-1/(4\pi^{2}s^{2})$ so that the integral in Eq.~(\ref{eq:transitionratesharp}) converges, except in the case where one point is on the horizon as already mentioned.

We also wish to examine the field in the Hartle-Hawking state. This is a thermal state whereby the field temperature is equal to the black hole temperature $T_{\textrm{H}}=1/(2\pi)$. The state is usually constructed by demanding that the propagator be periodic in imaginary time with periodicity equal to the inverse temperature of the black hole. In Ref.~\cite{TaylorOttewill2012}, the Feynman propagator for a massless scalar field satisfying the Dirichlet boundary conditions in the Hartle-Hawking state propagating in the Bertotti-Robinson spacetime was found to be
	\begin{align}
		\label{GHH}
		G^{\textrm{HH}}(x,x')=\frac{i}{8\pi^{2}}\Bigg(\mathcal{P}\frac{1}{\zeta^2}-i\,\pi\,\delta(\zeta^2)\Bigg),
	\end{align}
	where we have used the shorthand $\zeta^2=\cosh\lambda-\cos\gamma$, ${\cal P}(z)$ signifies the principal part and
	\begin{align}
		\cosh\lambda=\rho\,\rho'-(\rho^{2}-1)^{1/2}(\rho'^{2}-1)^{1/2}\cosh\Delta\mathsf{t}.
	\end{align}
	Using the general relationship between the Feynman and the Wightman Green functions \cite{Bire82},
	\begin{align}
		G(x,x')= i\,\Theta(\Delta\mathsf{t})W(x,x')+i\,\Theta(-\Delta\mathsf{t})W^{\dagger}(x,x'),
	\end{align}
	along with the distributional identity
	\begin{align}
		\lim_{\epsilon\to 0^{+}}\frac{1}{z^{2}+i\epsilon}=\mathcal{P}\frac{1}{z^{2}}-i\,\pi\,\delta(z^{2}),
	\end{align}
	we may glean from Eq.~\eqref{GHH} the closed-form representation of the Wightman function
	\begin{equation}
		\label{WightH}
		W_{\textrm{HH}}(x,x')=\frac{1}{8\pi^{2}}\frac{1}{\cosh\lambda-\cos\gamma},
	\end{equation}
where, as before, we have taken the limit $\epsilon\to 0^{+}$ explicitly since the pathology on the vertex of the lightcone is explicitly regularized by a counterterm in the integral expression for the transition rate. We also note that this propagator is equivalent to the propagator for the field in the Poincar{\'e} vacuum \cite{TaylorOttewill2012}. The Hartle-Hawking state is regular on the horizon and indeed this state would be appropriate for considering a particle detector's response across the horizon.

The Hartle-Hawking state is a thermal state at the Hawking temperature, but one could also consider arbitrary thermal states (also known as KMS states). The corresponding propagators are periodic in imaginary time with periodicity equal to the inverse of an arbitrary temperature. Such states will not be regular on the horizon unless the temperature is the Hartle-Hawking temperature. Nevertheless, a closed-form representation for the Wightman function for a scalar field satisfying the Dirichlet boundary conditions can be found in an analogous way to that above from the Feynman propagator given in Ref.~\cite{TaylorOttewill2012}. The result is
\begin{align}
\label{eq:WightT}
    W_{T}(x,x')=\frac{T}{4\pi\,R^{1/2}}\frac{\sinh(2 \pi T\,\eta)}{\cosh(2\pi T\,\eta)-\cosh(2\pi T\,\Delta \mathsf{t})}
\end{align}
where $\eta$ and $R$ are given by Eqs.~(\ref{eq:defetaR}).

In the next section, we employ the closed-form Wightman Green functions (\ref{WightB}), (\ref{WightH}) and (\ref{eq:WightT}) in our expression for the transition rate (\ref{eq:transitionratesharp}) in order to study the response of an Unruh-DeWitt detector near an extremal black hole.

	\section{Results}
		\label{sec:rates}
	In this section, we present results for the transition rate of an Unruh-DeWitt detector near an extremal charged black hole coupled to a quantum scalar field in a range of quantum states. We will consider the detector on both stationary and non-stationary trajectories for geodesic and accelerated motion. For each trajectory, we explore the full gamut of the parameter space (i.e. variations in the energy gap $\omega$, initial radius $\rho_0$, angular momentum $L$, detection time $\Delta\tau$, and field temperature $T$, where applicable). We also seek evidence of the so-called anti-Unruh or anti-Hawking effect \cite{Mann2020,Brenna2016} in the large parameter space. The analysis in this section comprises the main results of the paper.
	
\subsection{Geodesic Detectors}
	When the detector is on a geodesic trajectory, its motion is described by the equations
	\begin{align}
		\label{arbtraj}
		\dot{\mathsf{t}}&=\mathcal{E}/(\rho^{2}-1),\nonumber\\
		\dot{\rho}^{2}&=L^{2}(\rho_{0}^{2}-\rho^{2}),\nonumber\\
		\dot{\phi}&=h,
	\end{align}
	where $\rho_{0}$ is the initial radius and $\mathcal{E}$ and $h$ are the energy and angular momentum related by
	\begin{align}
	    \mathcal{E} = L\sqrt{\rho_0^2-1},\qquad L=\sqrt{ h^{2}+1}.
	\end{align}
	  Specifying $\rho_{0}$ and $L$ uniquely determines the trajectory subject to given initial conditions which we set below. As $\dot{\rho}^{2}>0$, we must have $\rho<\rho_{0}$, so that all geodesics are necessarily inbound. In this sense, we interpret the initial radius $\rho_0$ as the radius of farthest approach which is a consequence of being in the near-horizon regime. The solutions to the coupled equations are elementary and yield the following
	\begin{align}
	\label{eq:geodesics}
		\mathsf{t}(\tau)&=\artanh\left[\frac{\tan L \tau}{\sqrt{\rho_{0}^{2}-1}}\right],\nonumber\\
		\rho(\tau)&= \rho_{0}\cos L \tau,\nonumber\\
		\phi(\tau)&=h\,\tau,
	\end{align}
	where $\tau$ is the detector's proper time and we have assumed the initial conditions $\mathsf{t}(0)=0$, $\rho(0)=\rho_{0}$,  $\dot{\rho}(0)=0$ and $\phi(0)=0$.
	
	We see from these solutions that the proper time it takes the detector to reach the horizon $\rho=1$ is
	\begin{align}
	    \tau_{\mathrm{H}}=\frac{1}{L}\arccos\left(\frac{1}{\rho_{0}}\right),
	\end{align}
	which implies that the detection time in the exterior is short regardless of the geodesic trajectory. The maximum detection time along a geodesic in the exterior region is constrained by $|\tau_{H}|<\pi/2$ with the upper bound approached by a radially infalling geodesic ($L=1$) with large initial radius ($\rho_0\to\infty$). The short detection time suggests that geodesics are not the best candidate to investigate, at least within the near-horizon regime and sharp-switching limit we are considering, since the transition rate will likely be dominated by transient effects associated with switching the detector on sharply.
	
		Nevertheless, since the radial geodesics permit the longest detection time in the exterior, we focus on those. Other geodesics with $L>1$ approach the horizon in an arc rather than head-on but it is not possible to orbit the black hole without accelerating in the near-horizon regime. 
		
		The solutions to the geodesic equations for radially infalling detectors are obtained simply by taking $L=1$ in Eqs.~(\ref{eq:geodesics}). If we turn the interaction between the field and the detector on at $\tau=t=0$, then the total detection time is identical to the proper time $\tau$. Moreover, the radius $\rho(\tau)$ becomes a more intuitive measure of the detection time along our path. Hence, we find it convenient to express the integration variable in Eq.~(\ref{eq:transitionratesharp}) in terms of $\rho^\prime\equiv \rho(\tau-s)$ where the integration now runs over the history of the trajectory in a natural way from initial radius $\rho_{0}$ to final radius $\rho$. The result is 
	\begin{align}
		\label{transraterho}
		\dot{{\cal F}}_{\tau}(\omega)&=2\int_{\rho}^{\rho_{0}}\frac{d\rho^{\prime}}{\sqrt{\rho_{0}^{2}-\rho^{\prime2}}}\Bigg(\cos\left(\omega \sigma\right)W(\rho,\rho^{\prime})+\frac{1}{4\pi^{2}\sigma^{2}}\Bigg)
		\nn\\&-\frac{\omega}{4\pi}+\frac{1}{2\pi^{2}\arccos(\rho/\rho_{0})},
	\end{align}
	where $\sigma\equiv \arccos(\rho/\rho_{0})-\arccos(\rho^{\prime}/\rho_{0})$. The Wightman Green function $W(\rho,\rho')$ in each of the quantum states is given explicitly by Eqs. (\ref{WightB}), (\ref{WightH}) and (\ref{eq:WightT}) with
		\begin{align}
	    \eta&=\textrm{arcosh}\left(\frac{\rho\rho'-1}{\sqrt{\rho^{2}-1}\sqrt{\rho'^{2}-1}}\right)\nonumber\\
	    \Delta\mathsf{t}&=\frac{\rho\,\rho'(\rho_{0}^{2}-1)-\sqrt{\rho_{0}^{2}-\rho^{2}}\sqrt{\rho_{0}^{2}-\rho'^{2}}}{\rho_{0}^{2}\sqrt{\rho^{2}-1}\sqrt{\rho'^{2}-1}}.
	\end{align}

\begin{figure}[!htp]
	    \centering
	   \includegraphics[width=\linewidth]{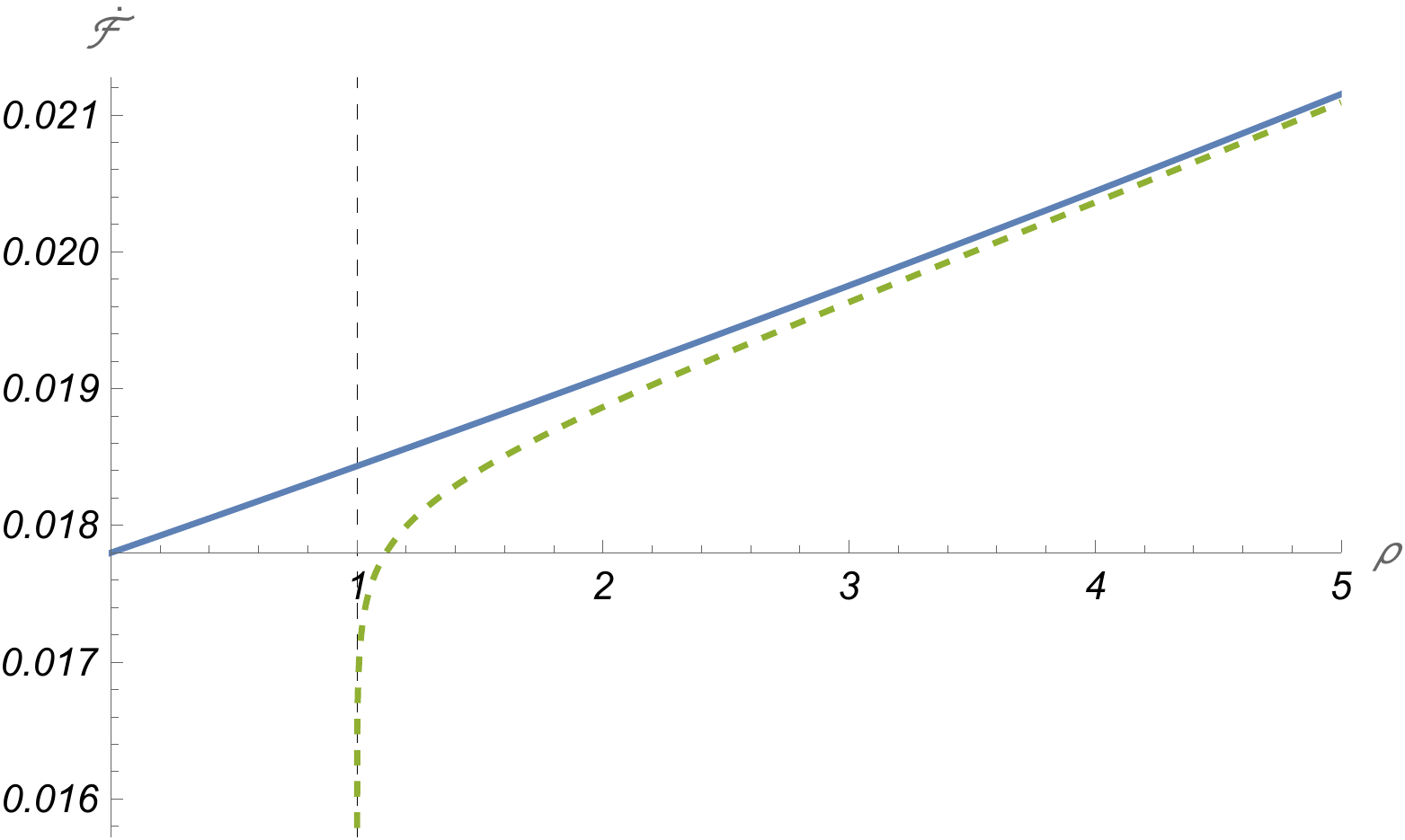}
		    \caption{Plot illustrating the transition rate for a particle detector approaching the horizon in radial free-fall, with the detector coupled to a quantum field in the Hartle-Hawking state (blue) and the Boulware state (green, dashed). The initial radius has been chosen to be $\rho_0=40$ and the energy gap $\omega=1/10$. 
		    }
		    \label{fig:Radial_rho}
	\end{figure}

		We begin in Fig.~\ref{fig:Radial_rho} by examining the behaviour in the near-horizon regime of a radially infalling detector coupled to a field in both the Boulware and Hartle-Hawking states. The first thing to note here is that the transition rate for the Hartle-Hawking state is regular across the horizon in contrast to the Boulware case. This is the expected behaviour since the propagator for a field in the Hartle-Hawking state is known to be Hadamard across the horizon while the propagator for the field in the Boulware state is not.
	
\begin{figure}[!htp]
	    \centering
	    \subfloat[$\omega=1/10$]{
	    \includegraphics[width=\linewidth]{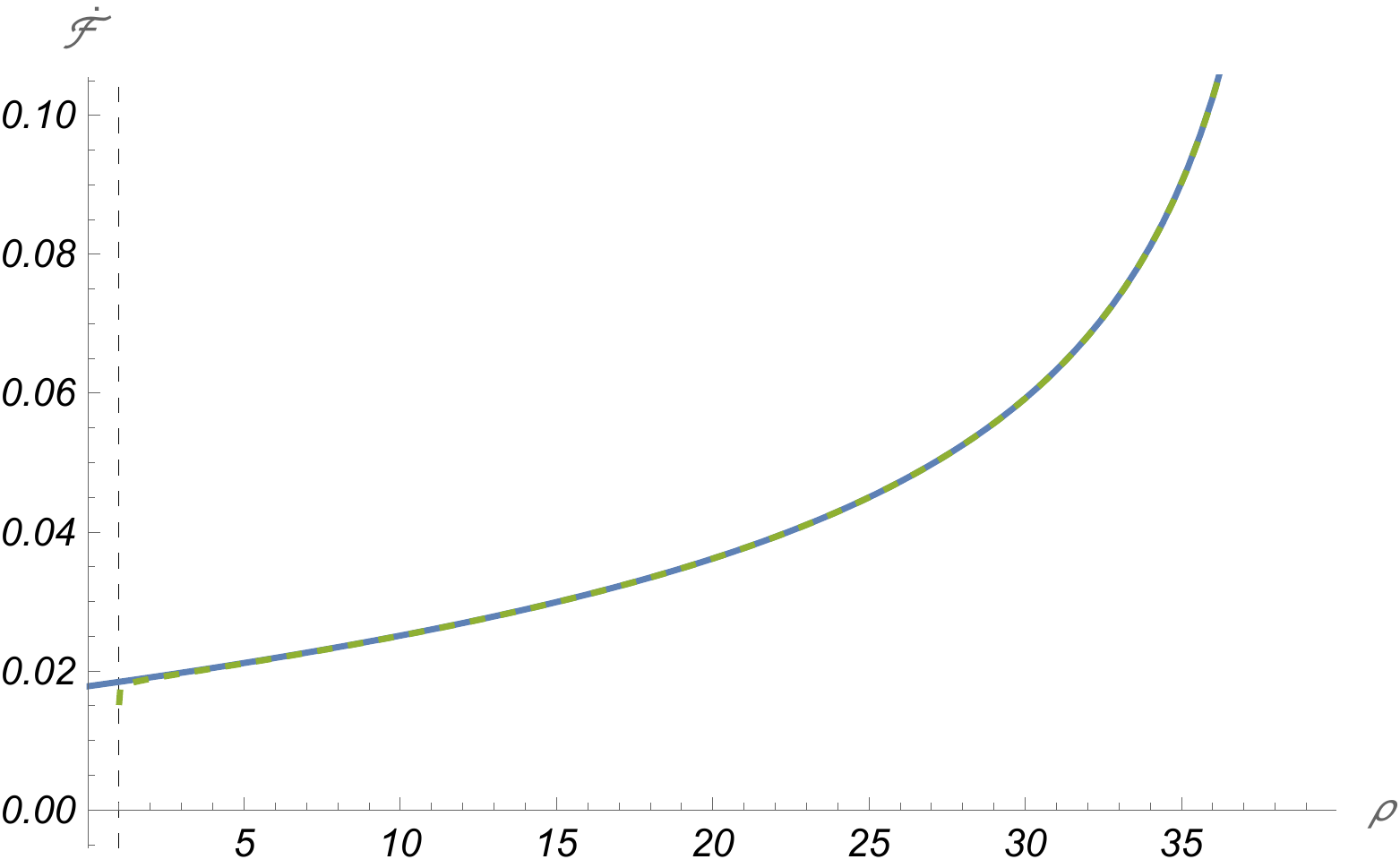}
	    	  	    }\\
	      \subfloat[$\omega=3$]{
	  \includegraphics[width=\linewidth]{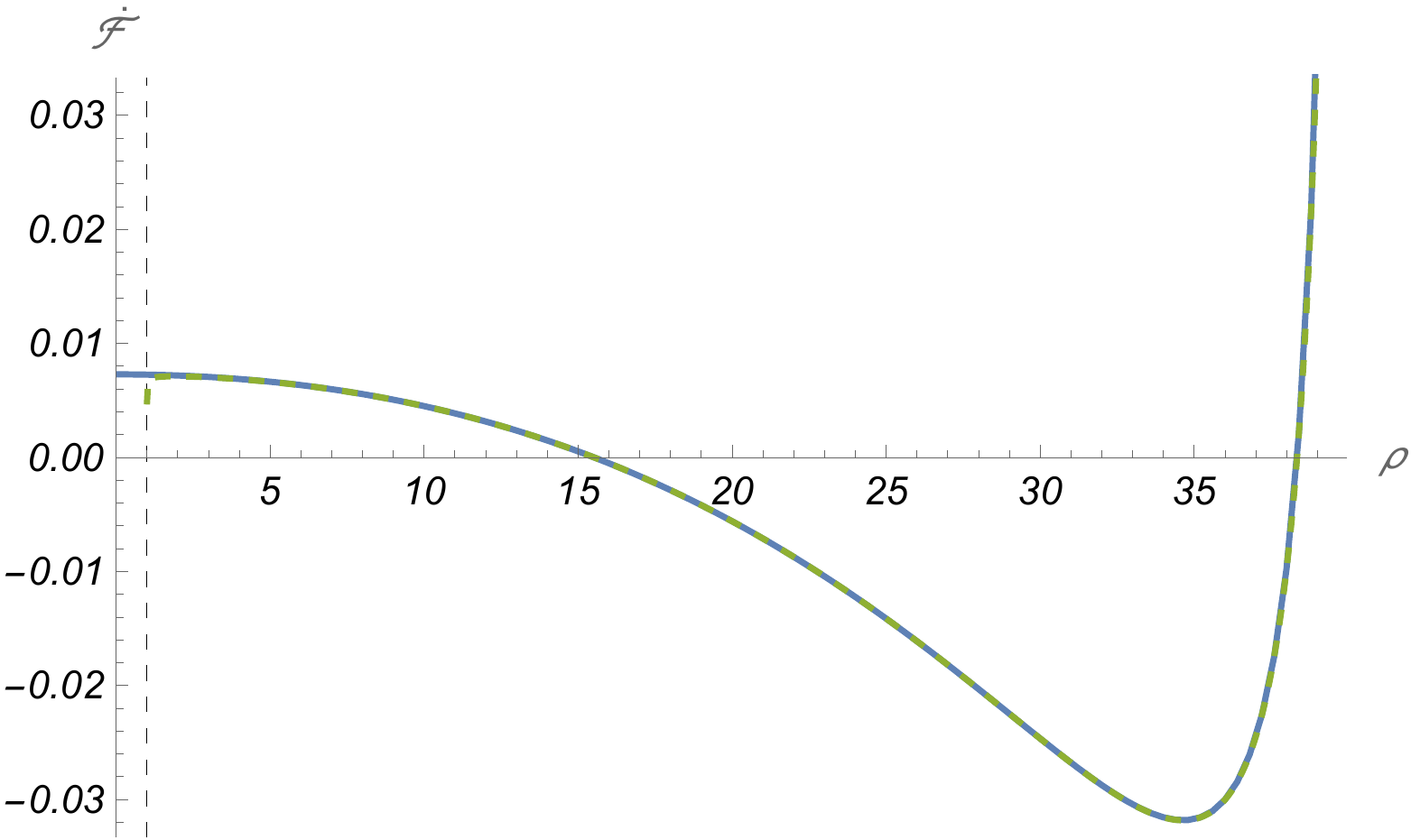}
		    }\\
	      \subfloat[$\omega=20$]{
	  \includegraphics[width=\linewidth]{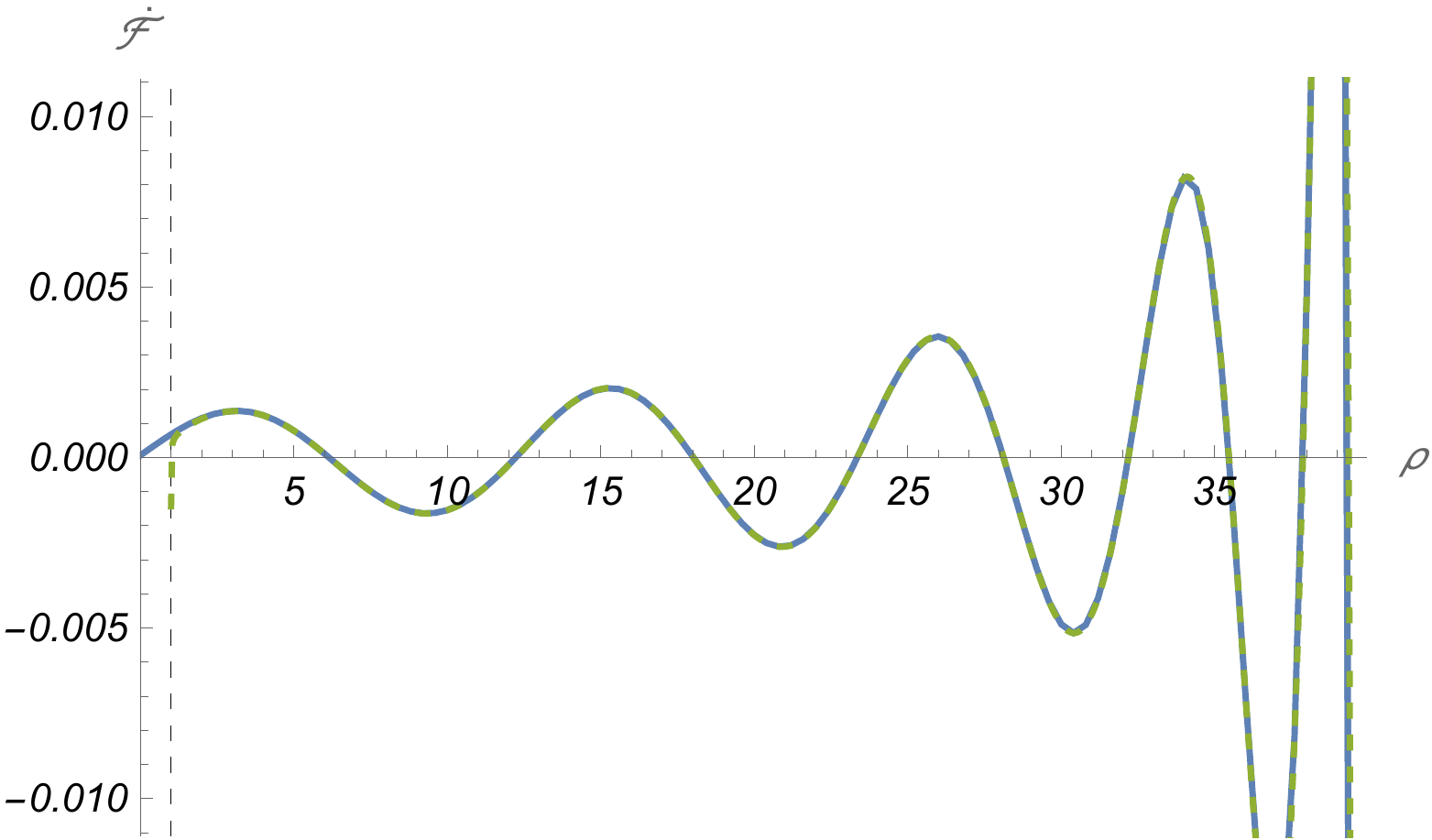}
		    }
		    \caption{Plots depicting the transition rate for a particle detector in radial free-fall as a function of the final radius $\rho$ for various energy gaps. With the choice of initial radius $\rho_0=40$, the transition rates for the Hartle-Hawking (blue) and Boulware (green, dashed) states are indistinguishable away from the horizon.}
		    \label{fig:Radial_rhoab}
	\end{figure}

		We see from Fig.~\ref{fig:Radial_rhoab} that the transition rates for both the Boulware and Hartle-Hawking states are largely indistinguishable from each other away from the horizon. In Fig.~\ref{fig:Radial_rhoab}~(a) we consider a transition rate with energy gap $\omega=1/10$ and initial radius $\rho_{0}=40$.	As we track from right to left, we find a decreasing transition rate as the horizon is approached. This is somewhat counter-intuitive since the local KMS temperature  $T_{\textrm{H}}/\sqrt{\rho^{2}-1}$ increases as the horizon is approached. One would therefore expect that the transition rate would also increase. By raising the energy gap in Fig.~\ref{fig:Radial_rhoab}~(b), we see a markedly different profile. In this case, we observe a sharp initial decline in the transition rate, followed by a turning point, and a monotonically increasing phase. Increasing the energy gap further as in Fig.~\ref{fig:Radial_rhoab}~(c), results in the emergence of damped oscillations, which helps clarify that the general behaviour in Fig.~\ref{fig:Radial_rhoab} shows oscillations in the transition rate which dampen as the horizon is approached, with the frequency of these oscillations growing as the magnitude of the energy gap increases. We keep in mind that the detection time involved here is very short and these oscillations are almost certainly a transient effect associated with turning the detector on sharply.
	
\begin{figure}[!htp]
    \centering
   \subfloat[$T=1$]{ \includegraphics[width=\linewidth]{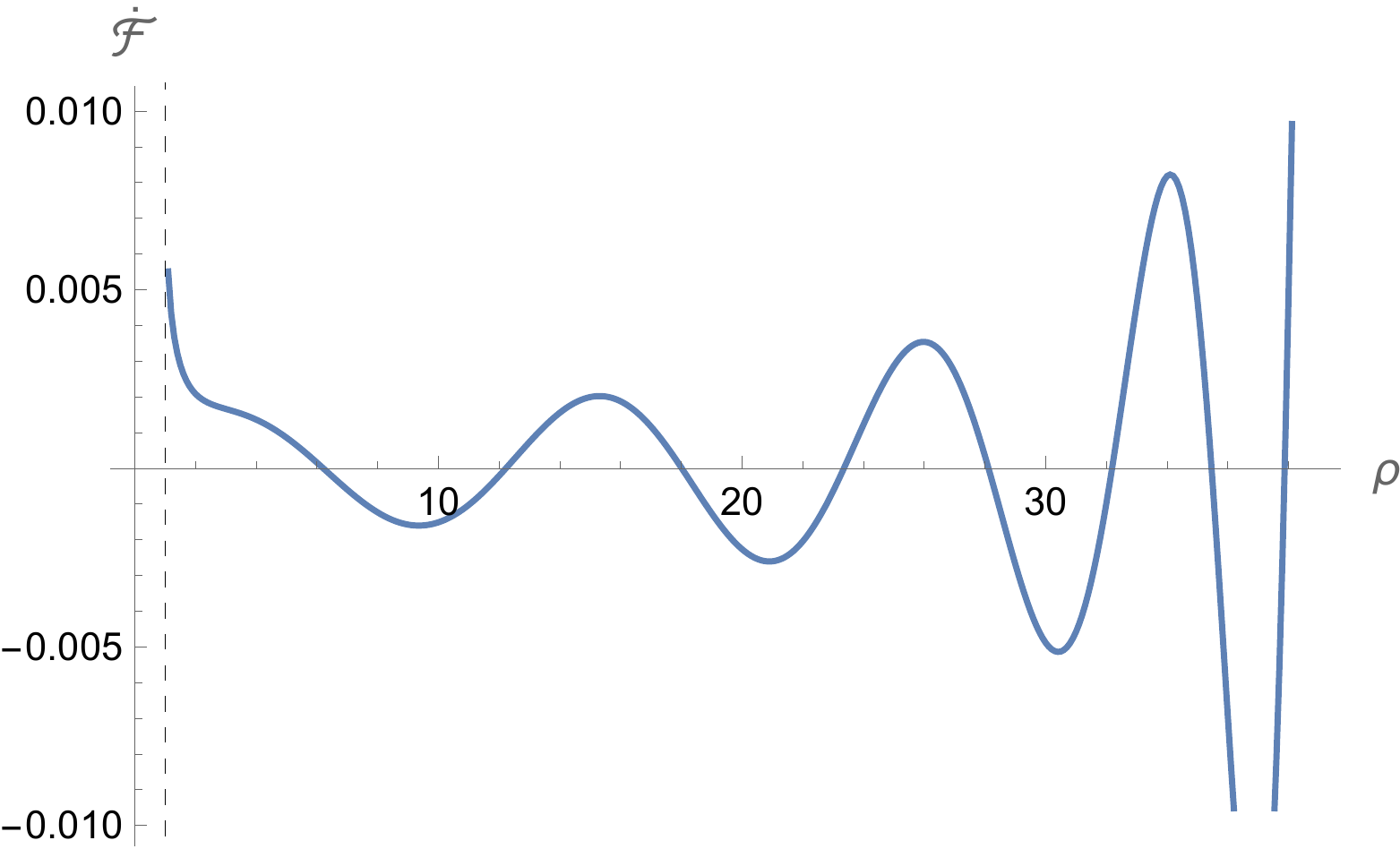}}\\
\subfloat[$T=10$]{    \includegraphics[width=\linewidth]{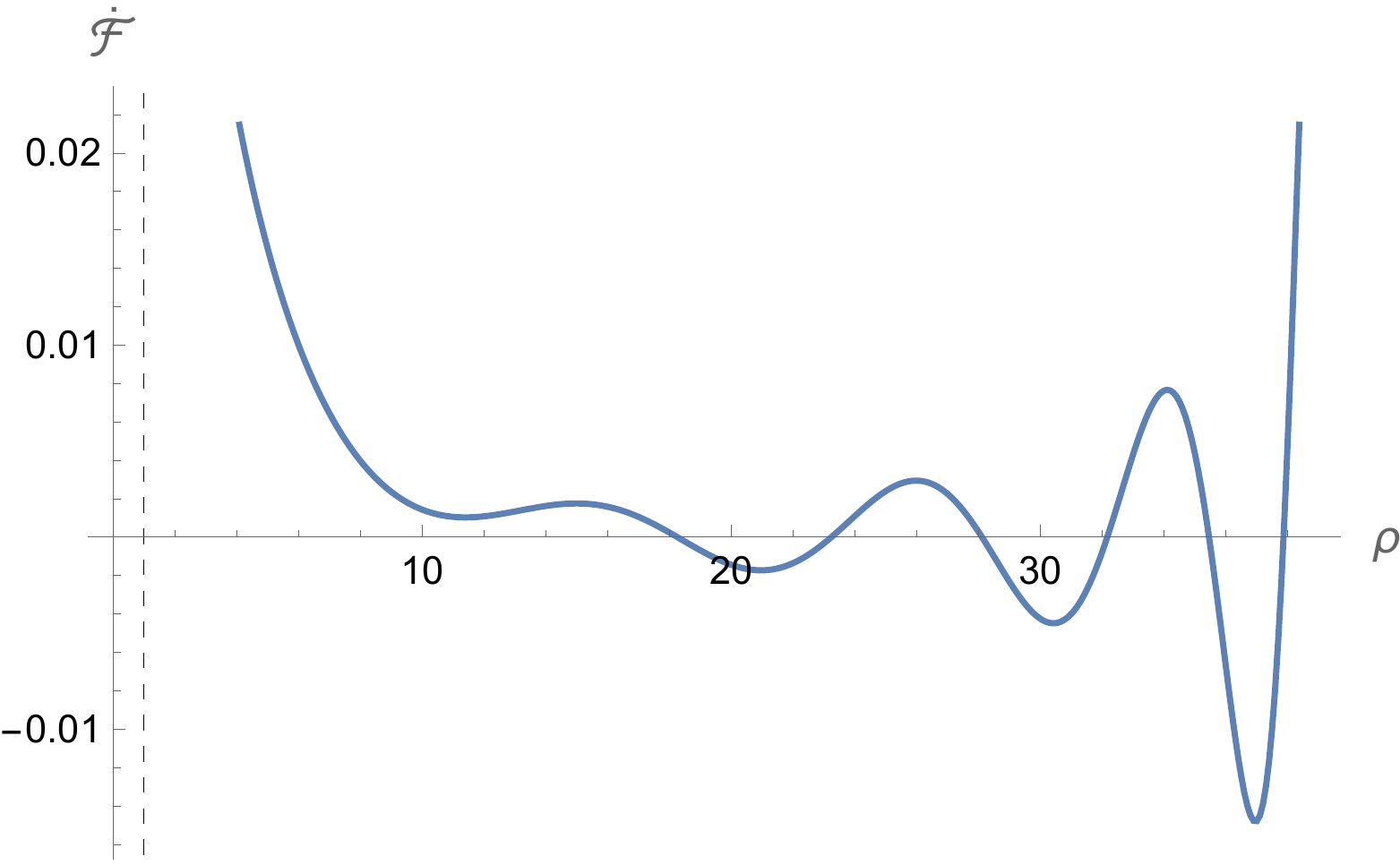}}\\
\subfloat[$T=25$]{    \includegraphics[width=\linewidth]{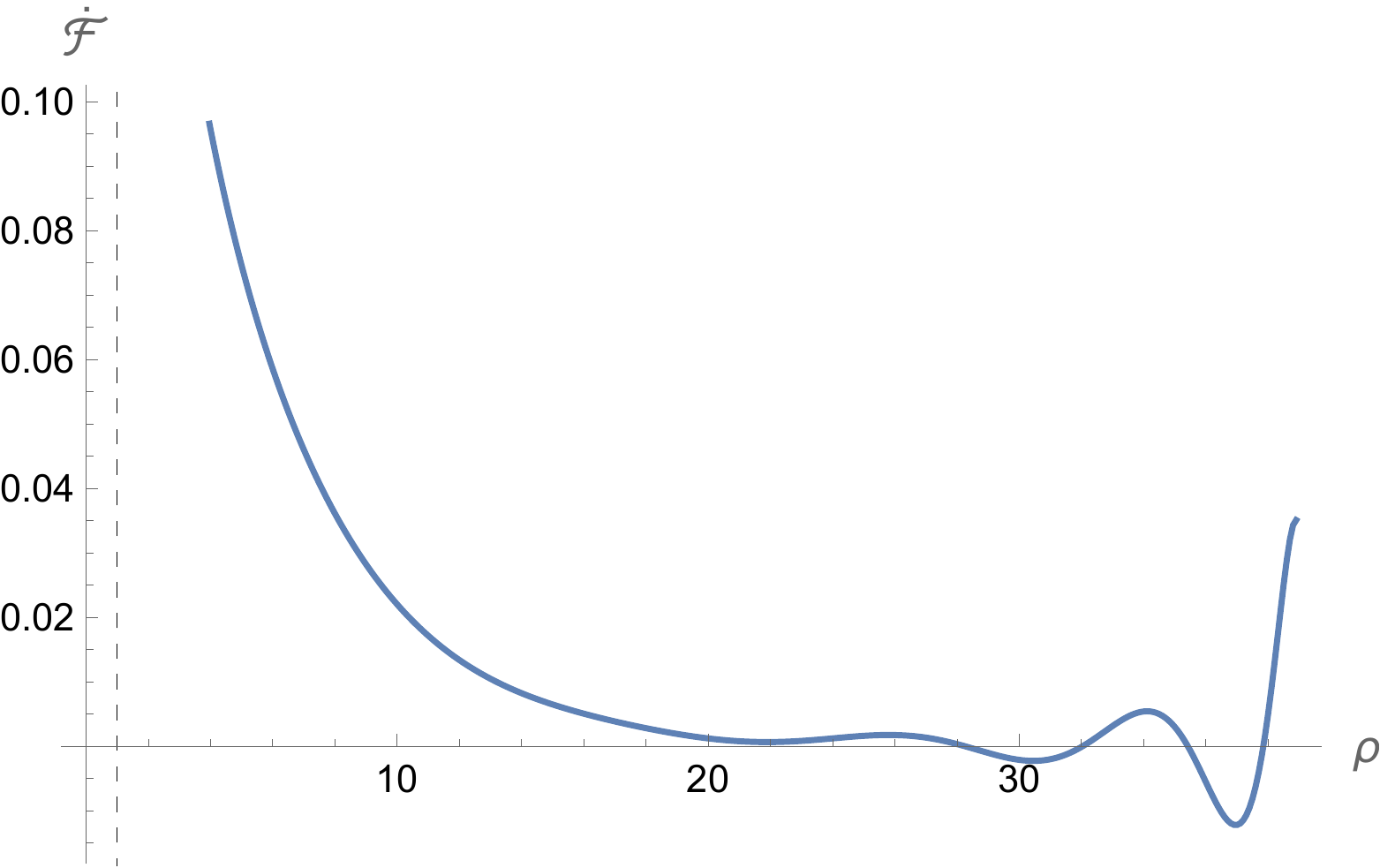}}\\
    \caption{Plots showing the transition rate for a radially-infalling detector coupled to a field in a thermal KMS state, with a selection of field temperatures $T$. In each case we have chosen an initial radius $\rho_0=40$ and an energy gap $\omega=20$. }
    \label{fig:Radial_therm}
\end{figure}
\begin{figure}[!htp]
    \centering
 \subfloat[$T=1$]{   \includegraphics[width=\linewidth]{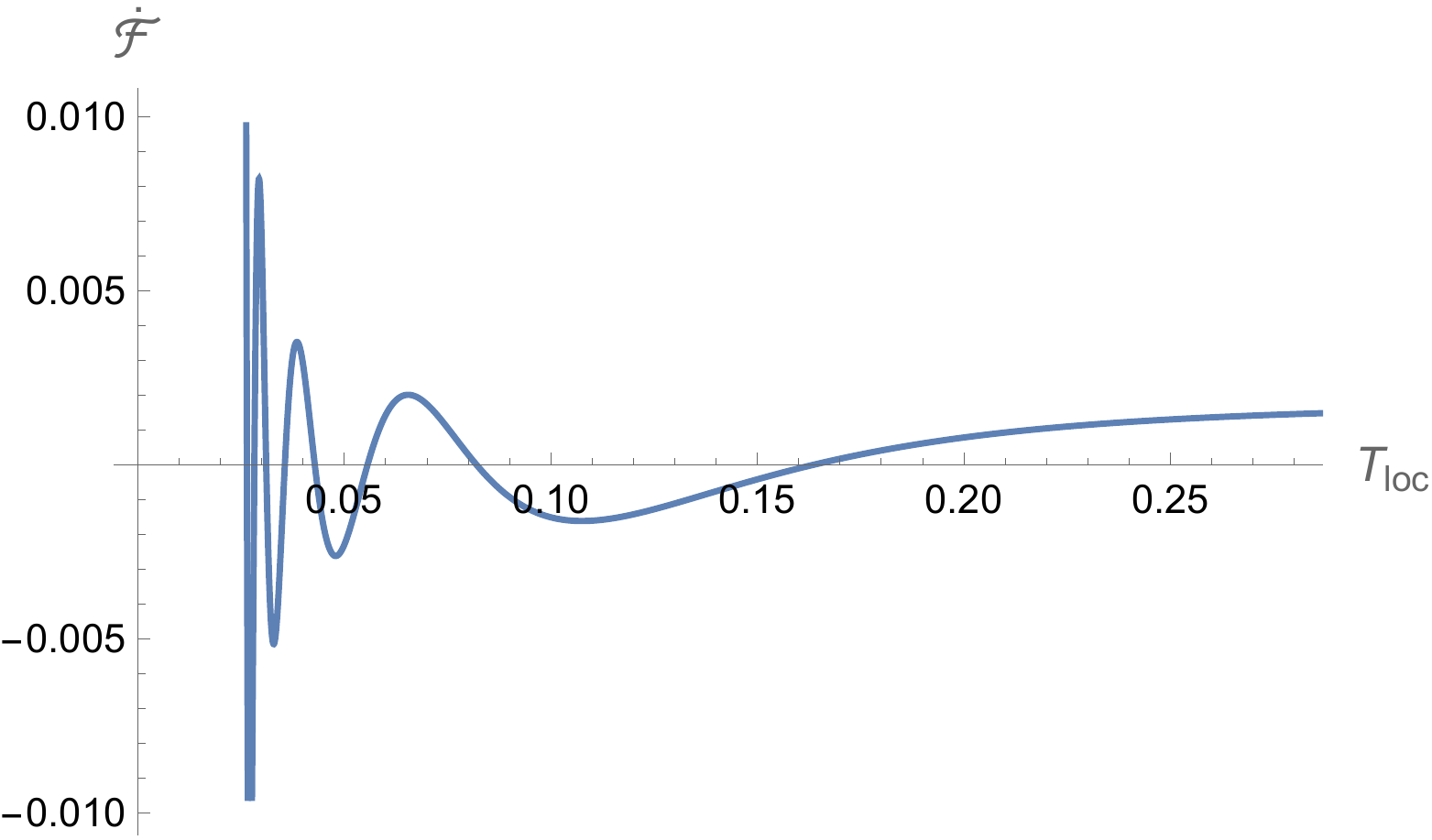}}\\
 \subfloat[$T=10$]{   \includegraphics[width=\linewidth]{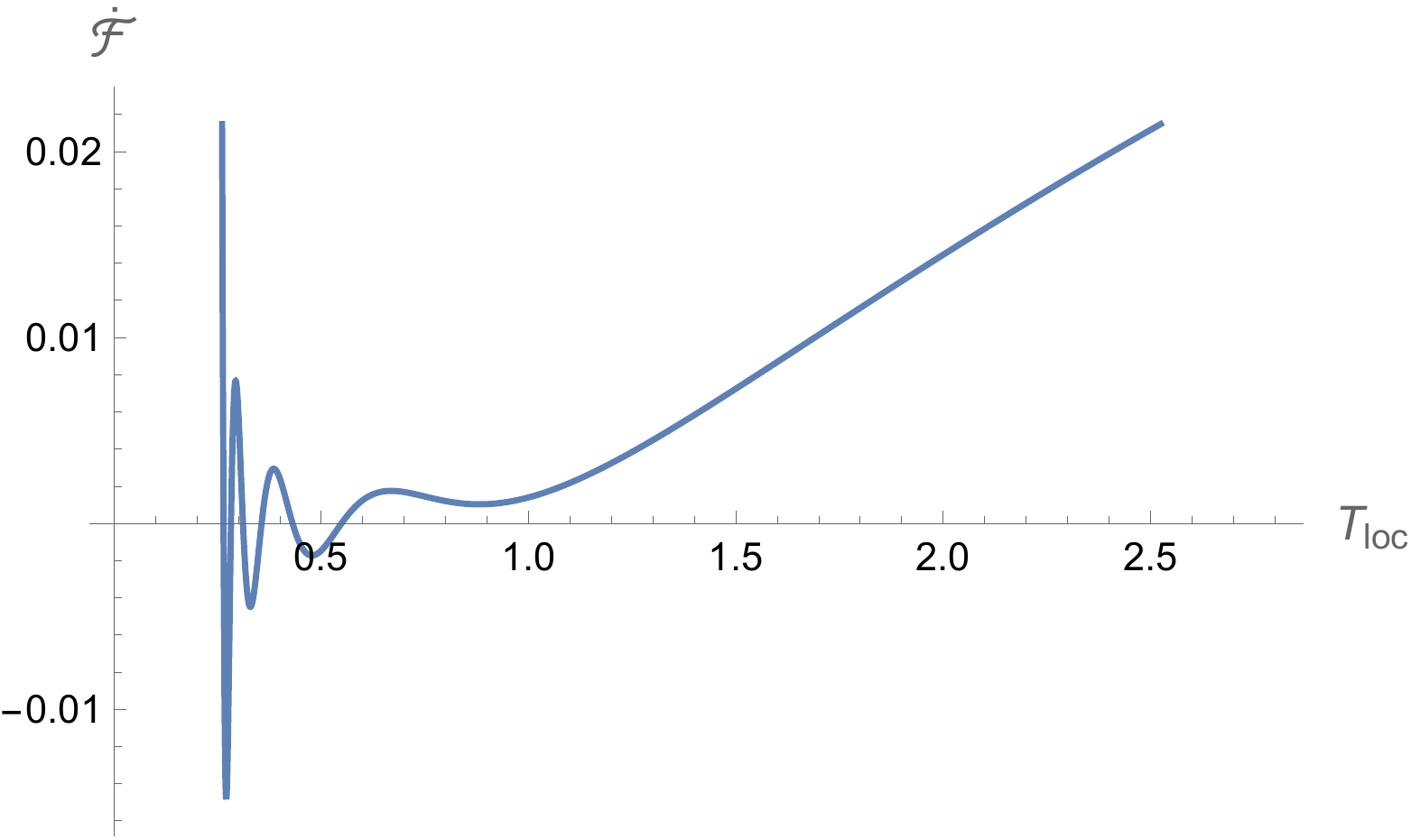}}\\
    \subfloat[$T=25$]{ \includegraphics[width=\linewidth]{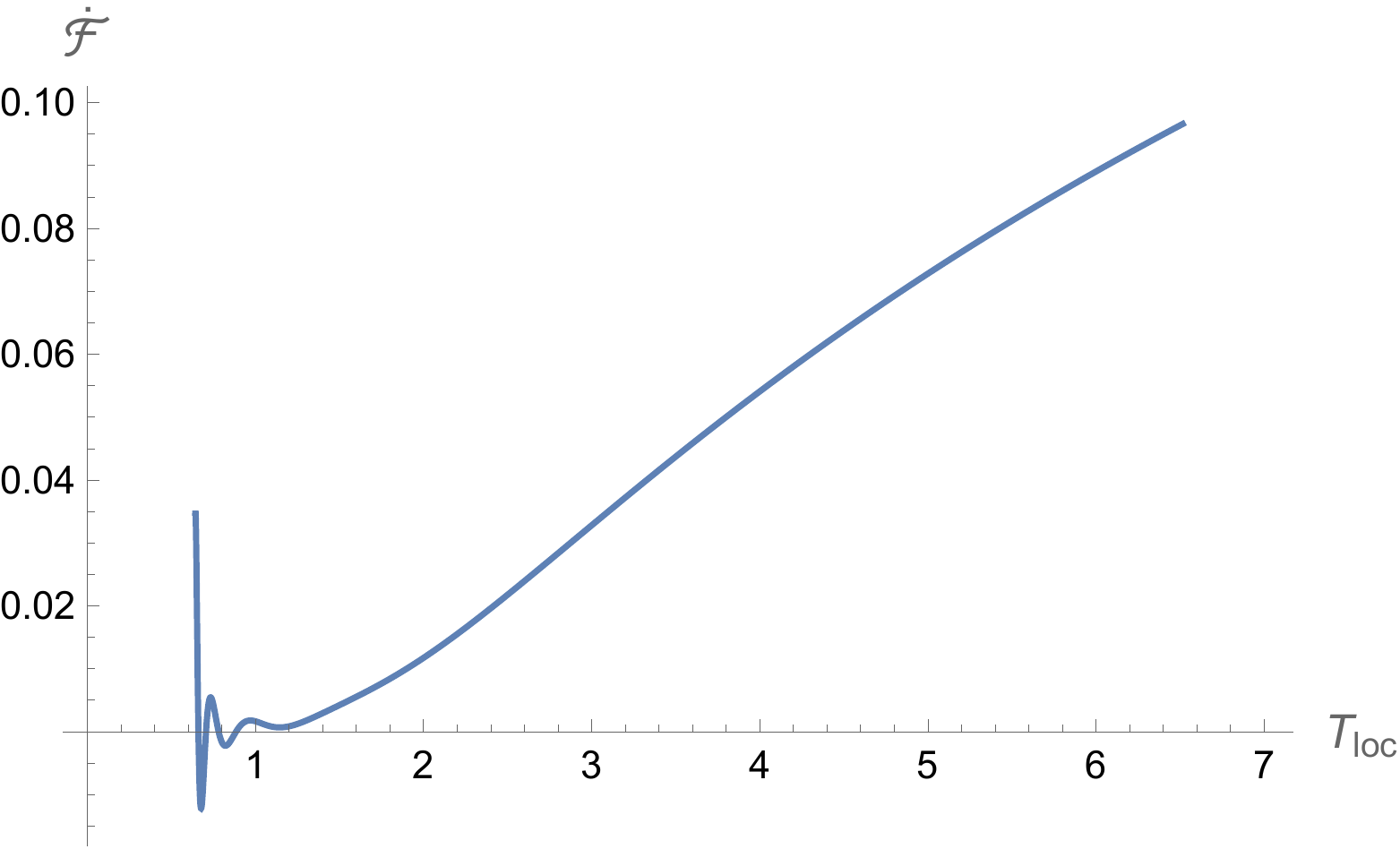}}\\
    \caption{Plots showing the transition rate as a function of locally-measured temperature for a radially-infalling detector coupled to a field in a selection of thermal states with field temperature $T$. In each case, we have chosen $\rho_0=40$ and $\omega=20$. 
    }
   \label{fig:Radial_Tloc}
\end{figure}

To try to distill this apparent transient effect from the non-transient behaviour of the detector, we cannot simply increase the detection time for the reasons already mentioned. We can, however, consider the detector coupled to a field in a hot thermal state. We expect the transition rate to increase as the field temperature increases and hence the effect of transience ought to be subdominant. This is precisely what we find in Fig.~\ref{fig:Radial_therm}. For small field temperatures $T\lesssim 1$, the profiles of the transition rate are dominated by the oscillations we observed in Fig.~\ref{fig:Radial_rhoab}. For larger field temperatures, however, a monotonic increase emerges from the oscillations as the detector approaches the horizon. This is precisely the behaviour we would expect since the local KMS temperature is increasing as the horizon is approached. This is seen most clearly in Fig.~\ref{fig:Radial_Tloc} where we plot the transition rates for the detector coupled to a field in an arbitrary KMS thermal state as a function of local KMS temperature. We observe transient oscillations for large $\rho$ (which is equivalent to small local KMS temperature or short detection times) giving way to a monotonic increase as the horizon is approached (as the local KMS temperature is increased). This monotonically increasing phase is reached earlier and is more dominant as the field temperature increases, cf. Fig.~\ref{fig:Radial_Tloc}~(a)-(c). With a sufficiently high field temperature, as in Fig.~\ref{fig:Radial_Tloc}~(c), the transient oscillations are only relevant for short detection times far from from the black hole whereas, in the case of the Hartle-Hawking state, the transition rate is so small that it is buried in transient noise over the entire trajectory.  

\begin{figure}
    \centering
    \includegraphics[width=\linewidth]{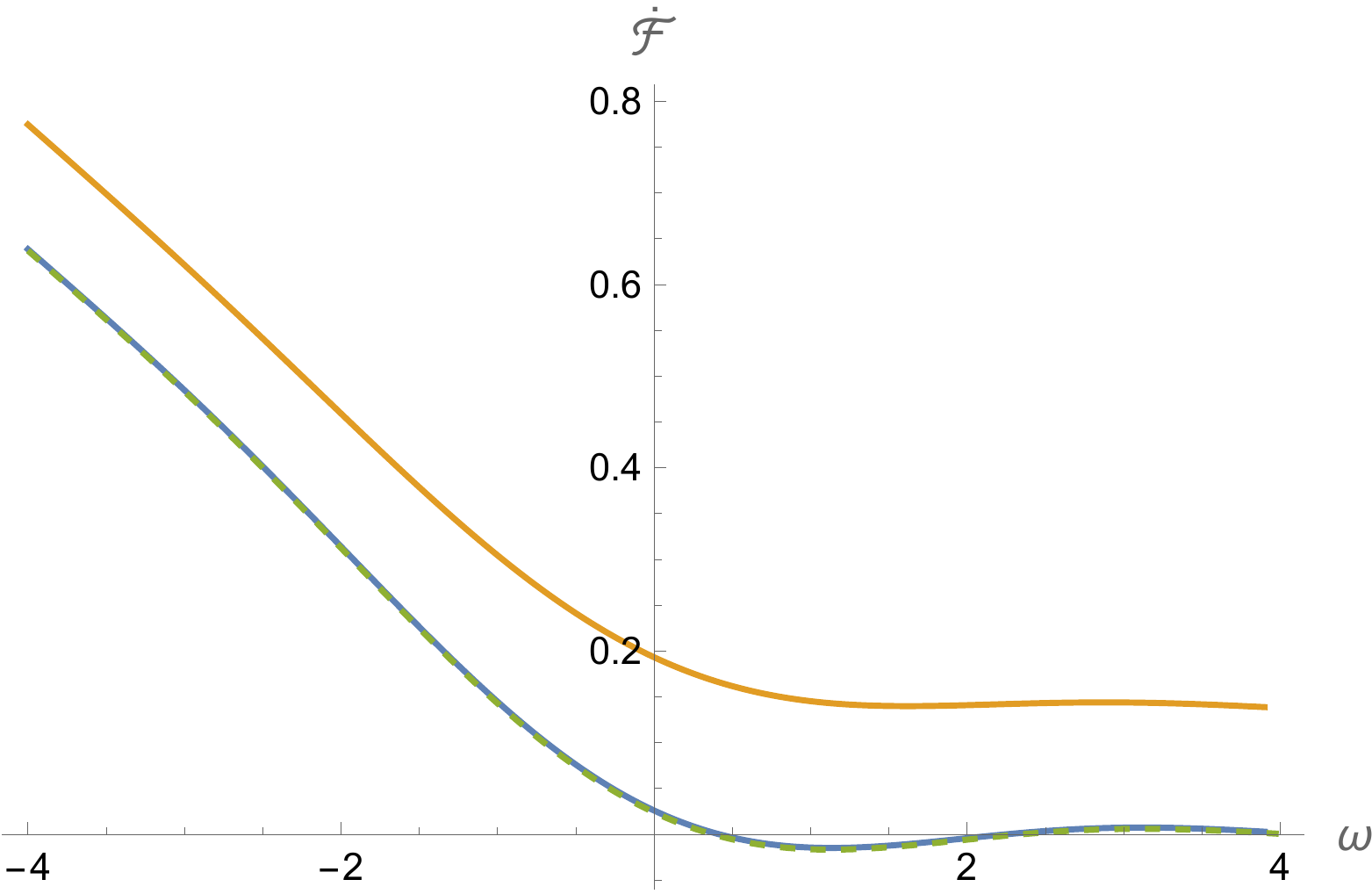}
    \caption{Figure showing the transition rate as a function of energy gap for a particle detector in radial free-fall. Plots for the Boulware vacuum (green, dashed) and Hartle-Hawking state (blue) are shown, though indistinguishable, for the choice of initial radius $\rho_{0}=40$. The profile is shifted upwards when the field temperature is increased to $T=40T_H$ (yellow). The detection time has been taken to be a time when the detector is near the horizon. }
    \label{fig:Radial_omega}
\end{figure}

While the plots shown in Figs.~\ref{fig:Radial_therm}-\ref{fig:Radial_Tloc} were for $\omega=20$, the conclusion holds more generally. When the energy gap is negative, the profiles are shifted upwards since the transition rate is considerably larger. This is expected since it is generically more likely for the detector to de-excite ($\omega<0$) than to excite ($\omega>0$), as seen in Fig.~\ref{fig:Radial_omega} where we plot the dependency of the transition rate on the energy gap. We further note that when the field temperature is increased beyond the Hartle-Hawking temperature (see yellow curve), the profiles in Fig .~\ref{fig:Radial_omega} are also shifted upwards. One final remark is that we observe oscillatory behaviour which is not present in the accelerated cases we turn to now. We put this behaviour down to transience as the detection time in the infalling case is necessarily short.

	%
	%
	
	\subsection{Accelerated Detectors}
For the remainder of our analysis, we focus on accelerated detectors. This is more interesting in the context of our near-horizon approximation since acceleration is required to prevent a rapid plunge into the black hole. With such a rapid plunge, the detector will only register transient effects unless the field temperature is very hot, as we saw in the previous subsection. In any case, there are many scenarios of interest to consider here, especially the static and the circular trajectories. But we can also consider more generic inspirals into the black hole.

	\subsubsection*{Static Detectors}
We begin with the static case, the simplicity of which makes it possible to perform the integral in Eq.~(\ref{eq:transitionratesharp}) explicitly. For a static detector at $\rho=\rho_{0}$ coupled to a field in the Boulware vacuum, we obtain
	\begin{align}
		\dot{\mathcal{F}}_{\tau}(\omega)&=2\int_{0}^{\Delta\tau}\frac{1-\cos (\omega s)}{4\pi^{2}s^{2}}ds-\frac{\omega}{4\pi}+\frac{1}{2\pi^{2}\Delta\tau}\nonumber\\
		&=\frac{\cos (\omega \Delta\tau)}{2\pi^{2}\Delta\tau}+\frac{\omega}{2\pi^{2}}\mathrm{Si}(\omega\,\Delta\tau)-\frac{\omega}{4\pi},
	\end{align}
	where $\textrm{Si}(z)$ is the sine integral function. This transition rate is identical to that of an inertial detector in Rindler spacetime coupled to a scalar field in the Minkowski vacuum \cite{LoukoSatz2008}. We can further compute the transition rate in the limit of infinite detection time using the well-known asymptotic expansions of the sine integral function \cite{DLMF}, yielding
	\begin{equation}
		\dot{\mathcal{F}}_{\tau}(\omega)\to \frac{-\omega+|\omega|}{4\pi}=\frac{|\omega|\Theta(-\omega)}{2\pi},\quad \Delta\tau\to\infty.
	\end{equation}
	When the energy gap is positive, the transition rate in this limit is zero for the Boulware vacuum. This, of course, makes sense since the field is in an empty state (in the detector's frame of reference) and so the detector cannot absorb any quanta from the field. This mirrors results for an inertial detector in flat spacetime \cite{HodgkinsonLoukoOttewill}.

	For a static detector coupled to a field in the Hartle-Hawking state, the transition rate (\ref{eq:transitionratesharp}) can be expressed in terms of incomplete beta functions (or their equivalent hypergeometric representations), the result is
	\begin{align}
	\label{eq:staticHHfinite}
		\dot{\mathcal{F}}_{\tau}(\omega)&=\frac{1}{4\pi^{2}\sqrt{\rho_{0}^{2}-1}}\cos (\omega\Delta\tau)\,\frac{z+1}{z-1}\nonumber\\
		& +\frac{i\,\omega}{8\pi^{2}}\Bigg[B_{z}(i\,\omega\sqrt{\rho_{0}^{2}-1},0)-B_{z}(-i\,\omega\sqrt{\rho_{0}^{2}-1},0)\nonumber\\
		& +B_{z}(1+i\,\omega\sqrt{\rho_{0}^{2}-1},0)-B_{z}(1-i\,\omega\sqrt{\rho_{0}^{2}-1},0)\Bigg]\nonumber\\
		& -\frac{\omega}{4\pi}\left(1+\coth\left(\pi\,\omega\,\sqrt{\rho_{0}^{2}-1}\right)\right),
	\end{align}
	where $z=\exp\{\Delta\tau/\sqrt{\rho_{0}^{2}-1}\}$. 	This form is useful for computing the limit of infinite detection time in that we can employ the asymptotic expansion of the beta function $B_{z}(a,b)$ for large $z$ \cite{DLMF}. One finds that the middle two lines of the above equation have the asymptotic expansion
	\begin{align}
		\frac{\omega}{2\pi}\coth\left(\pi\,\omega\sqrt{\rho_{0}^{2}-1}\right)-\frac{\cos (\omega\Delta\tau)}{4\pi^{2}\sqrt{\rho_{0}^{2}-1}}+\mathcal{O}(z^{-1}),
	\end{align}
	so that for large $\Delta\tau$, we have
	\begin{align}\dot{\mathcal{F}}_{\tau}(\omega)=\frac{\cos(\omega\Delta\tau)}{4\pi^{2}\sqrt{\rho^{2}-1}}\left(\frac{z+1}{z-1}-1\right)\nonumber\\
		+\frac{\omega}{2\pi}\frac{1}{e^{2\pi\omega\sqrt{\rho_{0}^{2}-1}}-1}+\mathcal{O}(z^{-1}).
	\end{align}
	In the limit $\Delta\tau\to\infty$, the first term vanishes and we are left with
	\begin{align}
		\dot{\mathcal{F}}_{\tau}(\omega)\to 
		\frac{\omega}{2\pi}\frac{1}{e^{2\pi \omega\sqrt{\rho_{0}^{2}-1}}-1},\qquad \Delta\tau\to\infty.
	\end{align}
Noting that the Hawking temperature is $T_{\textrm{H}}=1/2\pi$ on the event horizon, while the local KMS temperature is $T_{\textrm{loc}}=T_{\textrm{H}}/\sqrt{\rho_{0}^{2}-1}$, we can rewrite the limit above as
	\begin{align}
	\label{TransHHinfTl}
		\dot{\mathcal{F}}_{\tau}(\omega)\to 
		\frac{\omega}{2\pi}\frac{1}{e^{\omega/T_{\textrm{loc}}}-1},\qquad \Delta\tau\to\infty.
	\end{align}
	Hence the static detector coupled to a field in the Hartle-Hawking state registers an exactly Planckian distribution for a blackbody in thermal equilibrium in the limit of infinite detection time. Moreover, the detector temperature defined by Eq.~\eqref{TEDR} is in thermal equilibrium with the local KMS temperature $T_{\textrm{EDR}}=T_{\textrm{loc}}$. 

For a field in an arbitrary thermal state at temperature $T$, Eq.~(\ref{TransHHinfTl}) remains valid except that $T_{\textrm{loc}}=T/\sqrt{\rho_{0}^{2}-1}$. Similarly, the transition rate for finite detection time for a field in an arbitrary thermal state can be obtained from Eq.~(\ref{eq:staticHHfinite}) by the substitution $\sqrt{\rho_{0}^{2}-1}\to 1/(2\pi T_{\textrm{loc}})$ where, as before, the local KMS temperature is the red-shifted field temperature. That we can eliminate the explicit dependence of $\rho_{0}$ in the transition rate in favour of $T_{\textrm{loc}}$ implies there is a degeneracy in the static transition rate between hotter fields at a given radius and colder fields at a smaller radius.

\begin{figure}[h!]
    \centering
    \includegraphics[width=\linewidth]{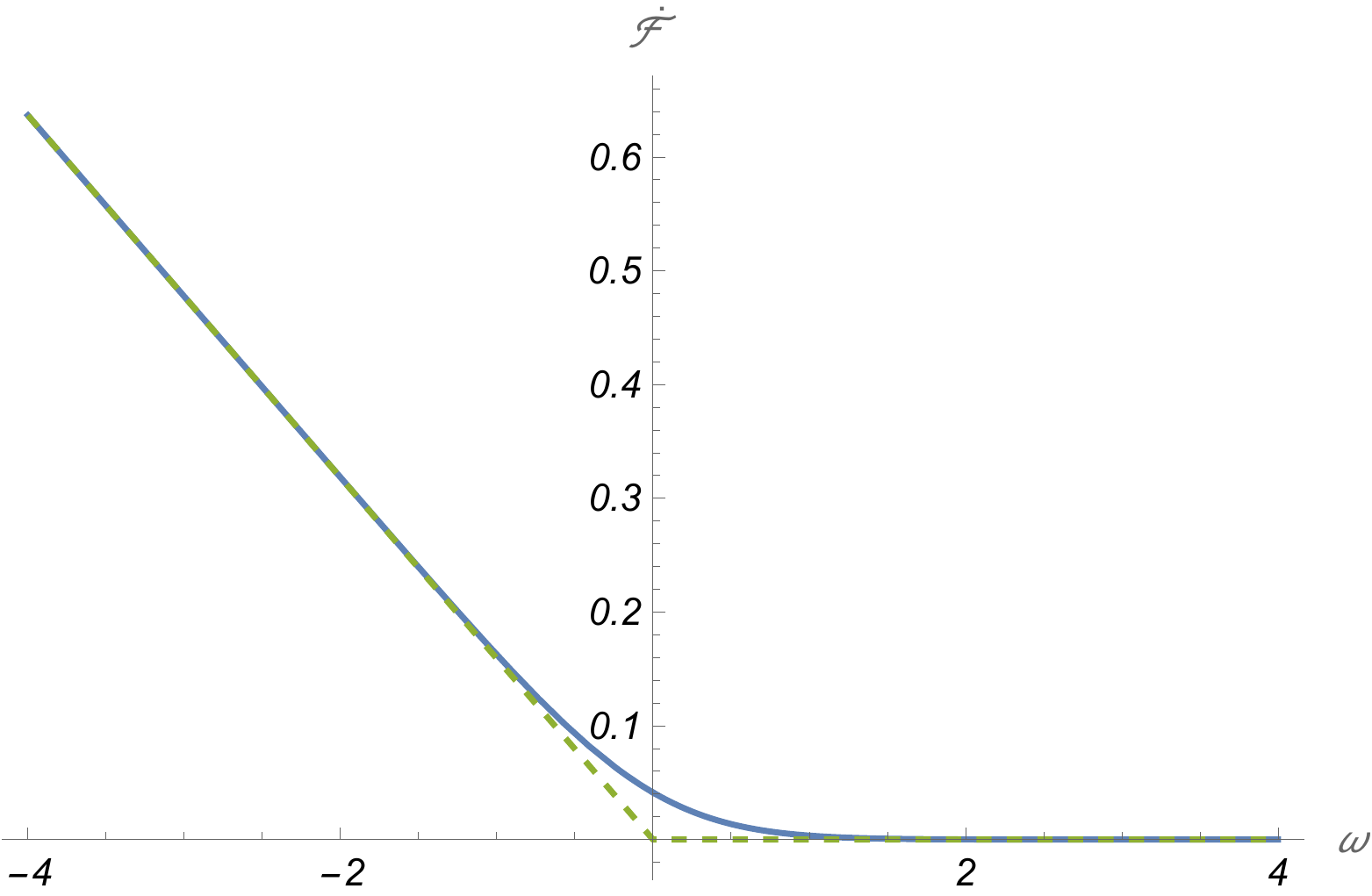}
    \caption{Plot depicting the transition rate as a function of energy gap for a static detector coupled to a field in the Boulware vacuum (green, dashed) and Hartle-Hawking state (blue). To accentuate the differences in the states' profiles near $\omega=0$, we have chosen a radius close to the horizon of $\rho_{0}=1.172$ which for the Hartle-Hawking state corresponds to a locally measured temperature of  $T_{loc}=0.26$. At greater distances from the black hole, the difference between the transition rates for these states decreases.}
    \label{fig:Static_omega}
\end{figure}

In Fig.~\ref{fig:Static_omega}, we examine how the transition rates for both the Boulware and Hartle-Hawking states respond to variations in energy gap. To avoid transient effects, we work in the limit of infinite detection time. We see here that the transition rate for the Hartle-Hawking state asymptotes to that of the Boulware state for large magnitude energy gap. However, the dependence on $\omega$ differs significantly between the states near $\omega=0$. This disparity becomes more pronounced as the local temperature increases (and the distance to the black hole decreases), with the transition rate for the Hartle-Hawking state increasing for small $\omega$ as the local temperature increases. 

We now turn our attention to the anti-Hawking effect. Recall that the expected behaviour, apropos the Hawking effect, is a transition rate that is monotonically increasing with local (KMS) temperature $T_{\textrm{loc}}$. In Ref.~\cite{Mann2020}, it was observed that the response function (not the transition rate) had a region of parameter space that decreased with respect to local temperature in the context of a BTZ spacetime. This was dubbed the ``anti-Hawking" effect and was later confirmed for a variety of boundary conditions by Campos and Dappiaggi in Refs. \cite{Campos2021,CamposDappiaggi2021}. On the other hand, they found no evidence of the effect for massless, topological black holes in four-dimensions. 
	\begin{figure}[!htp]
		\centering
		\includegraphics[scale=.55]{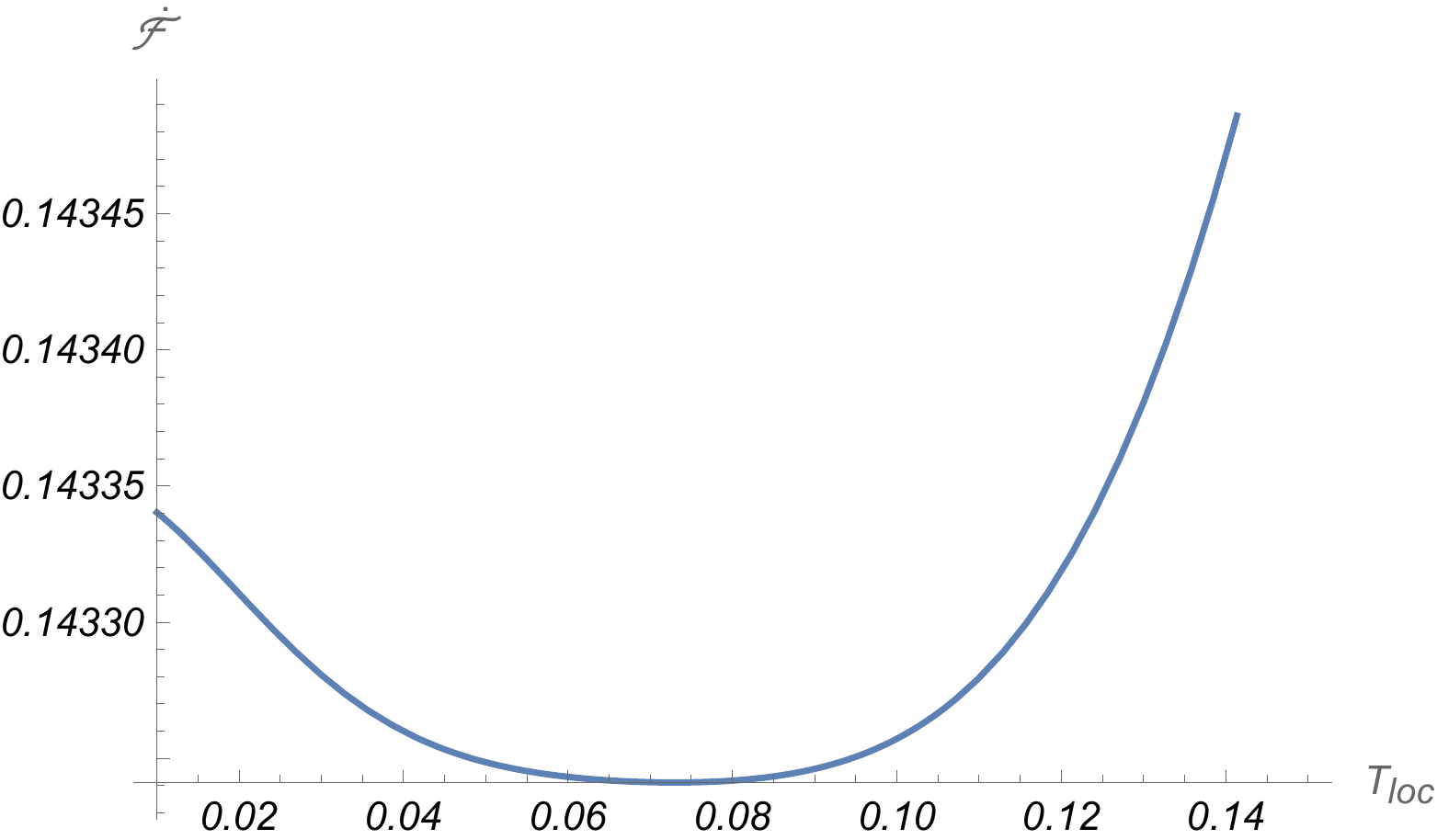}
		\caption{Plot of transition rate for a static detector coupled to a field in the Hartle-Hawking state. We have chosen the energy gap to be $\omega= -0.9$ with detection time $\Delta\tau=20$. Note that the transition rate decreases with small local Hawking temperature.}
		\label{fig:TlocStatic}
	\end{figure}
	
In the static case, it is certainly possible to find regions of the parameter space where the transition rate decreases with increasing local temperature, as seen in Fig.~\ref{fig:TlocStatic}. The effect is observed only when the local temperature is small. The real question, however, is whether these regions of parameter space where the transition rate is negatively correlated with the local KMS temperature is a transient effect, or a genuine physical phenomena we might call the anti-Hawking effect.
	\begin{figure}[!htp]
		\centering
\subfloat[]{\includegraphics[width=\linewidth]{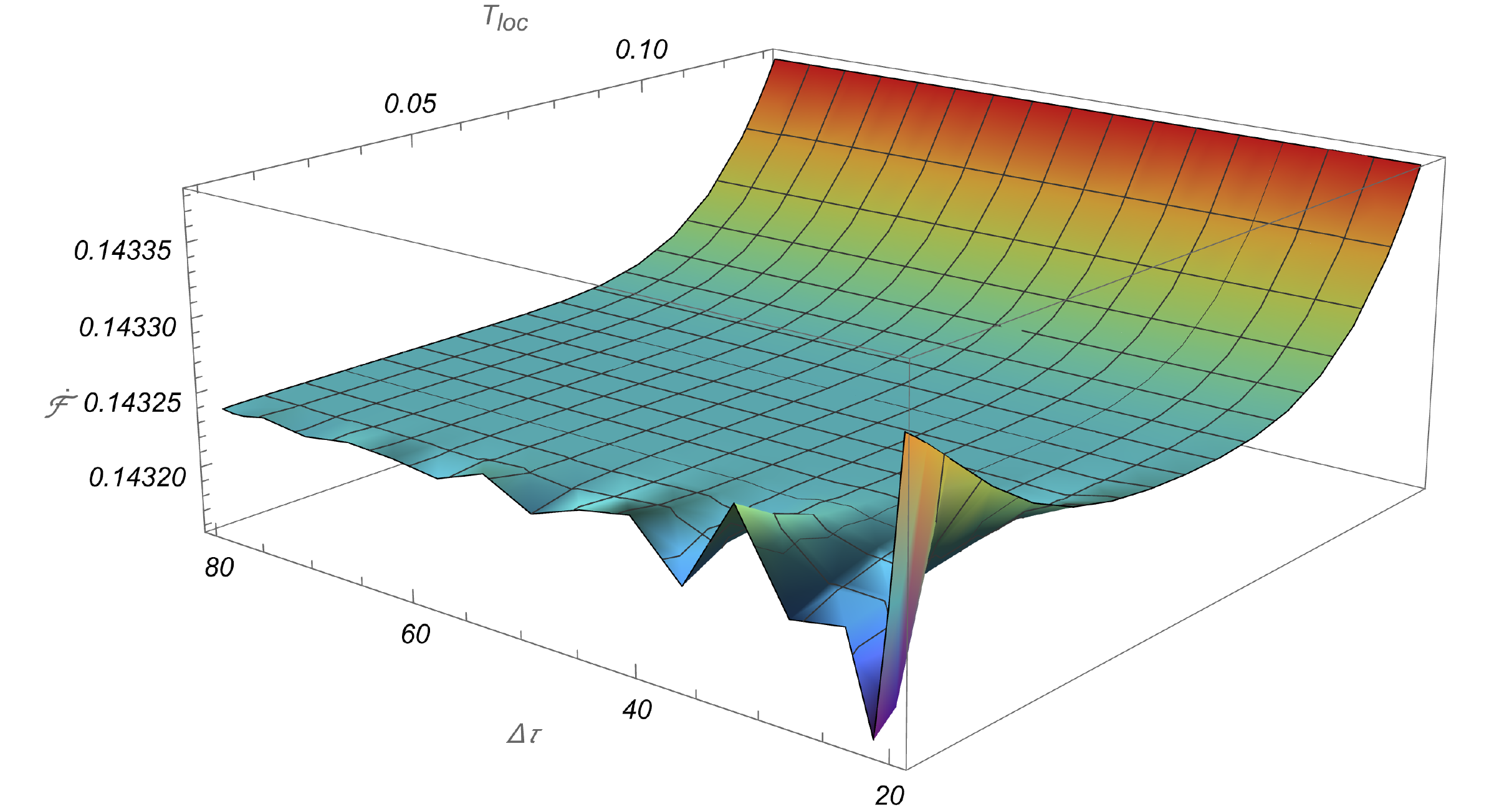}}\\
\subfloat[$T_{loc}=0.01$]{\includegraphics[width=\linewidth]{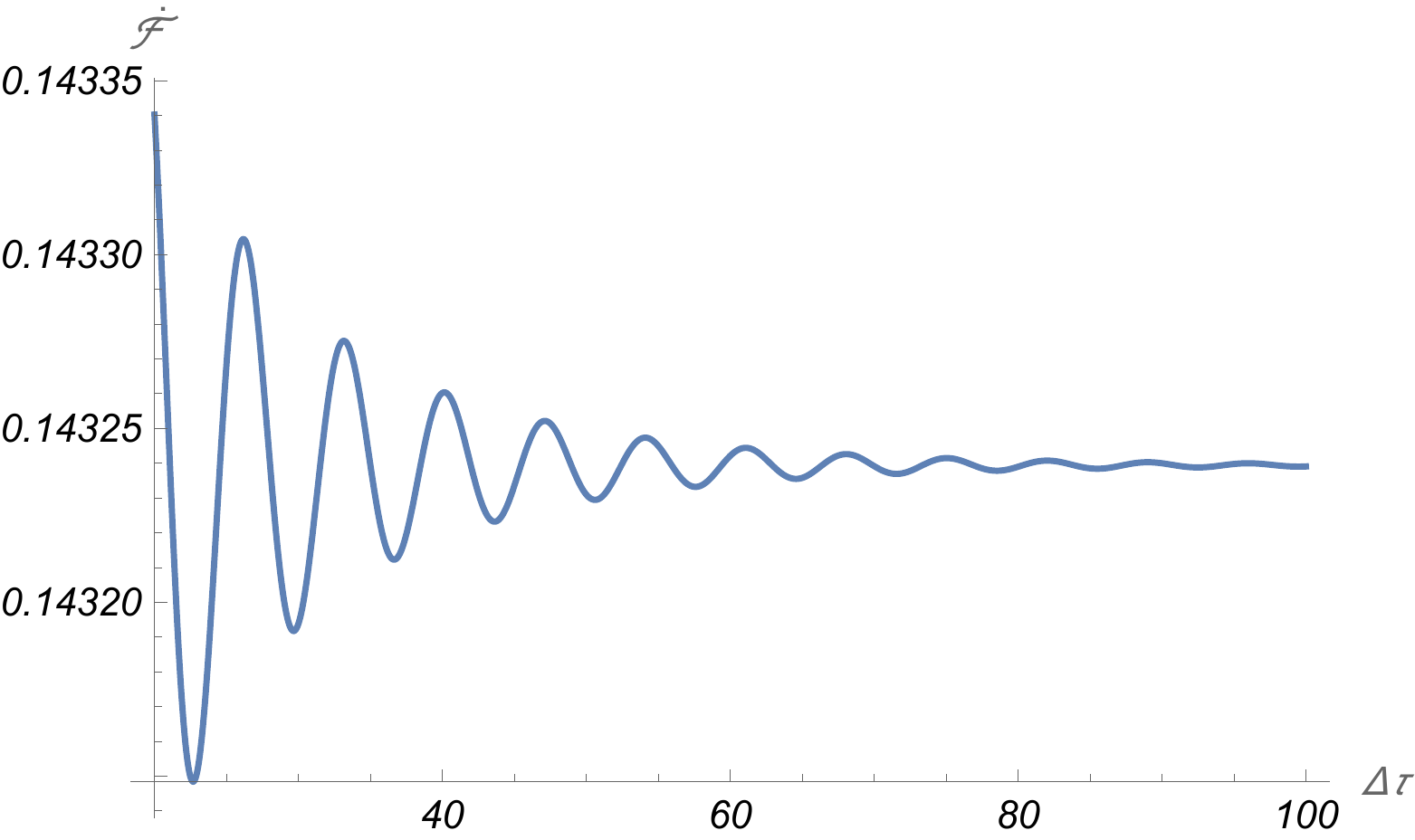}}
		\caption{Fig.~\ref{fig:Static3D}~(a) shows the transition rate for a static detector coupled to a field in the Hartle-Hawking state as a function of local temperature $T_{loc}$ and detection time $\Delta\tau$. In Fig.~\ref{fig:Static3D}~(b) we consider the $T_{loc}=0.01$ slice of the 3D plot in Fig.~\ref{fig:Static3D}~(a), which corresponds to a radius of approximately $\rho_0=15.9$. In each case, $\omega=-0.9$.}
		\label{fig:Static3D}
		\end{figure}	
		
To this end, we plot the transition rate as a function of both local temperature and detection time in Fig.~\ref{fig:Static3D}~(a). Here we can identify that the slices of constant small $T_{\textrm{loc}}$ have damped oscillations in detection time. Fig.~\ref{fig:Static3D}~(b) shows one of these slices for $T_{\textrm{loc}}=0.01$. If we choose the detection time $\Delta\tau$ to coincide with one of the peaks in Fig.~\ref{fig:Static3D}~(b) and plot the transition rate as a function of local temperature, the result will contain a region where the transition rate decreases as a function of local temperature. This can clearly be seen in the 3D plot in Fig.~\ref{fig:Static3D}~(a). The fact that these oscillations are damped with increased detection times suggests that this may be a transient effect. If the effect persists beyond the thermalization timescale however, we would interpret this effect as non-transient.
		
\begin{figure}[!htp]
		\centering
		\includegraphics[scale=.55]{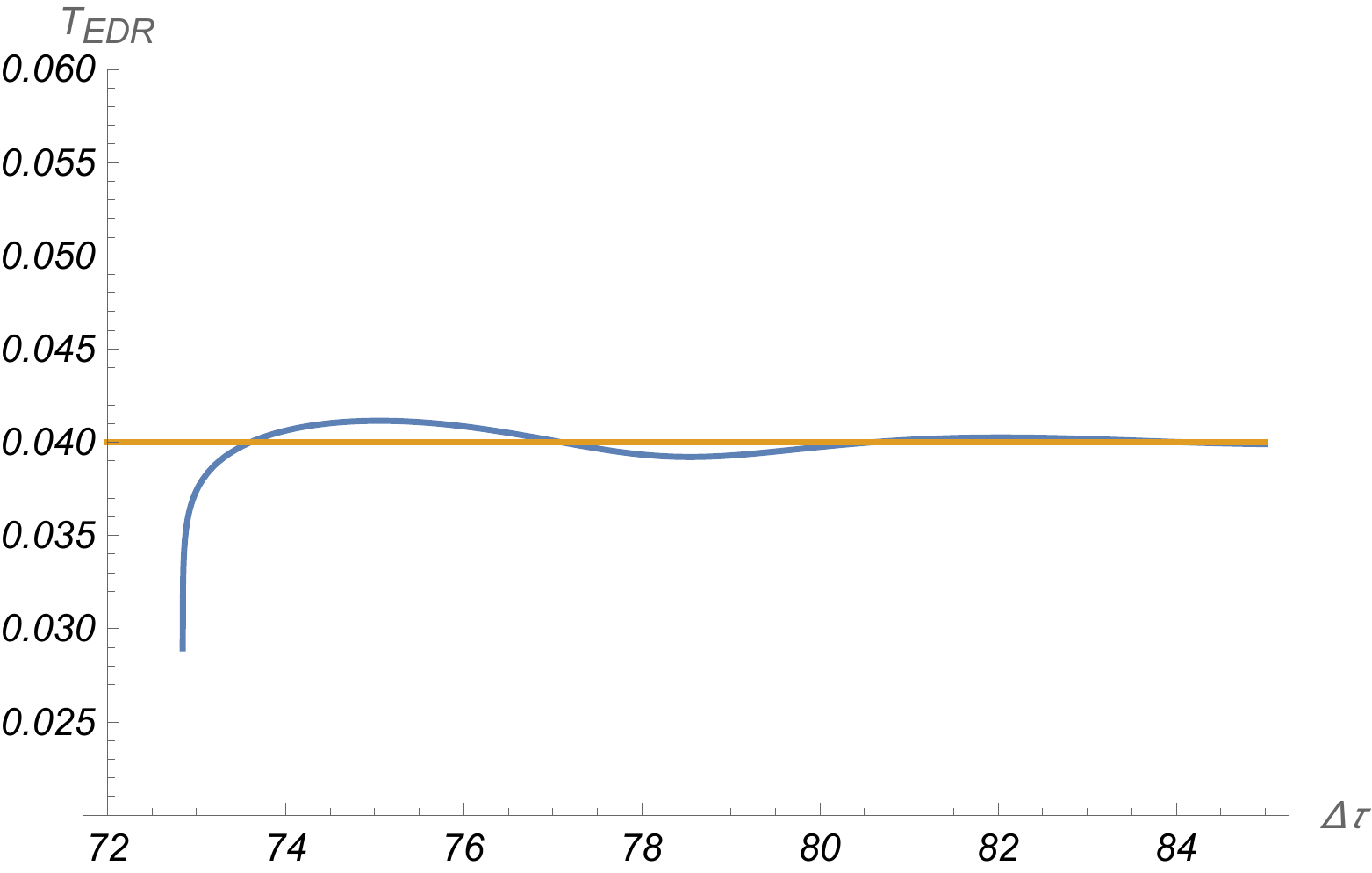}
		\caption{The temperature estimator $T_{\textrm{EDR}}$ (blue) defined in Eq.~(\ref{TEDR}) is plotted as a function of detection time $\Delta\tau$ where it is found to approximate locally-measured Hawking temperature (orange) when the detection time is large. Here, we have chosen $\omega=-0.9$ and $T_{loc}=0.04$, which corresponds to a radius of approximately $\rho_0\approx 4.1$.}
		\label{fig:StaticTEDR}
		\end{figure}

		We investigate this possibility in Fig.~\ref{fig:StaticTEDR} by plotting the temperature estimator $T_{\textrm{EDR}}$ as a function of detection time and comparing with local temperature. The figure shows that $T_{\textrm{EDR}}$ approximates local temperature when the detection time is large. Indeed they become approximately equal at around the same time that the oscillations in Fig.~\ref{fig:Static3D} become negligible. We infer from this that the oscillations in Fig.~\ref{fig:Static3D} will be negligible if the detection time is long enough for the detector and field to have thermalized, leaving only a transition rate that increases monotonically as a function of local temperature. Hence, the effect identified in Fig.~\ref{fig:TlocStatic} is transient and we find no evidence of the anti-Hawking effect for static detectors. However, we remind the reader that we have checked only for the field in Dirichlet boundary conditions. It might be the case that the anti-Hawking effect is present for other boundary conditions.

		\subsubsection*{Circular trajectories}
	While there are no circular geodesics in the spacetime, we can nevertheless consider a particle detector which is  accelerating along a circular orbit according to
	\begin{align}
	\label{eq:circtrajectory}
		\rho(\tau)&=\rho_{0}\nonumber\\
		\phi(\tau)&=h\,\tau\nonumber\\
		t(\tau)&= \frac{L}{\sqrt{\rho_{0}^{2}-1}}\tau,
	\end{align}
where as before, we have $h=\sqrt{L^{2}-1}$.	The acceleration required to ensure such a trajectory is given by
\begin{equation}
	\label{circacc}
	|a|=\frac{\rho_0 L^2}{\sqrt{\rho_0^2-1}}.
\end{equation}

For a circular trajectory, the Wightman propagator is independent of detection time so that Eqs. \eqref{WightB} and \eqref{WightH} have only $s$-dependency. It is helpful to redefine the variable of integration in the transition rate \eqref{eq:transitionratesharp} from $s$ to $\psi\equiv\Delta\phi$. This allows us to treat $\phi$ as a proxy for detection time with $\phi=2\pi$ corresponding to the time it takes the detector to complete one full revolution around the black hole at some orbital radius $\rho_0$. With this in mind, we write the transition rate as 
\begin{align}
\label{CircTransphi}
    \dot{\mathcal{F}}_{\tau}(\omega)&=\frac{2}{h}\int_{0}^{\phi}d\psi\left[\cos\left(\frac{\omega\psi}{h}\right)W(\psi)+\frac{h^{2}}{4\pi^{2}\psi^{2}}\right]\nn\\
    &-\frac{\omega}{4\pi}+\frac{h}{2\pi^{2}\phi},
\end{align}
where $W(\psi)$ represents the Wightman propagator defined in Eqs. \eqref{WightB}, \eqref{WightH}, and \eqref{eq:WightT} for the Boulware, Hartle-Hawking and arbitrary KMS state, respectively. In these expressions, for the circular trajectory (\ref{eq:circtrajectory}), we have
\begin{align}
    \eta=\mathrm{arccosh}\left(\frac{\rho_{0}^{2}-\cos\psi}{\rho_{0}^{2}-1}\right),\quad \Delta\mathsf{t}=\frac{L\psi}{h\sqrt{\rho_{0}^{2}-1}}.
\end{align}

\begin{figure}[!htp]
    \centering
  \subfloat[$\omega=-0.2$]{  \includegraphics[width=\linewidth]{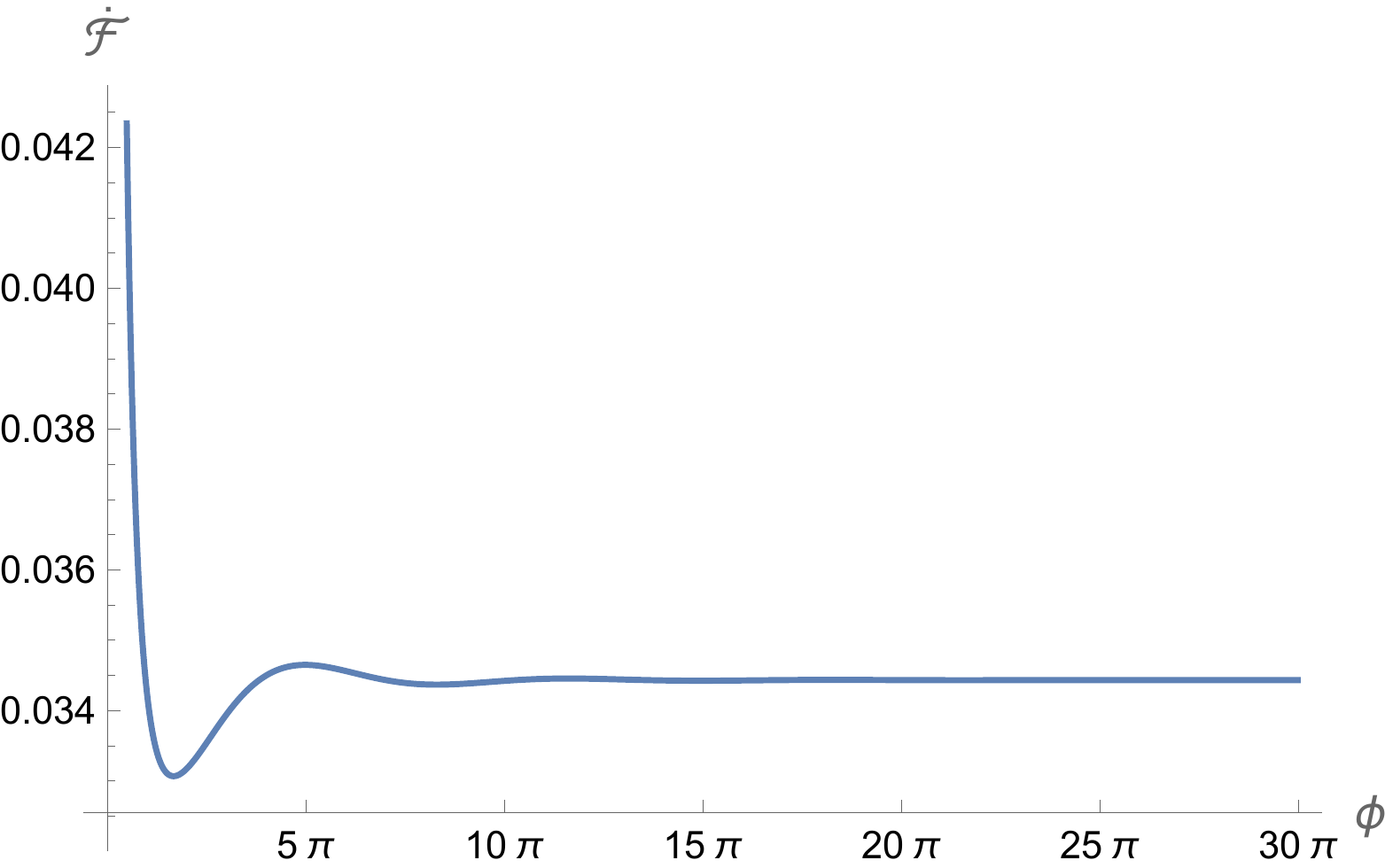}}\\
  \subfloat[$\omega=-2$]{  \includegraphics[width=\linewidth]{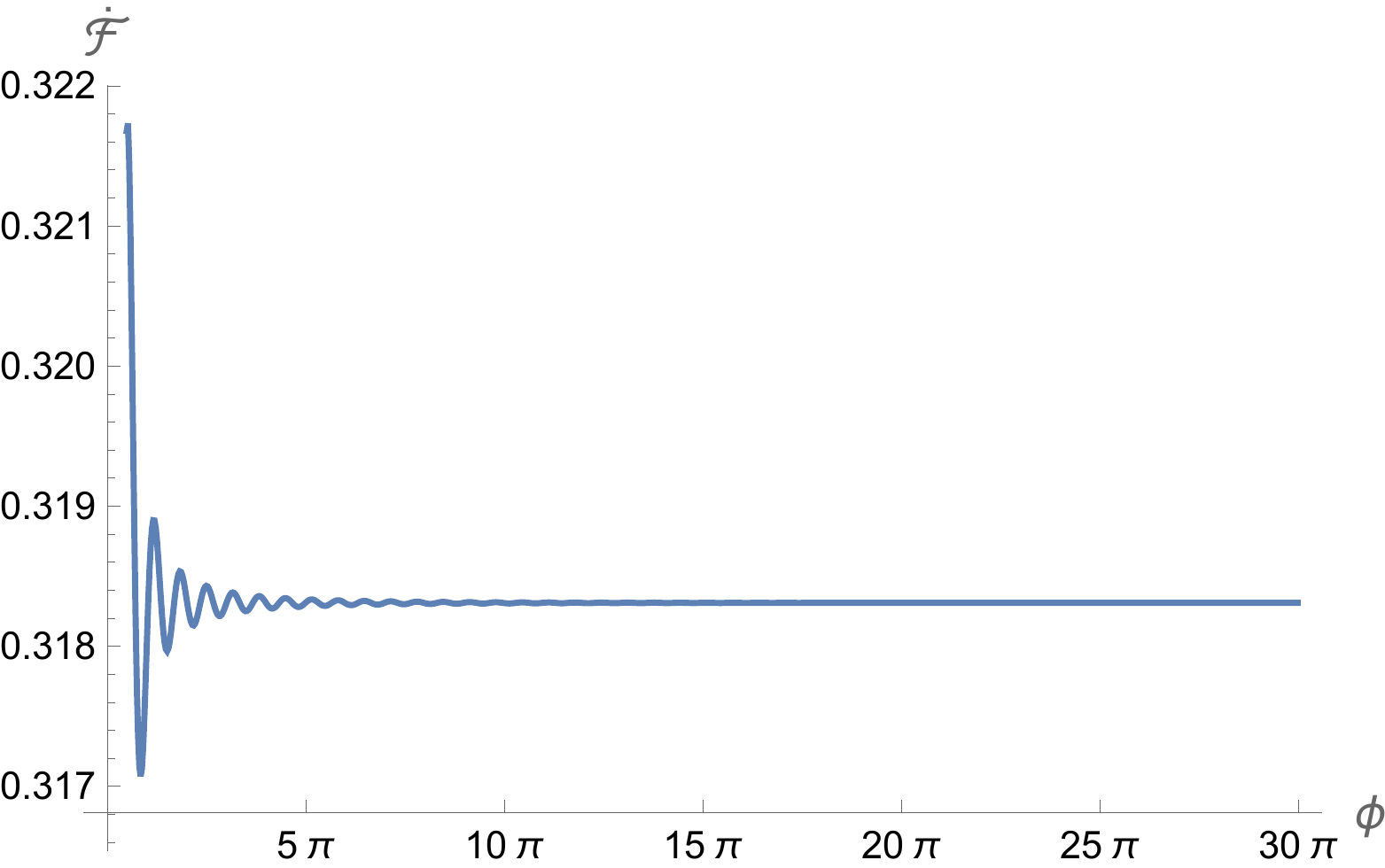}}
    \caption{Plot of the transition rate for the Hartle-Hawking state as a function of $\phi$. In Fig.~\ref{fig:Circ_phi}~(a) we have chosen $\omega=-0.2$. In Fig.~\ref{fig:Circ_phi}~(b), we have $\omega=-2$. In both cases we have chosen $\rho_0=20$ and $L=1.2$.}
    \label{fig:Circ_phi}
\end{figure}

We begin by investigating how the transition rate depends on the detection time which is measured by $\phi$. Examples of the profiles we obtain for two different values of energy gap are plotted in Fig.~\ref{fig:Circ_phi}. We observe that for the first few orbits around the black hole, the profiles are dominated by transient oscillations with a frequency proportional to the magnitude of the energy gap. These oscillations dampen as the transition rate asymptotes to an approximately constant value. The memory of the early transient effects is extraordinarily long however and tiny oscillations about the approximately constant value can be observed even after hundreds of orbits.

\begin{figure}[!htp]
    \centering
    \includegraphics[width=\linewidth]{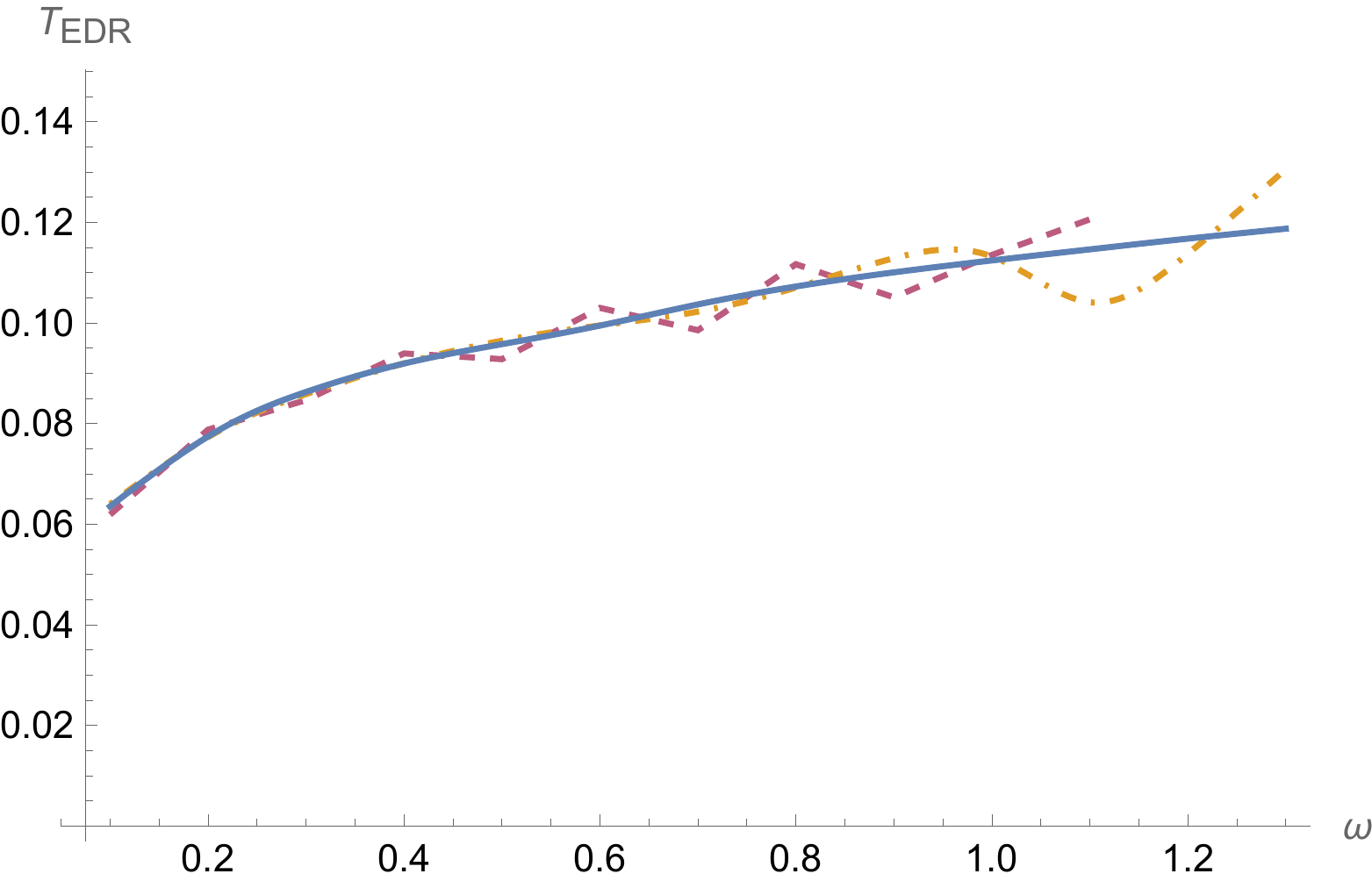}
    \caption{Plot of $T_{\textrm{EDR}}$ as a function of energy gap $\omega$ for a detector in a circular orbit coupled to a field in the Hartle-Hawking state. Here we have chosen an orbital radius of $\rho_0=20$, the angular momentum to be $L=1.2$, and we considered detection times of $\phi=6\pi$ (purple, dashed), $\phi=10\pi$ (yellow, dot-dashed), and $\phi=300\pi$ (blue).
    }
    \label{fig:Circ_TEDR}
\end{figure}

We wish also to clarify whether the detector on a circular orbit thermalizes in any region of the parameter space. To this end we consider the temperature estimator $T_{\textrm{EDR}}$ defined in Eq.~\eqref{TEDR}. First we examine $T_{\textrm{EDR}}$ as a function of $\omega$ for various detection times for a field in the Hartle-Hawking state. One only expects $T_{\textrm{EDR}}$ to be a meaningful measure of the detector's temperature in the limit of long detection time. In Fig.~\ref{fig:Circ_TEDR}, we see that the plots for $T_{\textrm{EDR}}$ for shorter detection times ($\phi=6\pi$ and $\phi=10\pi$) are dominated by transient oscillations whereas for sufficiently long detection times (we take $\phi=300\pi$ in Fig.~\ref{fig:Circ_TEDR}) these oscillations are no longer present and we observe a slow monotonic increase with increasing energy gap. Similar behaviour is reported for a detector on a circular geodesic in the Schwarzschild black hole spacetime in Ref.~\cite{HodgkinsonLoukoOttewill}. Though it is difficult to discern from the plot, it appears that for sufficiently long detection times, these curves asymptote to a constant temperature for large $\omega$, i.e., the detector thermalizes for large energy gap provided the detection time is sufficiently long. However, as in the case of a detector on a circular geodesic in Schwarzschild, the temperature that the detector thermalizes to is hotter than the locally measured Hartle-Hawking temperature or indeed the Doppler shifted local temperature.

\begin{figure}[!htp]
    \centering
      \includegraphics[width=\linewidth]{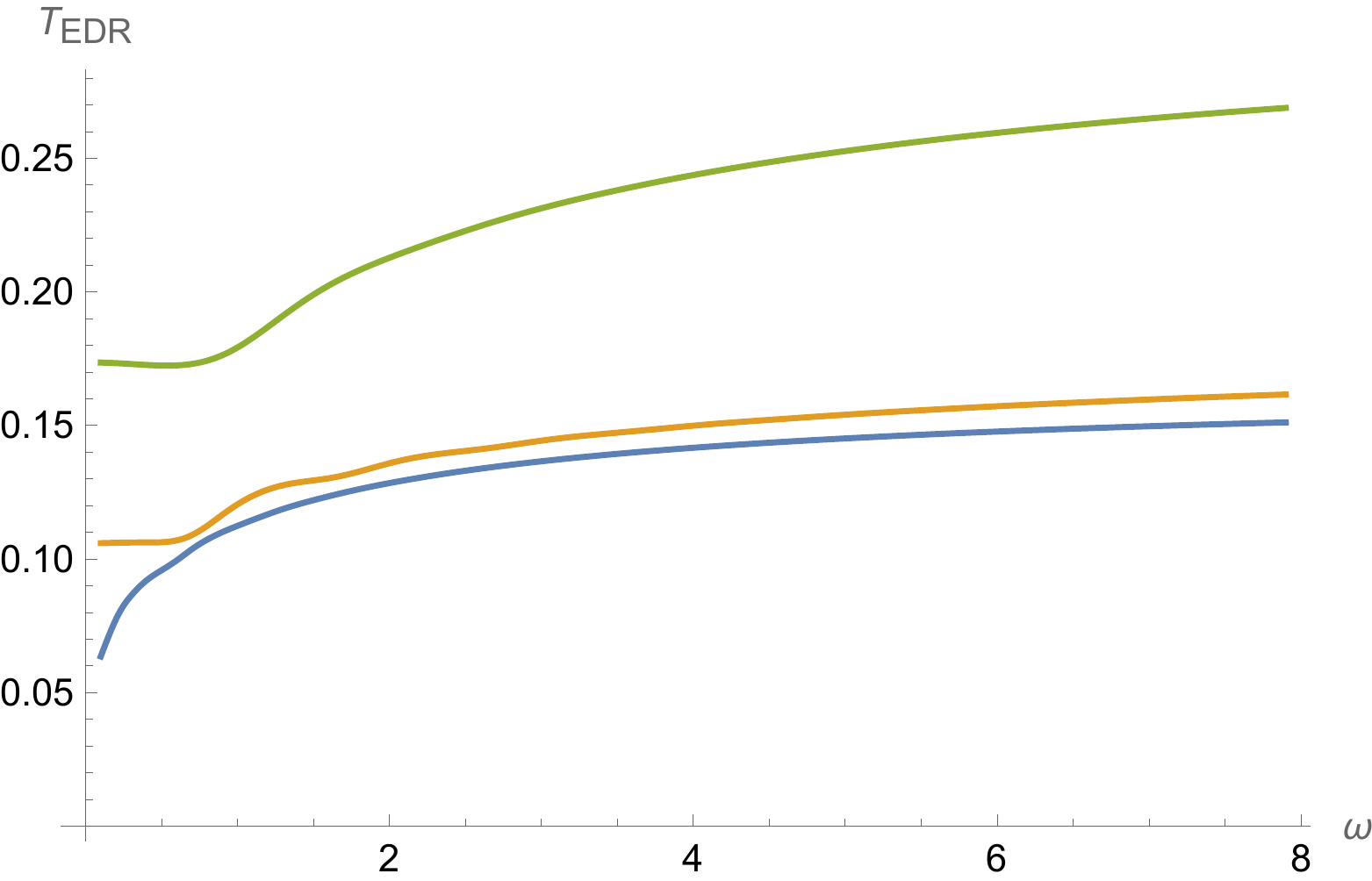}
    \caption{Plot of $T_{\textrm{EDR}}$ as a function of energy gap for a detector on a circular orbit coupled to a field in various KMS states. We have chosen $L=1.2$ and $\rho_0=20$ and $\phi=300\pi$. We consider states with temperature $T= T_{\textrm{H}}$ (blue), $T=10\,T_{\textrm{H}}$ (yellow), and $T=20\,T_{\textrm{H}}$ (green).}
    \label{fig:Circ_TEDR_b}
\end{figure}

In Fig.~\ref{fig:Circ_TEDR_b}, we examine $T_{\textrm{EDR}}$ for a detector coupled to a field in other KMS states. It is challenging to compute $T_{\textrm{EDR}}$ for large energy gap since the transition rate tends to zero exponentially with increasing $\omega$. This means that to compute Eq.~(\ref{TEDR}) for larger $\omega$ requires tremendous accuracy. The plots in Fig.~\ref{fig:Circ_TEDR_b} are obtained by using the built-in numerical integrator in Mathematica with an accuracy goal of 350 decimal places and a working precision of 400 decimal places. We find that for each KMS state, $T_{\textrm{EDR}}$ appears to asymptotes to a constant value for large $\omega$ (provided the detection time is long). As expected, the temperature that $T_{\textrm{EDR}}$ limits to for large energy gap increases with increasing field temperature. A surprising feature of these plots is that for a range of field temperatures higher than the Hartle-Hawking temperature, $T_{\textrm{EDR}}$ can actually decrease for small increasing $\omega$. As we increase the field temperature in Fig.~\ref{fig:Circ_TEDR_b}, we see the profiles flatten out for small $\omega$, eventually developing into a small region where $T_{\textrm{EDR}}$ decreases as a function of $\omega$. This only appears to occur for a finite range of field temperature, that is, if we continue to increase the field temperature, the monotonically increasing behavior reemerges.

To analyse whether the detector on a circular trajectory thermalizes at large energy gap for long detection times, let us start by considering the limit of infinite detection time, ignoring the physical impossibility of accelerating indefinitely. Following Ref.~\cite{HodgkinsonLoukoOttewill}, we can rearrange the terms in the expression for the transition rate so we have
\begin{align}
\label{eq:transitionrearranged}
    \dot{\mathcal{F}}_{\tau}(\omega)=\dot{\mathcal{F}}_{\tau}^{\textrm{corr}}(\omega)+\frac{|\omega|\,\Theta(-\omega)}{2\pi},
\end{align}
with
\begin{align}
\label{eq:transitioncorrection}
    \dot{\mathcal{F}}_{\tau}^{\textrm{corr}}(\omega)=2\int_{0}^{\infty}ds\,\cos(\omega\,s)\Big(W(\tau,\tau-s)+\frac{1}{4\pi^{2}s^{2}}\Big).
\end{align}
The last term in Eq.~(\ref{eq:transitionrearranged}) is just the transition rate for a static detector coupled to a field in the Boulware vacuum in the limit of infinite detection time. Hence, we can view Eq.~(\ref{eq:transitioncorrection}) as the correction to this resulting from the temperature of the state and the dynamics of the detector. From the Riemann-Lebesgue lemma, $\dot{\mathcal{F}}_{\tau}^{\textrm{corr}}(\omega)$ will tend to infinity in the limit $\omega\to\infty$. More than that, this ought to fall off faster than $\omega^{-n}$ for any positive integer $n$. To estimate this more precisely, we write the integral explicitly as
\begin{align}
    \dot{\mathcal{F}}_{\tau}^{\textrm{corr}}(\omega)&=\frac{1}{4\pi^{2}}\int_{-\infty}^{\infty}ds\,e^{i\,|\omega|\,s}\Bigg(\frac{1}{s^{2}}\nonumber\\
    &+\frac{\pi\,T\,\sinh(2\pi\,T\,\eta)}{\sinh\eta\,\sinh(\pi T(\eta+\Omega\,s))\,\sinh(\pi T(\eta-\Omega\,s))}\Bigg)
\end{align}
where we have defined $\Omega\equiv L/\sqrt{\rho_{0}^{2}-1}$ and we have used the fact that the integrand is an even function of $s$. Now, this integrand is regular everywhere along the real $s$-axis including $s=0$. We can recast this integral into a complex one with a contour that is closed in the upper-half plane. Deforming the contour in a small semi-circle centred on $s=0$ means that the contribution coming from the $1/s^{2}$ term vanishes in the limit where the radius tends to zero. Similarly, the contribution coming from the large semi-circle vanishes in the limit of large semi-circle radius. Hence, the integral we wish to evaluate can be computed from a standard application of Cauchy's residue theorem,
\begin{align}
\label{eq:residues}
    \dot{\mathcal{F}}_{\tau}^{\textrm{corr}}(\omega)&=2\pi\,i \sum_{k}\textrm{Res}(W, z_{k})
\end{align}
where $z_{k}$ are the residues in the upper-half plane of the propagator
\begin{align}
\label{eq:Wcomplex}
W=\frac{T}{4\pi}\frac{e^{i\,|\omega|\,z}\sinh(2\pi\,T\,\eta)}{\sinh\eta\,\sinh(\pi T(\eta+\Omega\,z))\,\sinh(\pi T(\eta-\Omega\,z))},
\end{align}
and $\eta$ is given by
\begin{align}
    \eta=\textrm{arccosh}\left(\frac{\rho_{0}^{2}-\cos(z\,h)}{\rho_{0}^{2}-1}\right).
\end{align}

Since we are only interested in the behaviour as $|\omega|\to\infty$, the dominant contribution from Eq.~(\ref{eq:residues}) comes from the residue with smallest imaginary part, those with larger imaginary parts being exponentially suppressed by the factor of $e^{i\,|\omega| z}$ in the numerator of Eq.~(\ref{eq:Wcomplex}). The poles are simple and come from the $\sinh(\pi T(\eta\pm\Omega\,z)$ terms in the denominator. When the arguments here are zero or integer multiple of $\pi \,i$ then we have a pole. If we take $z=i Z$ and look for the poles with the smallest imaginary part, then the result depends on the temperature of the field, but is always either $Z_{+}$ or $Z_{-}$ satisfying the transcendental equations
\begin{align}
    \cosh(h\,Z_{+})&=1-(\rho_{0}^{2}-1)(\cos(\Omega\,Z_{+})-1)\nonumber\\
    \cosh(h\,Z_{-})&=1-(\rho_{0}^{2}-1)(\cos(\Omega\,Z_{-}-1/T)-1).
\end{align}
The critical temperature that delineates whether it is $Z_{+}$ or $Z_{-}$ that is the dominant contribution is a solution to the transcendental equation
\begin{align}
     \cosh\left(\frac{h}{2\Omega\,T_{\textrm{crit}}}\right)=1-(\rho_{0}^{2}-1)\left[\cos\left(\frac{1}{2 T_{\textrm{crit}}}\right)-1\right].
\end{align}
For $T<T_{\textrm{crit}}$, we obtain
\begin{align}
    \dot{\mathcal{F}}_{\tau}^{\textrm{corr}}(\omega)\sim \frac{1}{4\pi }\frac{e^{-|\omega|\,Z_{+}}}{h\,\sinh(h\,Z_{+})-(\rho_{0}^{2}-1)\Omega\,\sin(\Omega\,Z_{+})},\nonumber\\
    |\omega|\to\infty,
\end{align}
while for $T>T_{\textrm{crit}}$, we have
\begin{align}
    \dot{\mathcal{F}}_{\tau}^{\textrm{corr}}(\omega)\sim \frac{1}{4\pi }\frac{e^{-|\omega|\,Z_{-}}}{h\,\sinh(h\,Z_{-})-(\rho_{0}^{2}-1)\Omega\,\sin(\Omega\,Z_{-}-\frac{1}{T})}\nonumber\\
    |\omega|\to\infty.
\end{align}
In either case, upon reference to Eq.~(\ref{eq:transitionrearranged}), one can explicitly show that the detailed balance form of the KMS condition (\ref{TEDR}) is satisfied at the temperature
\begin{align}
    T_{\textrm{circ}}=\begin{cases}
   \displaystyle{T_{+}=\frac{1}{Z_{+}}},\quad & T<T_{\textrm{crit}}\\
   \displaystyle{T_{-}=\frac{1}{Z_{-}}},\quad & T>T_{\textrm{crit}}.
    \end{cases}
\end{align}

\begin{figure}[!htp]
    \centering
    \subfloat[]{\includegraphics[width=\linewidth]{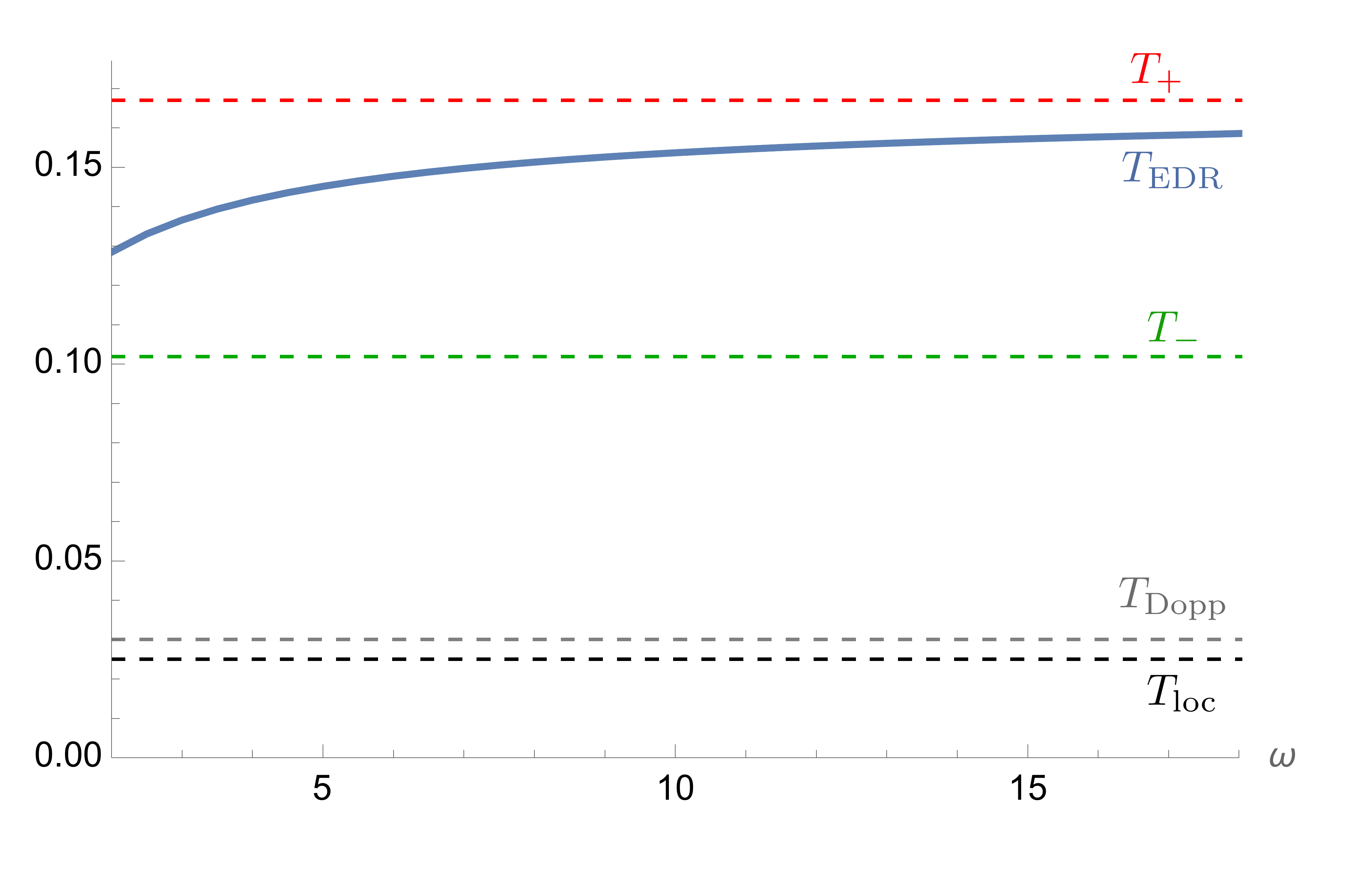} } 
    
    \subfloat[]{\includegraphics[width=\linewidth]{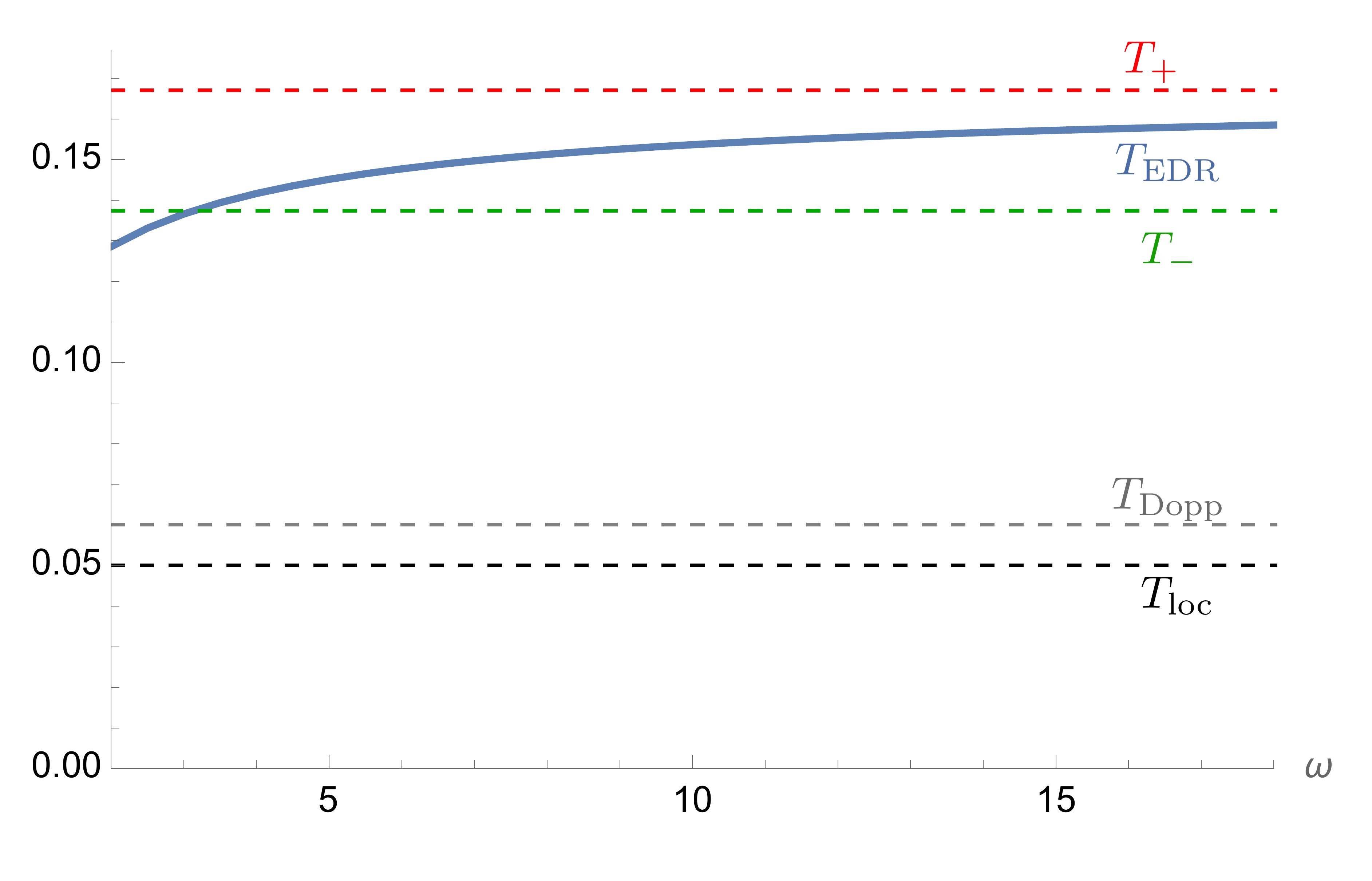} }
    
    \subfloat[]{\includegraphics[width=\linewidth]{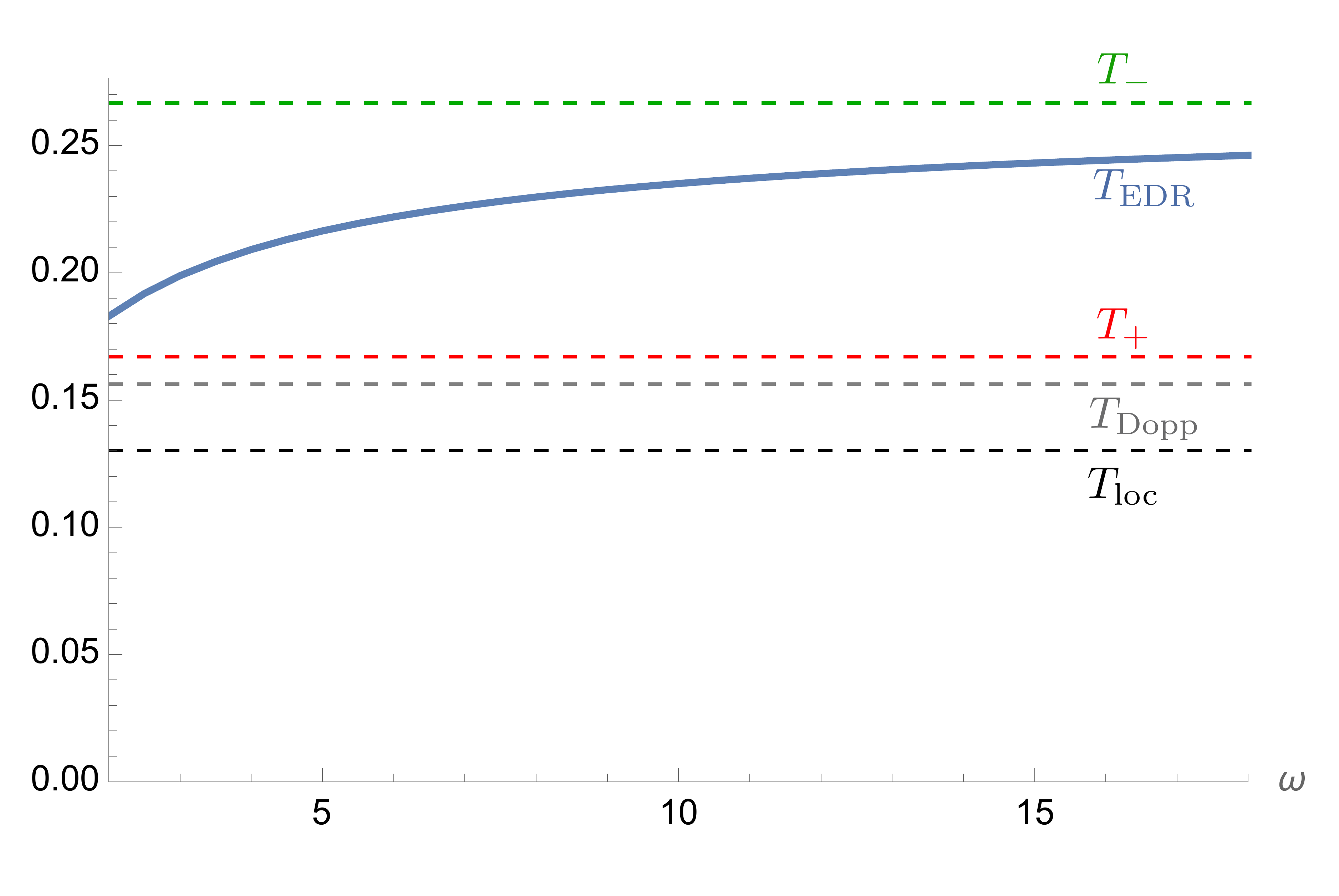} }
    \caption{Plot showing the behaviour of $T_{\textrm{EDR}}$ for large energy gap for a circular detector. The detector has parameters $\rho_{0}=20$ and $L=1.2$. The critical temperature for these parameters is approximately $T_{\textrm{crit}}\approx 1.39$. In plot (a), the field is in a KMS state at temperature $T=0.5$ while in plot (b), the field temperature is $T=1$. Since both are less than the critical temperature, $T_{\textrm{EDR}}$ asymptotes to $T_{+}$ independently of the field temperature. Plot (c) has field temperature $T=2.6>T_{\textrm{crit}}$ and hence $T_{\textrm{EDR}}$ asymptotes to $T_{-}$ which depends on $T$.}
    \label{fig:TEDR_Large_Omega}
\end{figure}

We note that for a field in a KMS state below the critical temperature, the detector thermalizes at large energy gap to a temperature that does not depend on the temperature of the quantum state since $Z_\pm$ is independent of $T$. We see this in the first two plots of Fig.~\ref{fig:TEDR_Large_Omega}. Figures (a) and (b) show $T_{\textrm{EDR}}$ for a circular detector, with large $\omega$, coupled to a scalar field in a KMS state at temperatures below the critical temperature. In this case the detector does not see the ``ambient'' temperature of the scalar field, with its temperature influenced only by its acceleration. For temperatures greater than the critical temperature, the detector's temperature has contributions from both the acceleration and the ambient field temperature. The dependence on the field temperature is wholly contained in $Z_{-}$. This can be seen in Fig.~\ref{fig:TEDR_Large_Omega}~(c). In all cases, the detector's temperature is greater than the Doppler shifted local temperature $T_{\textrm{Dopp}}=-(u_{\textrm{static}}\cdot u_{\textrm{circ}}) T_{\textrm{loc}}=L\,T_{\textrm{loc}}$. This is analogous to the case of a circular geodesic detector in Schwarzschild \cite{HodgkinsonLoukoOttewill}. 

\begin{figure}[!htp]
    \centering
\includegraphics[width=\linewidth]{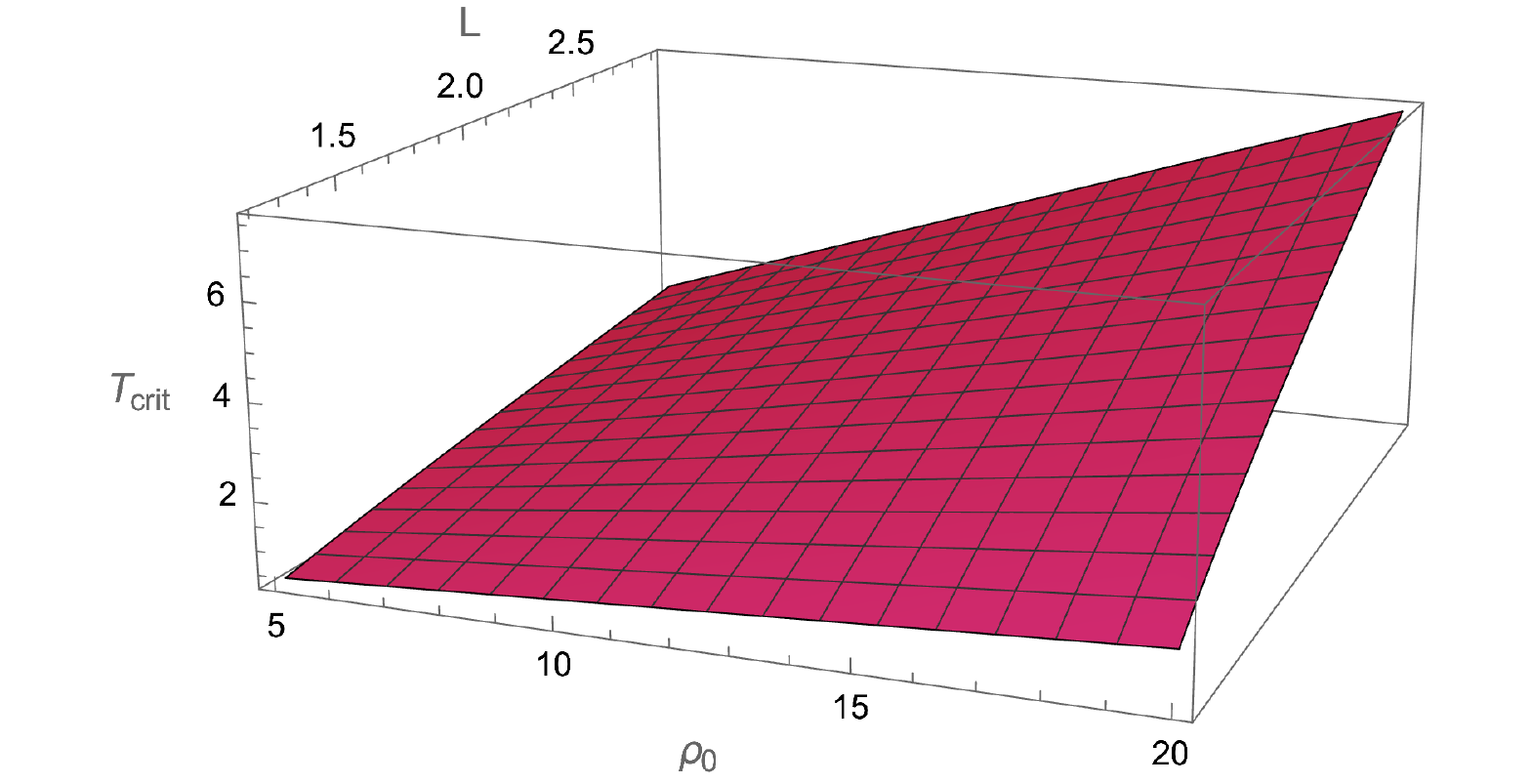}
    \caption{Plot showing $T_{\textrm{crit}}$ as a function of $\rho_{0}$ and $L$.}
    \label{fig:Tcrit}
\end{figure}

The critical temperature is an increasing function of both the radius of the circular orbit and the angular momentum. We plot $T_{\textrm{crit}}$ in Fig.~\ref{fig:Tcrit}. For orbits close to the black hole, the critical temperature is low and hence the detector (in the large $\omega$ limit) will only be sensitive to the ambient field temperature for very small field temperatures. On the other hand, for detectors with large angular momentum, the critical temperature is greater and hence the detector is only sensitive to the ambient temperature for very high field temperatures.

\begin{figure}[!htp]
    \centering
\subfloat[$L=1.01$]{    \includegraphics[width=\linewidth]{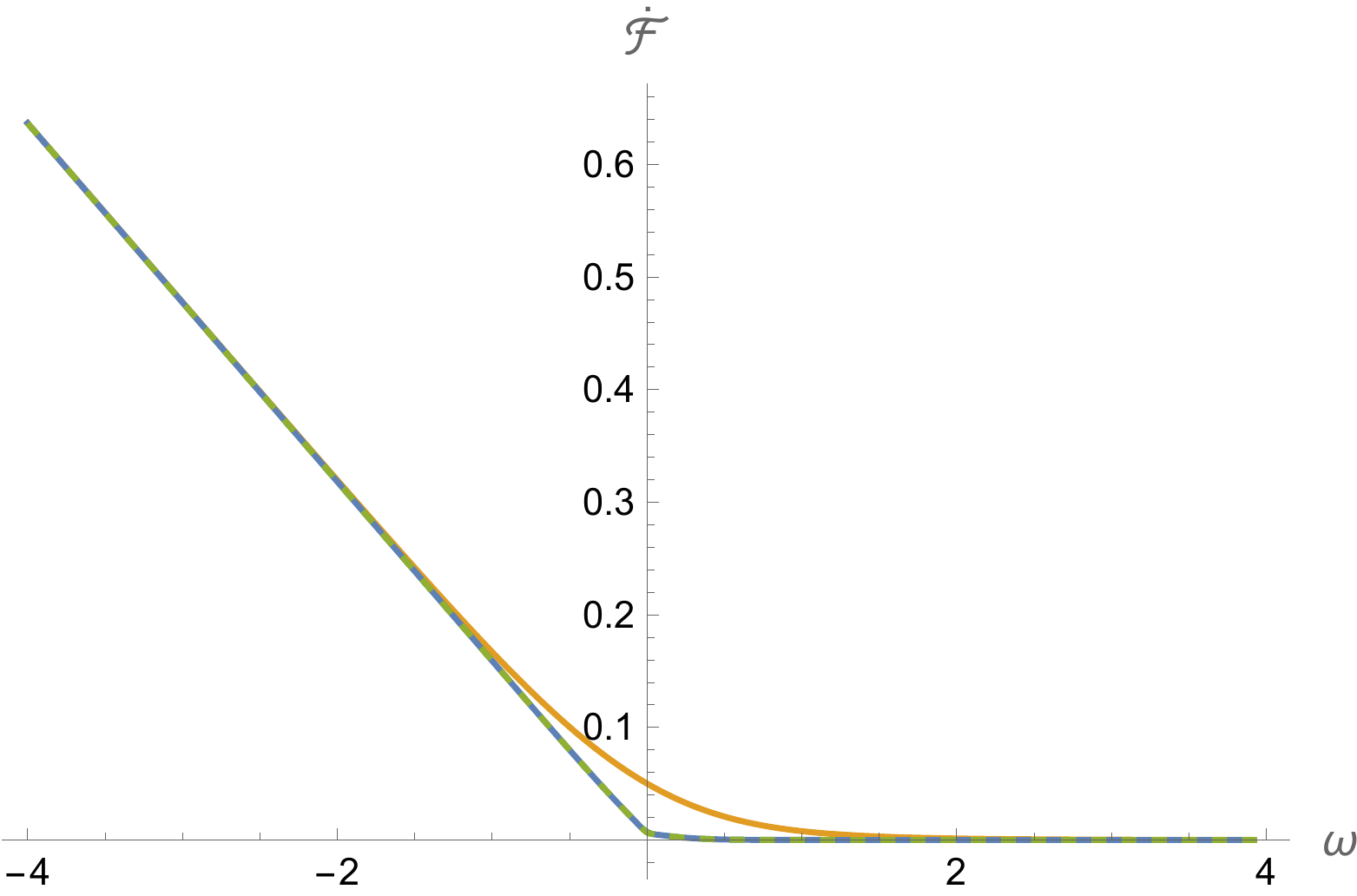}}\\
\subfloat[$L=2.25$]{    \includegraphics[width=\linewidth]{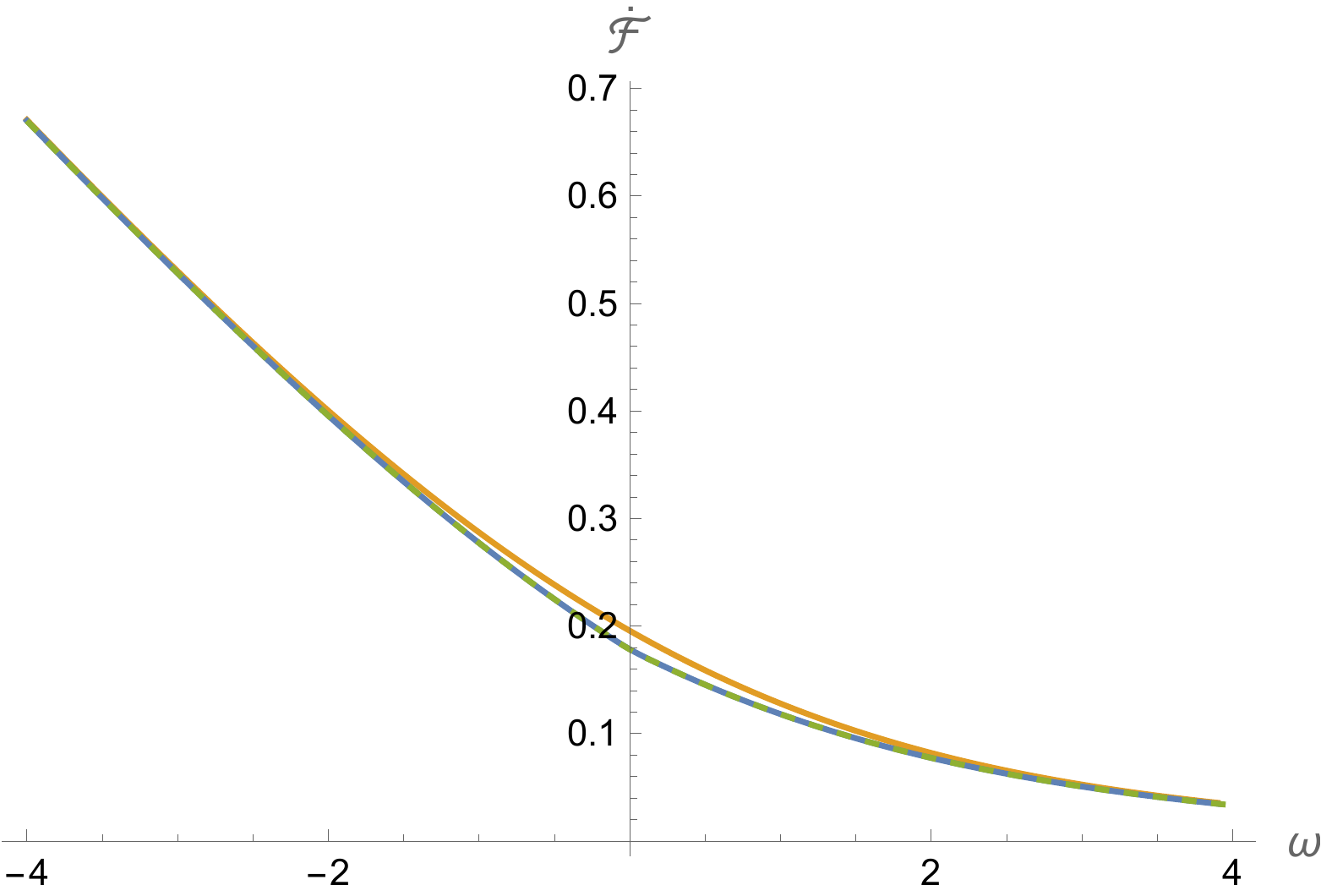}} 
    \caption{Figure (a) shows the transition rate for a particle detector in circular orbit with $L=1.01$ as a function of energy gap for various quantum states. Figure (b) shows the same plot but with higher angular momentum $L=2.25$. The plots show transition rates for the Boulware (green, dashed) and Hartle-Hawking (blue) states, though these are indistinguishable, as well as a KMS state with $T=40T_H$. In each case we have chosen a detection time $\phi=20\pi$.}
    \label{fig:Circ_omega}
\end{figure}

\begin{figure}[!htp]
    \centering
    \includegraphics[width=\linewidth]{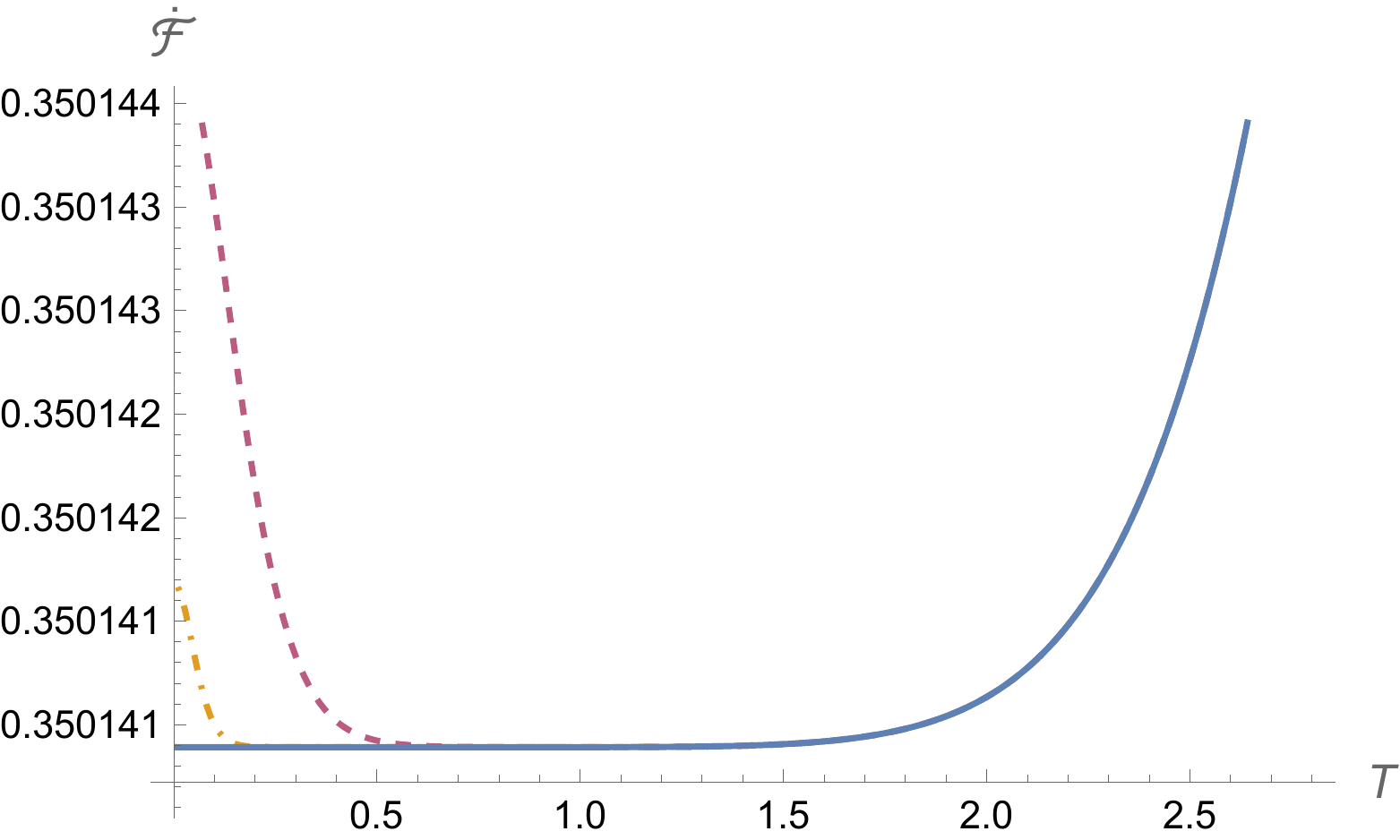}     
    \caption{Plot of transition rate for a detector on a circular trajectory as a function of KMS field temperature for various detection times. Here we have chosen $\omega=-2.2$, $\rho_0=20$, $L=1.2$ and detection times of $\phi=10\pi$ (purple, dashed), $\phi=30\pi$ (yellow, dot-dashed) and $\phi=300\pi$ (blue).}
    \label{fig:Circ_therm}
\end{figure}

Having established that the detector thermalizes in the limit of large energy gap, we continue to probe the parameter space of a detector in circular orbit. In Fig.~\ref{fig:Circ_omega}~(a), we plot the transition rate as a function of energy gap for different quantum states and different angular momentum (which is tantamount to different accelerations). For small angular momentum ($L\approx 1$), we see profiles reminiscent of those for the static detector. In particular, we observe a sharp transition from negative to positive energy gaps for the Boulware state, with the transition rate for positive energy gap approaching zero. For thermal states, the transition rate near $\omega=0$ increases with increasing temperature. Increasing the angular momentum changes the profiles considerably. The distinction between the transition rates for different quantum states diminishes and is only appreciable for small energy gaps. The sharp transition observed for small acceleration in the Boulware case is no longer present for larger accelerations.
In Fig.~\ref{fig:Circ_therm}, we plot the transition rate as a function of field temperature for various detection times. The plot shows that the transition rate is anti-correlated with the field temperature for shorter detection times. However, after a detection time long enough for transient effects to be negligible, we find that the transition rate is monotonically increasing with field temperature, as seen by the blue curve in Fig.~\ref{fig:Circ_therm}. Hence, we find no evidence of the (weak) anti-Hawking effect.

\begin{figure}[!htp]
    \centering
    \includegraphics[width=\linewidth]{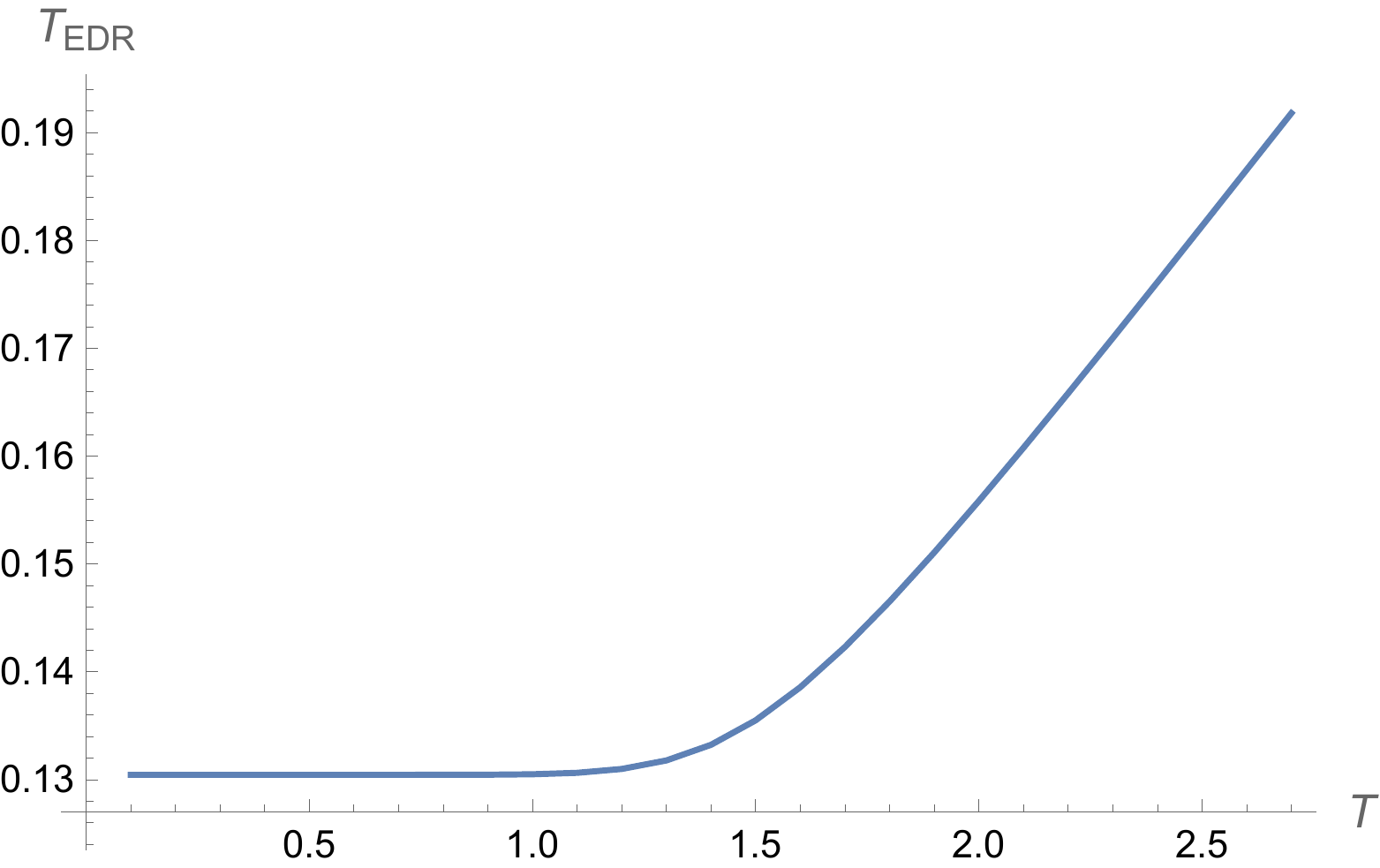}     
    \caption{Figure showing $T_{\textrm{EDR}}$ for a detector in a circular orbit as a function of KMS field temperature $T$. We have $\omega=-2.2$, $\rho_0=20$, $L=1.2$ and have taken the limit of large detection time.}
    \label{fig:Circ_TEDR_T}
\end{figure}

One could also search for evidence of the strong anti-Hawking effect, that is, non-transient anti-correlation between the the detector's temperature and the field temperature. In Fig.~\ref{fig:Circ_TEDR_T}, we plot the temperature estimator $T_{\textrm{EDR}}$ as a function of the states KMS temperature $T$. Again we observe only the expected monotonic increase once a detection time greater than the (approximate) thermalization timescale has been chosen. Hence, as the quantum field becomes hotter so too does the temperature of the detector, as we would expect intuitively. Of course, it is possible to find regions where $T_{\textrm{EDR}}$ decreases with increasing field temperature but only in the region of parameter space where the detector has not even approximately thermalized. Likewise, following a detailed search through the parameter space, we find no region of negative correlation that we could call the anti-Unruh effect, notwithstanding regions of negative correlation that can be attributed to transience.

\subsubsection*{Inspiral trajectories}
\begin{figure}[!htp]
\centering
\includegraphics[scale=.5]{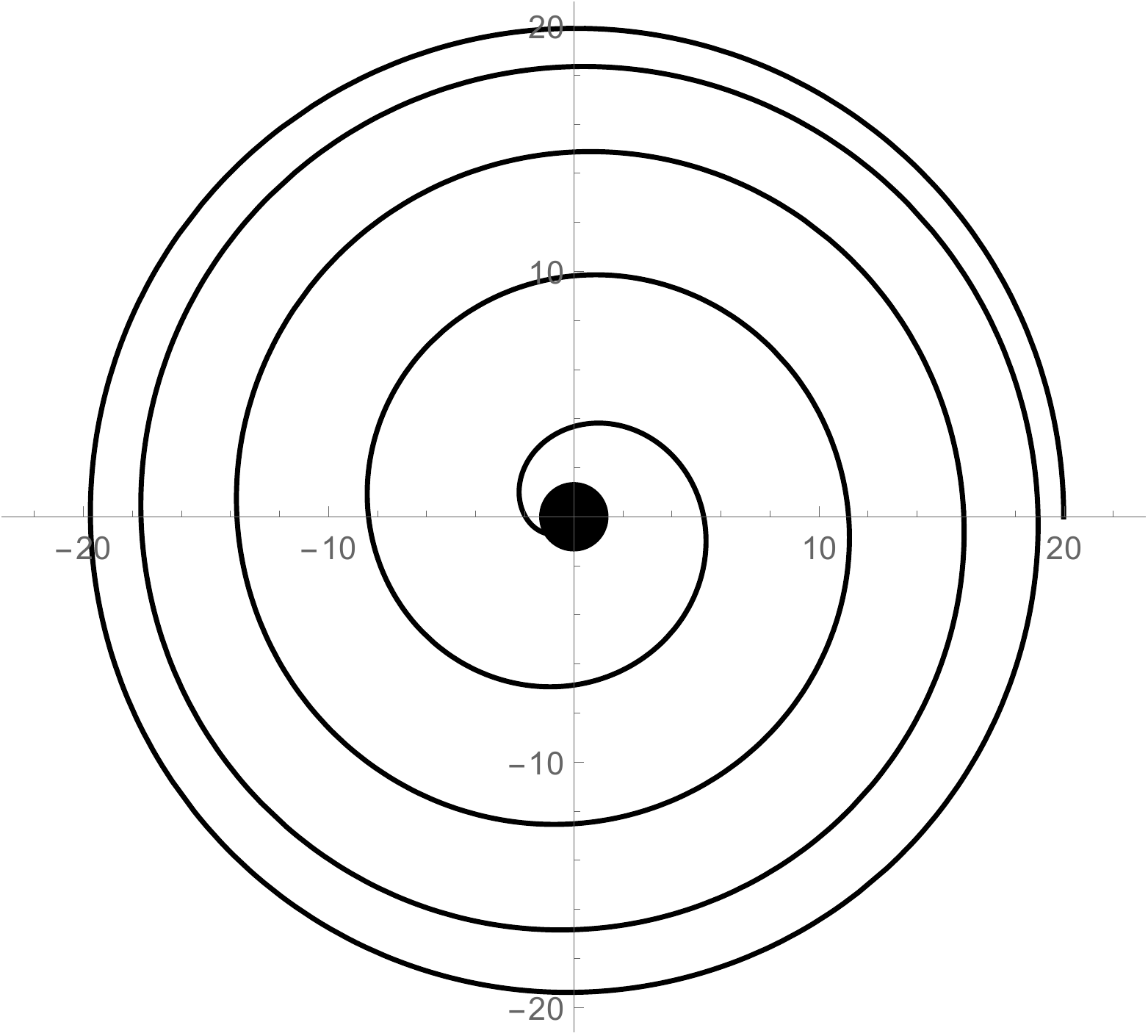}
\caption{Plot of an inspiral trajectory satisfying Eqs.~(\ref{inspiralrho}) with parameters $\rho_0=20$ and $L=4$.}
\label{inspiraltraj}
\end{figure}
We now turn our attention to a class of inspiral trajectories described by
\begin{align}
\label{inspiralrho}
\rho(\tau)&=\rho_{0}\cos(L\,\tau/\rho_{0})\nonumber\\
\phi(\tau)&=h\,\tau\nn\\
\mathbf{t}(\tau)&=\text{artanh}\left(\frac{\text{\ensuremath{\rho_{0}}}\sin\left(\frac{L\tau}{\text{\text{\ensuremath{\rho_{0}}}}}\right)}{\sqrt{\text{\ensuremath{\rho_{0}}}^{2}-1}}\right).
\end{align}
These trajectories have constant acceleration given by
\begin{equation}
\label{inspacc}
|a|=\frac{L^2\sqrt{\rho_0^2-1}}{\rho_0},
\end{equation}
where we're assuming the initial position is in the exterior, $\rho_0>1$. An example of a trajectory satisfying Eq.~(\ref{inspiralrho}) is given in Fig.~\ref{inspiraltraj}.

As in the circular case, we re-express the detection time in terms of the azimuthal angle $\phi$ since this gives a more intuitive sense of long detection times in terms of revolutions around the black hole. The detector will reach the black hole horizon after traversing an angle of
\begin{align}
\label{eq:phimax}
    \phi_{\rho=1}=\frac{\sqrt{L^{2}-1}}{L}\rho_{0}\arccos\left(1/\rho_{0}\right).
\end{align}
In an attempt to extract non-transient effects, we wish to examine cases where the azimuthal angle is large, or equivalently, where the detector has circumnavigated the black hole many times before plunging into it. We can see from Eq.~(\ref{eq:phimax}) that this requires a sufficiently large $\rho_{0}$ and an $L$ that is not close to one. In other words, we want the detector to begin its approach at a distance far from the black hole and to have sufficiently large acceleration so as to maximise detection time.

\begin{figure}[!htp]
\centering
\includegraphics[scale=.35]{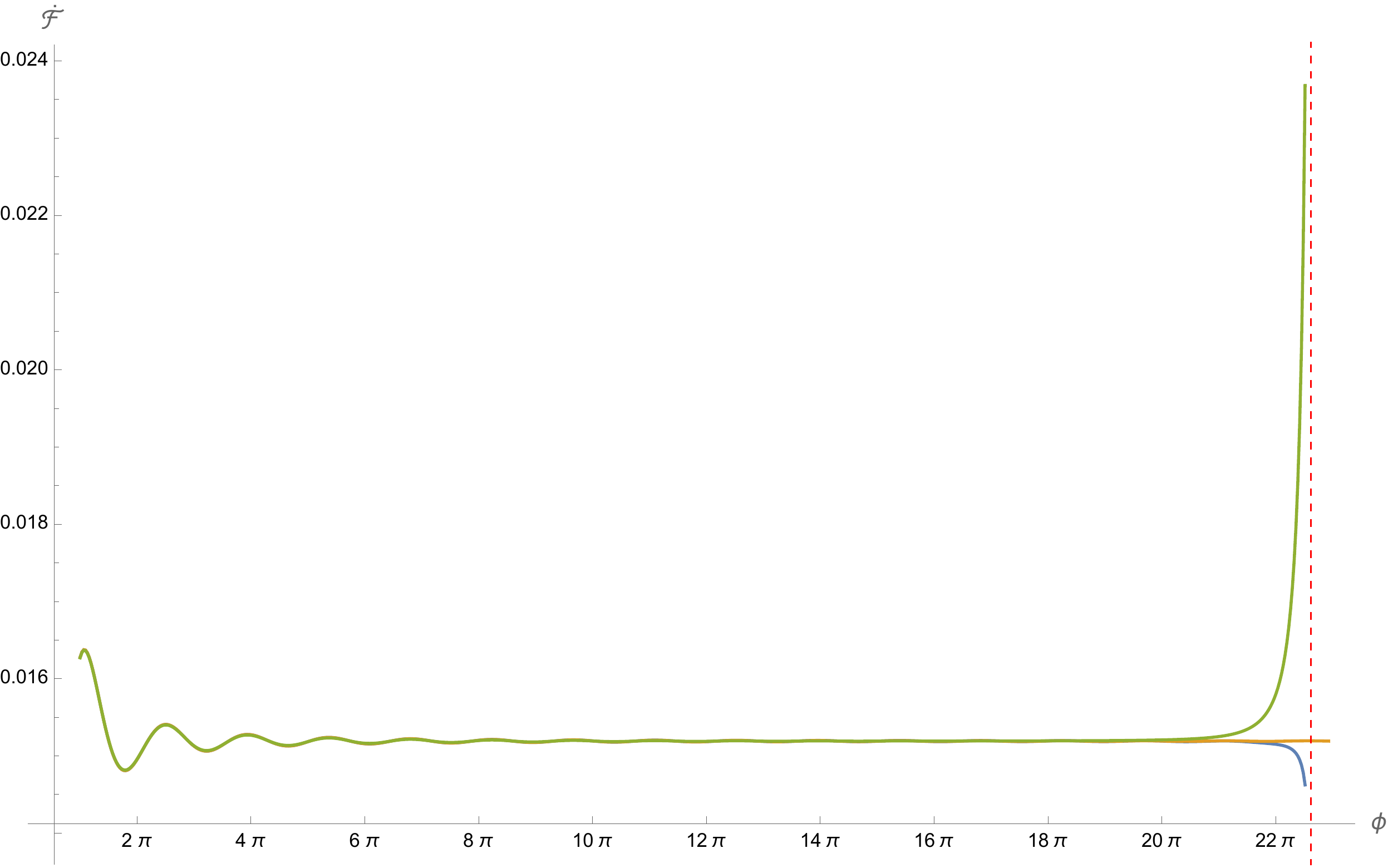}
\caption{Plot of the transition rate for the detector on an inspiral trajectory coupled to a quantum field in a thermal state at various temperatures. The states considered have temperatures $T=0.01$ (blue), $T=T_{\textrm{H}}$ (orange) and $T=0.6$ (green). The angular momentum is $L=2.5$ and energy gap is $\omega=3.2$ for each graph.}
\label{fig:inspiralstates}
\end{figure}

\begin{figure}[!htp]
\centering
\includegraphics[scale=.35]{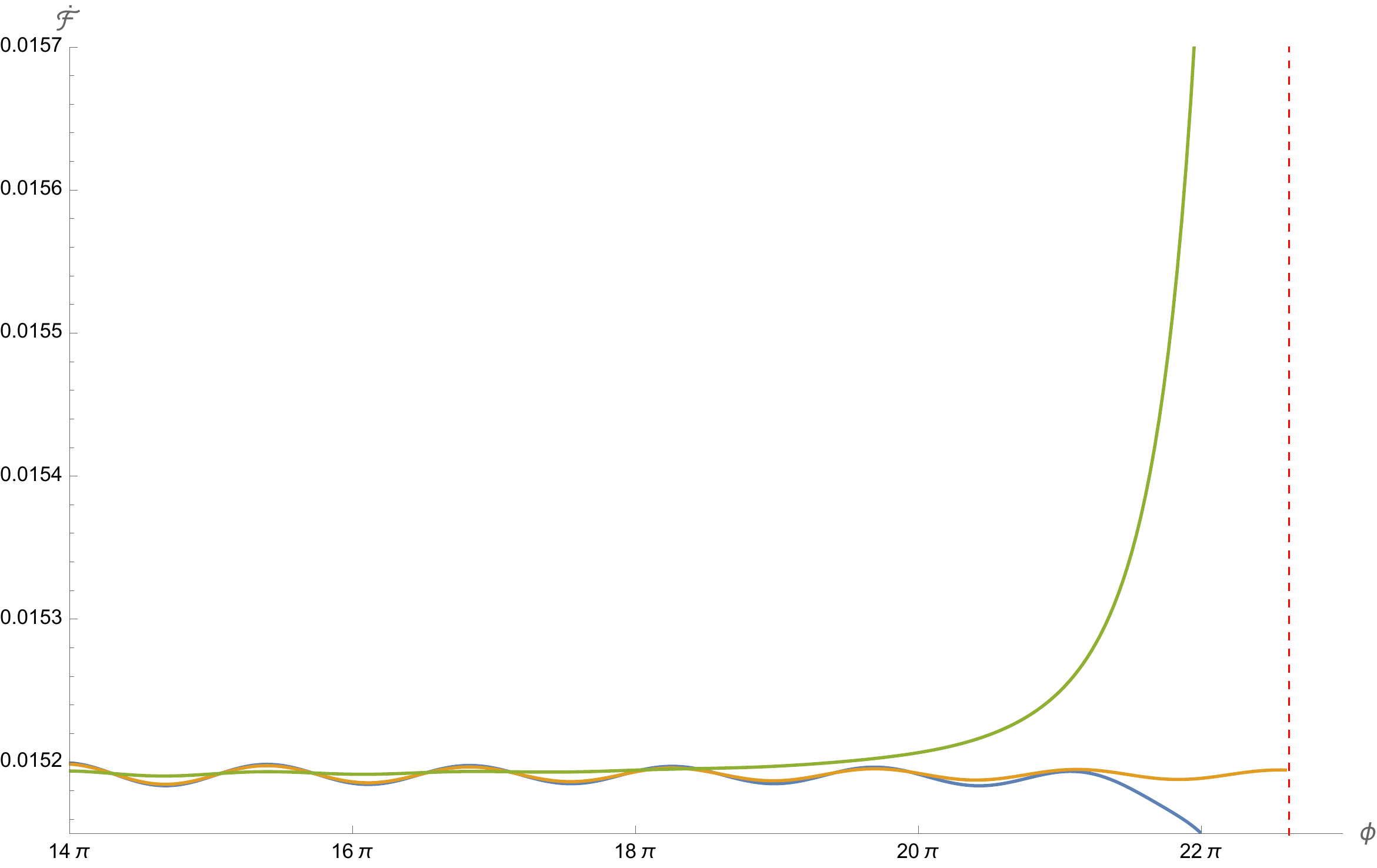}
\caption{Plot of the transition rate for the detector on an inspiral trajectory coupled to a quantum field in a thermal state at various temperatures. The states considered have temperatures $T=0.01$ (blue), $T=T_{\textrm{H}}$ (orange) and $T=0.6$ (green). The angular momentum is $L=2.5$ and energy gap is $\omega=3.2$ for each graph.}
\label{fig:inspiralstateszoom}
\end{figure}

The first figure we will examine is a comparison between the quantum states. The transition rates as a function of detection time (rather azimuthal angle) has very similar profiles for the field in different quantum states except when the detector approaches the horizon. In Fig.~\ref{fig:inspiralstates}, we plot the detection rate for a detector coupled to a field in the Hartle-Hawking state as well as two other thermal states. The profile for a field in the Boulware state is very similar. These profiles only diverge significantly near the horizon indicated by the red dashed line. We see this more clearly in Fig.~\ref{fig:inspiralstateszoom} which shows the latter half of the inspiral. Here, we also see the expected behaviour that the Hartle-Hawking state is smooth across the horizon whereas the the other states diverge. What is interesting to note, however, is that the sign of the divergence for these states differs for those with $T<T_{\textrm{H}}$ compared with those for $T>T_{\textrm{H}}$. The latter states produce a transition rate which increases without bound while the former decreases without bound.

Another key feature of Fig.~\ref{fig:inspiralstateszoom} is that it shows that the transient oscillations have a very long memory, as we saw in the circular case. However, for fields with higher temperatures, these oscillations dampen quicker and become negligible before the horizon is reached. One would hope that the non-transient effects can be easily discerned in these cases. In general, one can identify three phases in the profiles of the transition rates: (i) large transient oscillations; (ii) an approximately constant phase; and (iii) a near-horizon phase which is highly dependent on the choice of quantum state. For smaller temperatures, the transient oscillations are still important in the approximately constant phase. This fact is relevant when examining regions where the transition rate is anti-correlated with the temperature.

\begin{figure}[!htp]
\centering
\includegraphics[scale=.46]{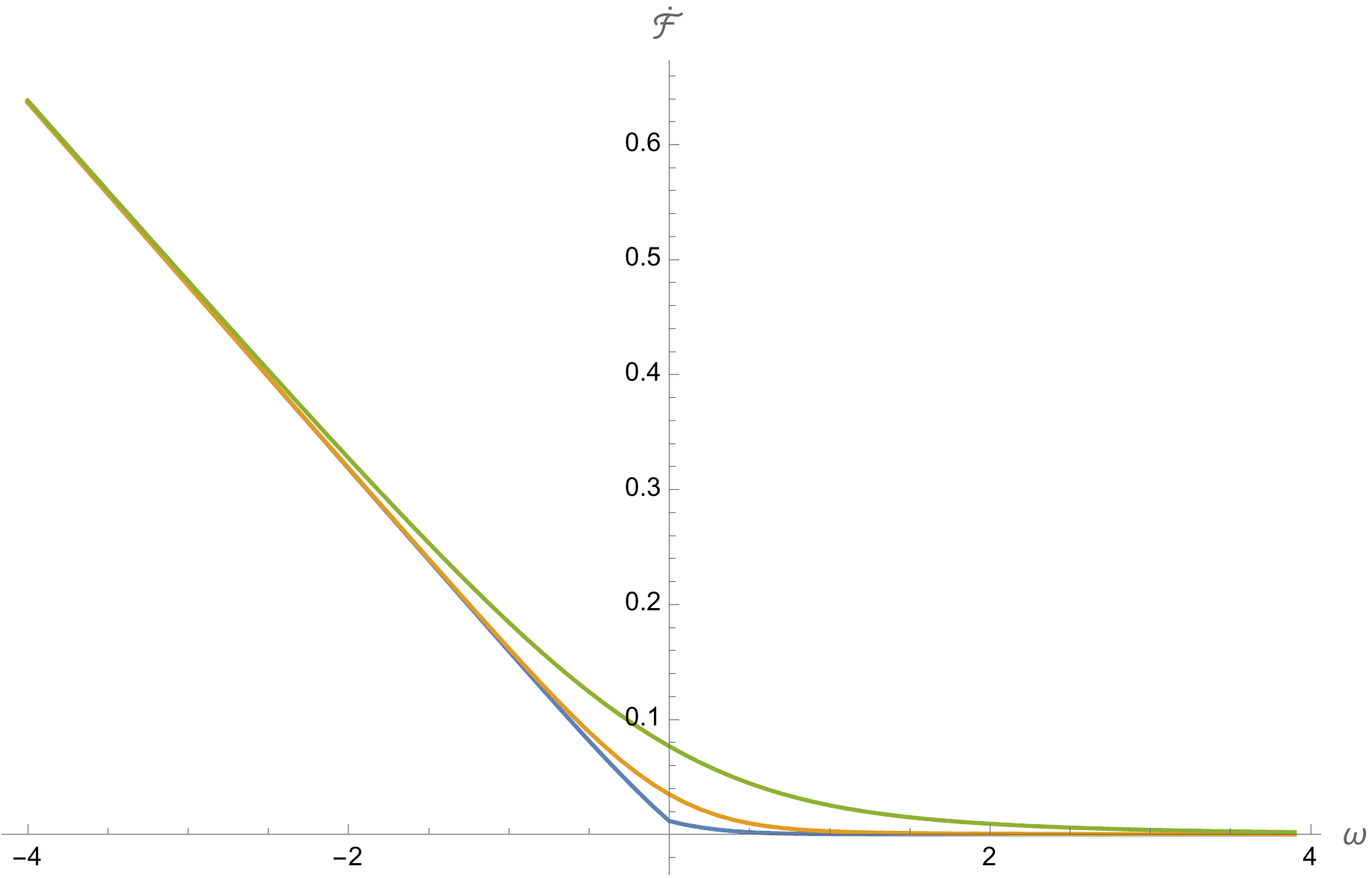}
\caption{Plot of the transition rate as a function of energy gap $\omega$ for the detector on an inspiral trajectory coupled to a quantum field in a thermal state at various temperatures. The states considered have temperatures $T=T_{\textrm{H}}=1/(2\pi)$ (blue), $T=2$ (orange) and $T=5$ (green). The angular momentum is $L=1.3$, the initial radius $\rho_{0}=50$ and the detection time corresponds to the time it takes for 7 revolutions of the black hole ($\phi=14 \pi$).}
\label{fig:inspiralomega}
\end{figure}

\begin{figure}[!htp]
\centering
\includegraphics[scale=.46]{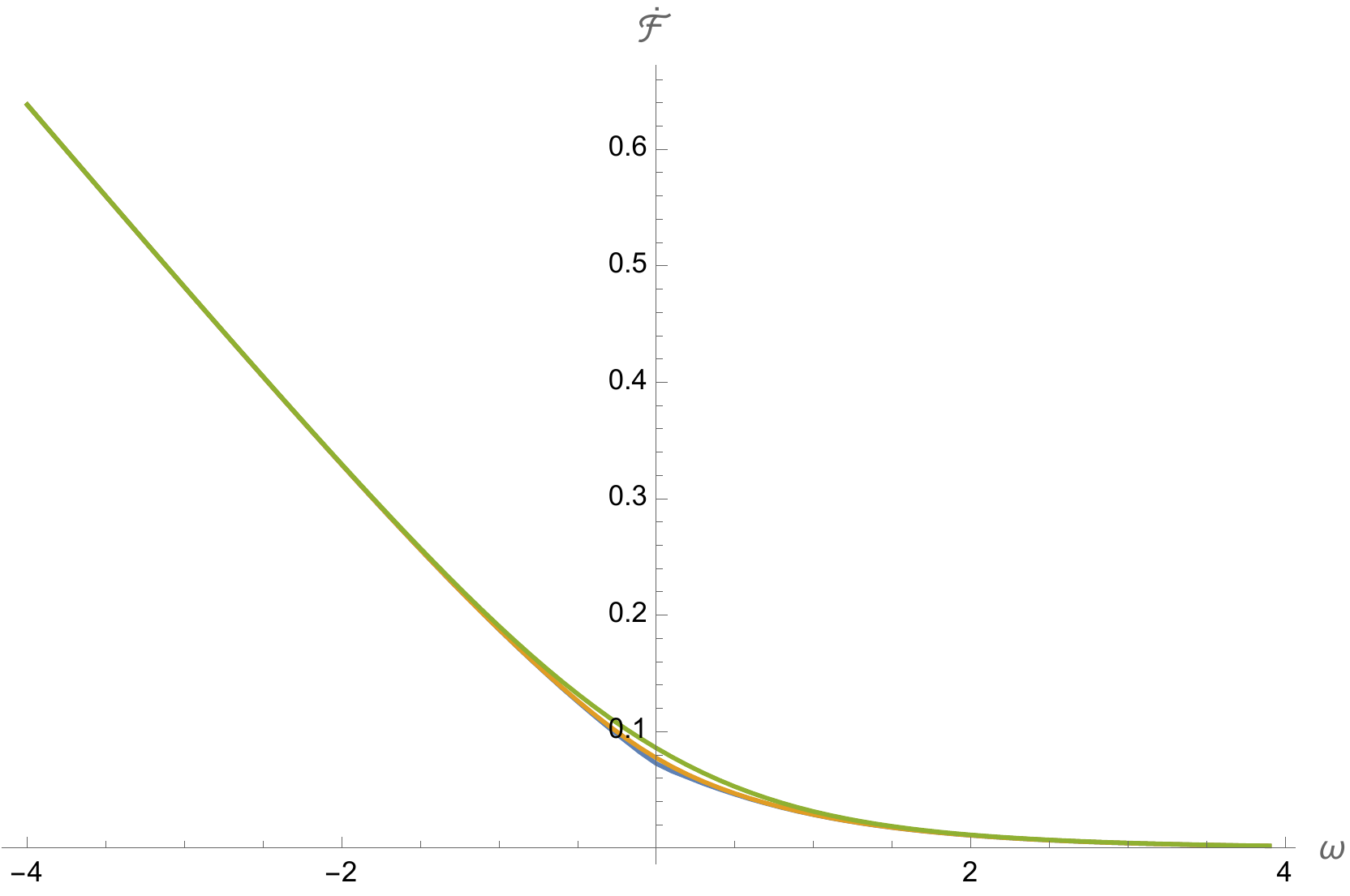}
\caption{Plot of the transition rate as a function of energy gap $\omega$ for the detector on an inspiral trajectory coupled to a quantum field in a thermal state at various temperatures. The states considered have temperatures $T=T_{\textrm{H}}=1/(2\pi)$ (blue), $T=2$ (orange) and $T=5$ (green). The angular momentum is $L=2.1$, the initial radius $\rho_{0}=50$ and the detection time corresponds to the time it takes for 7 revolutions of the black hole ($\phi=14 \pi$).}
\label{fig:inspiralomegalargeL}
\end{figure}

In Figs.~\ref{fig:inspiralomega}-\ref{fig:inspiralomegalargeL}, we plot the dependence of the transition rate for an inspiralling detector on the energy gap for a number of thermal states including the Hartle-Hawking state. We find profiles which are qualitatively very similar to those for the circular detector. In particular, we see that for smaller accelerations, the difference between the quantum states is accentuated, with lower temperature states displaying a sharp transition from negative to positive energy gaps (Fig.~\ref{fig:inspiralomega}). For larger accelerations, the difference between the temperature of the quantum field only appears to be relevant for very small magnitude energy gaps (see Fig.~\ref{fig:inspiralomegalargeL}).

\begin{figure}[!htp]
\centering
\includegraphics[scale=.48]{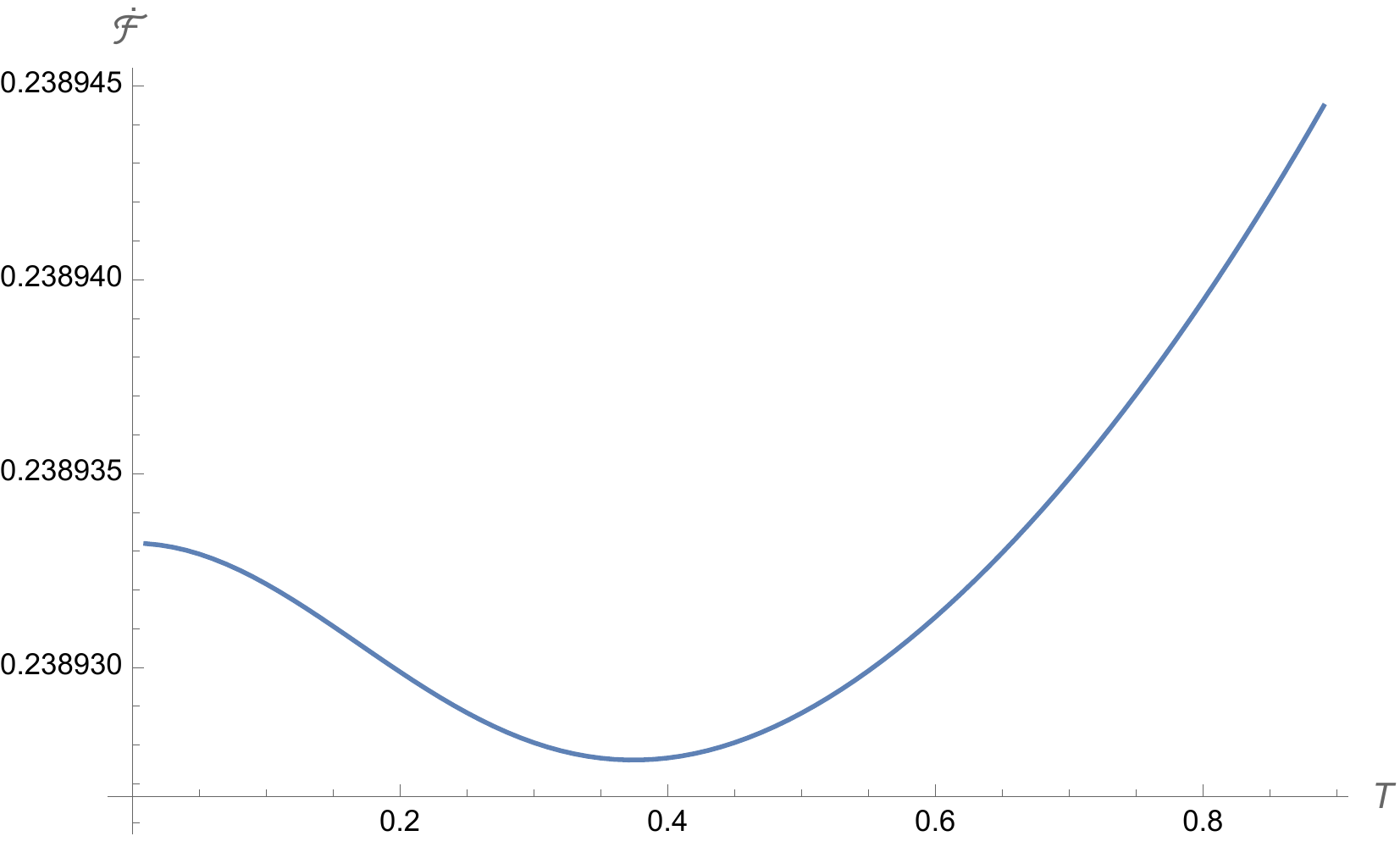}
\caption{Plot of the transition rate of an inspiralling detector as a function of temperature of the quantum KMS state. The angular momentum is $L=1.4$, the initial radius $\rho_{0}=50$ and the energy gap is $\omega=-1.5$.}
\label{fig:inspiralantiHawking}
\end{figure}

\begin{figure}[!htp]
\centering
\includegraphics[scale=.53]{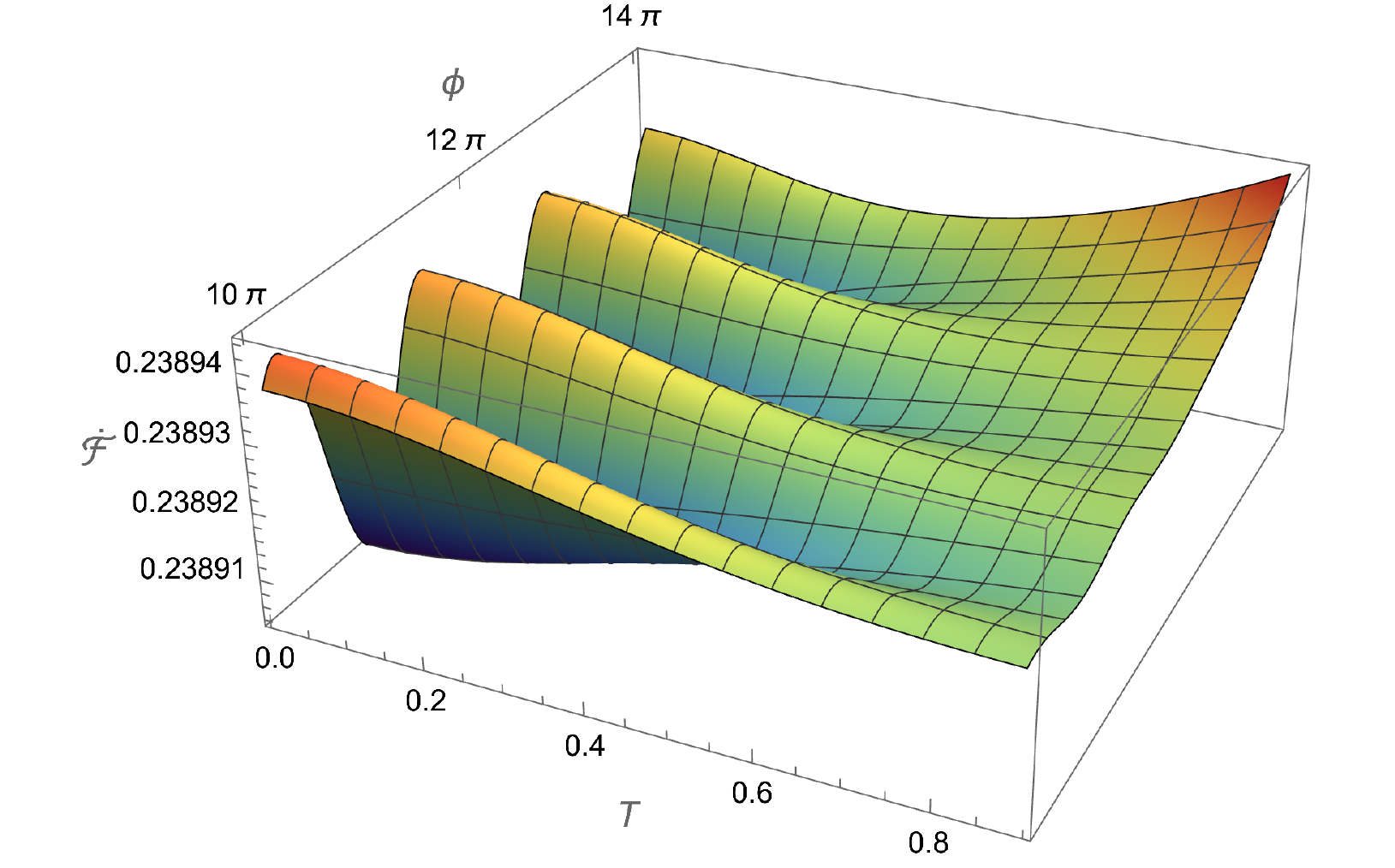}
\caption{3D plot of the transition rate of an inspiralling detector as a function of both temperature of the quantum state and detection time expressed in terms of the azimuthal angle. The angular momentum is $L=1.4$, the initial radius $\rho_{0}=50$ and the energy gap is $\omega=-1.5$.}
\label{fig:inspiralantiHawking3D}
\end{figure}

Turning now to the question of whether an effect analogous to the anti-Hawking effect is present in the parameter space. First we point out that the detector falls into the black hole in a finite proper time and so we cannot consider the limit of long interaction time. Hence, the detector and the field cannot be in thermal equilibrium. Indeed it appears that this is not even approximately the case for long-lasting inspirals. Nevertheless, much of the intuition remains the same as in the circular case. In particular, the transition rate is positively correlated with the field temperature in any region of the parameter space where transient effects are unimportant. Fig.~\ref{fig:inspiralantiHawking} shows an example of anti-correlation between the transition rate and the field temperature. This anti-correlation only appears to be present for small temperatures. While this plot is for a detection time corresponding to the time it takes the detector to orbit the black hole 7 times, we recall that the transient oscillations have a long memory for small temperatures. Hence, when the transient oscillations in the transition rate are no longer important for a detection time of the same duration, for example a KMS state with a higher temperature, then these regions of anti-correlation are no longer present. This can be seen in Fig.~\ref{fig:inspiralantiHawking3D} where the 3D plot shows that moving to hotter temperatures removes the oscillations present at lower temperatures, giving way to a monotonically increasing function of temperature. Our conclusion is therefore that these regions of anti-correlation between the transition rate and temperature are a transient effect and not something analogous to the anti-Hawking effect.

\begin{figure}[!htp]
\centering
\includegraphics[scale=.48]{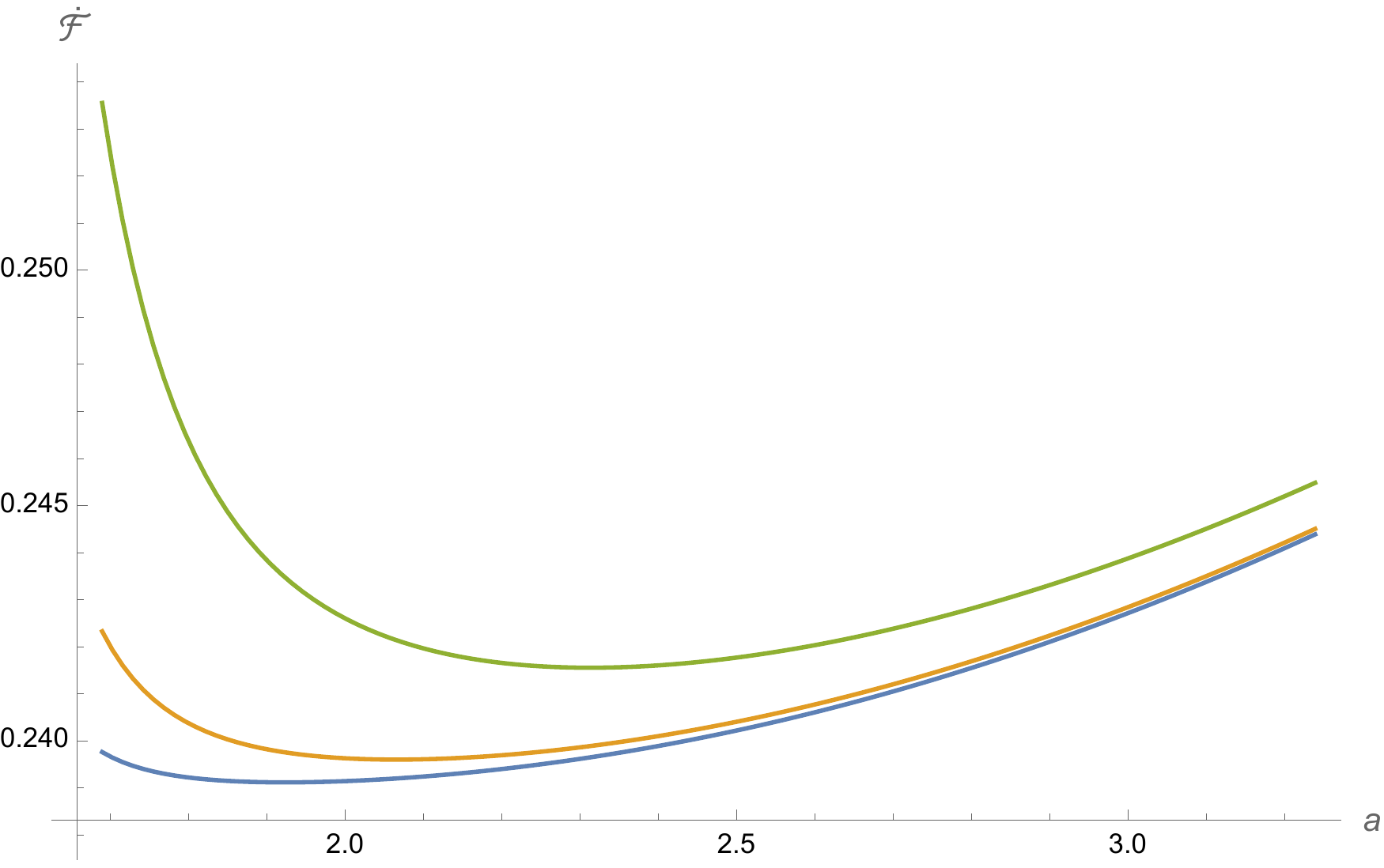}
\caption{Plot of the transition rate of an inspiralling detector as a function of the detector's acceleration for the detector coupled the a thermal scalar field at temperatures $T=2$ (blue), $T=3$ (orange) and $T=5$ (green). The initial radius is $\rho_{0}=50$, the energy gap is $\omega=-1.5$ and the detection time corresponds to the time it takes to orbit the black hole 7 times.}
\label{fig:inspiralantiUnruh}
\end{figure}

\begin{figure}[!htp]
\centering
\includegraphics[scale=.58]{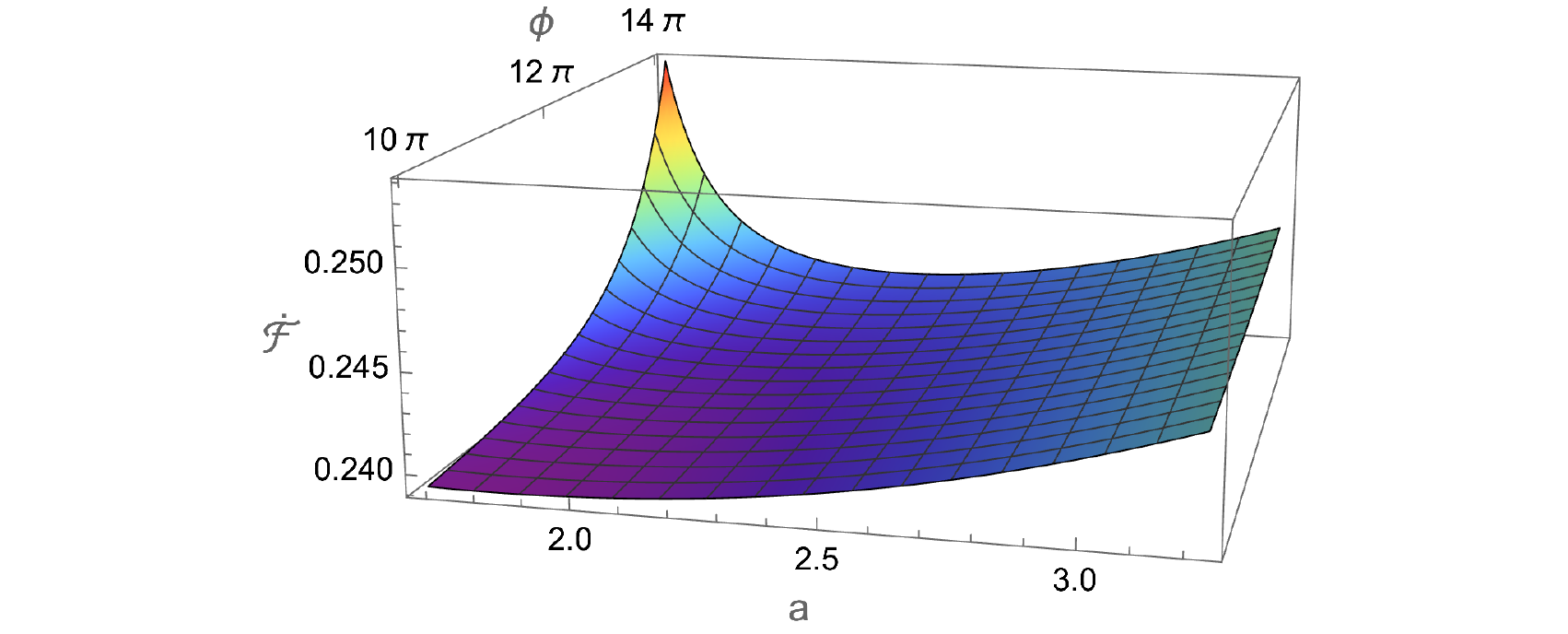}
\caption{3D plot of the transition rate of an inspiralling detector as a function of both the detector's acceleration and detection time (measured in terms of the azimuthal angle). The initial radius is $\rho_{0}=50$, the energy gap is $\omega=-1.5$ and the temperature of the state is $T=5.1$.}
\label{fig:inspiralantiUnruh3D}
\end{figure}

Turning our attention now to the relationship between the transition rate and the detector acceleration. Again, it is not difficult to find regions of the parameter space where the transition rate is anti-correlated with the acceleration. However, in this case, the evidence suggests the effect is analogous to the anti-Unruh effect in the sense that it does not appear to be attributable to transience. First, the anti-correlation effect is accentuated for states with hotter temperatures, as in Fig.~\ref{fig:inspiralantiUnruh}. If the effect was a transient one, we would expect the opposite since we know that the transient oscillations are longer lived for small temperatures. Second, the anti-correlation is present only for longer timescales, as seen in Fig.~\ref{fig:inspiralantiUnruh3D}. Hence, notwithstanding the fact that the detector is not in thermal equilibrium, we conclude that this anti-correlation between the transition rate and the inspiralling detector's acceleration is analogous to the anti-Unruh effect in the sense that it is an anti-correlation that is not associated with sharply switching on the detector.

\section{Summary}
\label{sec:conclusions}
In this paper, we have examined how an Unruh-DeWitt detector, modeled as a simple two-state quantum mechanical system coupled to a massless quantum scalar field, responds near an extremal black hole. In the limit where the interaction between the detector and the field is switched on and off instantaneously, the transition probability for the detector to be measured in its final state is ill-defined. However, the transition rate is finite in this limit so we choose to focus on this quantity. In either case, the computation is afflicted by some subtle technical challenges in black hole spacetimes associated with the fact that the Wightman function is not usually known in closed form. We circumvent these difficulties here by considering a certain near-horizon limit of the extremal charged black hole which yields a spacetime with enhanced symmetry; in fact the result is a simple direct product spacetime known as Bertotti-Robinson. We give closed-form representations for the Wightman function for a massless scalar field in several quantum states, assuming the field satisfies Dirichlet boundary conditions. This renders the computation of the transition rate quite straightforward in this limit facilitating an exploration of the complete parameter space. This is a huge advantage since the parameter space is very large. In general, the transition rate depends sensitively on the detector's trajectory, the energy gap between the detector's states, the detection time and the state of the quantum field. It would be a significant numerical undertaking to probe a large patch of this parameter space in the case where the Wightman function is known only in terms of its normal modes.

Equipped with closed-form representations, we considered the response of a detector moving on several different trajectories, both geodesic and accelerating. For all geodesic trajectories, the time until the detector crosses the horizon is very short since we are in a near-horizon throat. The profiles of the transition rate are therefore dominated by transient effects, even when the initial radius is large. We look only at a detector in radial free-fall; the results for other geodesics are broadly similar. We find, as expected, that the transition rate of a detector coupled to a field in the Hartle-Hawking state is regular across the horizon $\rho=1$, while it diverges for the Boulware vacuum and for any thermal state with temperature $T\ne T_{\textrm{H}}$. We find damped oscillations in the transition rate as the detector approaches the horizon with the frequency of oscillation proportional to the magnitude of the energy gap. One would expect instead to observe a transition rate that increases monotonically as the horizon is approached (since the red-shifted Hawking temperature increases) and we deduce that the observed behaviour to the contrary is a transient effect. To distil the non-transient effects, we consider fields in a thermal state with a large temperature. In this case the transient oscillations are subdominant compared with the contribution to the transition rate from the interaction with a field in the hot thermal state and the expected monotonic behaviour emerges with the transition rate increasing as the horizon is approached. We expect these profiles to be indicative of those for plunges in more generic black hole spacetimes.

The more interesting cases are the accelerated detectors. We considered first a static detector (which is accelerated since the spacetime is not ultrastatic) coupled to a field in both the Boulware vacuum and thermal states. For a field in the Boulware vacuum, the transition rate is identical to that of an inertial detector in Rindler spacetime coupled to a scalar field in the Minkowski vacuum. For thermal states (including the Hartle-Hawking state) in the limit of infinite detection time, we get a precisely Plankian distribution for the transition rate at the local KMS temperature $T/\sqrt{\rho_{0}^{2}-1}$, where $T$ is the field temperature and $\rho_{0}$ the position of the static detector. For these thermal states, we expect the transition rate to increase with increasing local temperature, similarly we expect the detector's temperature to increase with increasing KMS temperature. Violation of this expectation in black hole spacetimes is evidence of the anti-Hawking effect. We examined how the transition rate depends on increasing local temperature at a fixed radius by varying the temperature of the field's quantum state. While we do indeed find regions of the parameter space where the transition rate decreases as local temperature increases, a closer examination revealed that this effect was an artefact of switching the detector on sharply, that is a transient effect. After a detection time comparable with the thermalization timescale, the transition rate is an increasing function of local temperature. Hence we found no evidence of the anti-Hawking effect in this case. 

Next. we considered accelerated circular trajectories. The transition rate for circular trajectories is dominated by transient oscillations for the first few orbits, but this gives way to an approximately constant transition rate for longer times. We found that, for sufficiently long detection times, the temperature estimator of the detector defined in Eq.~(\ref{TEDR}) increases very slowly and appears to asymptote to a constant for large energy gap, suggestive of a detector which thermalizes at this limit. We show analytically that this is indeed the case and that the temperature  that  the  detector  thermalizes  to  is  always hotter  than  the  Doppler-shifted local field temperature. Moreover, for field temperatures below a certain critical temperature, the detector's temperature is insensitive to the ambient field temperature and sees only the temperature arising from its acceleration. For temperatures above this critical temperature, the detector's temperature sees both the ambient field temperature and the contribution from its acceleration. A similar conclusion was drawn for circular geodesic detectors in Schwarzschild \cite{HodgkinsonLoukoOttewill}. 

We also conducted a detailed search for anti-correlations in the parameter space of the circular detector's transition rate. We found that it is not difficult to find regions of parameter space where the transition rate is anti-correlated with either the field temperature or the detector's acceleration, but in either case this only appears to occur for shorter detection times. Hence we attribute this to transience and conclude that we find no evidence of the (weak) anti-Hawking or anti-Unruh effect. Similarly, we find no evidence of the strong anti-Hawking effect in that, after a detection time long enough for the transient oscillations to be negligible, we find that the detector's temperature is a monotonically increasing function of the field temperature.

Finally, we consider accelerated inspiralling detectors. The profiles of the transition rate tend to have three distinct phases as a function of detection time, a highly oscillatory transient phase, an approximately constant phase and a near-horizon phase. The near-horizon phase is regular for a detector coupled to a field in the Hartle-Hawking state, increases without bound for KMS states with $T>T_{\textrm{H}}$, and decreases without bound for $T<T_{\textrm{H}}$. For small temperatures, the transient oscillations are long-lived whereas for higher temperatures they are less important for the later stages of the inspiral. These orbits are non-stationary so we wouldn't expect the detector to thermalize, though one might expect that for long-lived inspirals, the detector reaches an approximate thermal equilibrium for large energy gaps analogous to the circular case. However, we find no evidence of this, not even in an approximate sense. Nevertheless, the relationship between the transition rate and field temperature obeys the same positive correlation as in the stationary cases, with the usual caveat that when transience effects are non-negligible, we find regions of anti-correlation. On the other hand, we do find an anti-correlation between the transition rate and the detector's acceleration that is not attributable to transience. While this effect is not quite the anti-Unruh effect since there is no sense in which we have thermal equilibrium, it is still a violation of what one might expect intuitively. It may be the case that the anti-Unruh effect is only present in lower spacetime dimensions, but here in this four-dimensional setting, we still observe a non-transient anti-correlation between the transition rate and the detector acceleration, analogous to the anti-Unruh effect.

There are a number of directions in which this work could be further developed. The most obvious is to examine how robust these results are when compared with the transition probability with a smooth switching function. This is still very doable owing to the fact that we have the propagator in closed form, albeit numerically more challenging. Indeed one could explore to what extent the profile of the switching function affects the results. Second, it would be interesting to compare some subset of these results with a numerical calculation of the transition rate near a Reissner-Nordstr{\"o}m black hole, not relying on the Bertotti-Robinson approximation. This is a more significant undertaking since the propagator would only be obtainable as an infinite mode-sum involving radial modes which need to be solved numerically. Nevertheless, there is much work in these directions in the context of computing vacuum polarization on black hole spacetimes, and these methods ought to be straightforwardly imported into the present context. Finally, in this paper we have focused only on the field satisfying Dirichlet boundary conditions. Previous work, \cite{Henderson2020, Campos2021, CamposDappiaggi2021, robbins2021antihawking} has indicated that boundary conditions play a very important role in whether or not anti-correlation effects are present. Moreover, other studies on the vacuum polarization \cite{MTW1, MTW2} suggest that the Dirichlet boundary conditions on quantum fields in asymptotically AdS spacetimes are special in that the vacuum polarization asymptotes to the same value on the spacetime boundary in all Robin boundary conditions except for the Dirichlet case. In other words, we should not expect the phenomenology of quantum fields satisfying the Dirichlet boundary conditions to be representative of more generic boundary conditions. We leave this for future work.

\acknowledgements{A.C. is supported by the Irish Research Council Postdoctoral Fellowship GOIPD/2019/536.} 
\bibliography{allcitationsBH}
\bibliographystyle{apsrev4-2}
\end{document}